\documentclass[12pt,twoside]{article}

\usepackage{epsf} 
\usepackage{colordvi}

\newcommand{ \be  }{\begin{equation}}
\newcommand{ \ee  }{\end{equation}} 
\newcommand{ \bea }{\begin{eqnarray}}
\newcommand{ \eea }{\end{eqnarray}}
\newcommand{ \beas }{\begin{eqnarray*}}
\newcommand{ \eeas }{\end{eqnarray*}}

\newcommand{ \NPB    }[3]{Nucl. Phys.      {\bf B#1} (#2) #3}
\newcommand{ \PLB    }[3]{Phys. Lett. B    {\bf #1}  (#2) #3}
\newcommand{ \PLBold }[3]{Phys. Lett.      {\bf #1B} (#2) #3}
\newcommand{ \PRD    }[3]{Phys. Rev. D     {\bf #1}  (#2) #3}

\newcommand{ \PRL    }[3]{Phys. Rev. Lett. {\bf #1}  (#2) #3}
\newcommand{ \PREP   }[3]{Phys. Rep.       {\bf #1}  (#2) #3}
\newcommand{ \ZPC    }[3]{Z. Phys. C       {\bf #1}  (#2) #3}

\newcommand{ \centeron }[2]{{\setbox0=\hbox{#1}\setbox1=\hbox{#2}\ifdim
                             \wd1>\wd0\kern.5\wd1\kern-.5\wd0\fi \copy0
                             \kern-.5\wd0\kern-.5\wd1\copy1\ifdim\wd0>\wd1
                             \kern.5\wd0\kern-.5\wd1\fi}}
\newcommand{ \ltap }{\;\centeron{\raise.35ex\hbox{$<$}}
                     {\lower.65ex\hbox{$\sim$}}\;}
\newcommand{ \gtap }{\;\centeron{\raise.35ex\hbox{$>$}}
                     {\lower.65ex\hbox{$\sim$}}\;}

\newcommand{ \slashchar }[1]{\setbox0=\hbox{$#1$}   
   \dimen0=\wd0                                     
   \setbox1=\hbox{/} \dimen1=\wd1                   
   \ifdim\dimen0>\dimen1                            
      \rlap{\hbox to \dimen0{\hfil/\hfil}}          
      #1                                            
   \else                                            
      \rlap{\hbox to \dimen1{\hfil$#1$\hfil}}       
      /                                             
   \fi}                                             %

\newcommand{ \NI         }{ \tilde{N}_1 }
\newcommand{ \NII        }{ \tilde{N}_2 }
\newcommand{ \NIII       }{ \tilde{N}_3 }
\newcommand{ \NIIII      }{ \tilde{N}_4 }

\newcommand{ \CI         }{ \tilde{C}_1 }
\newcommand{ \CII        }{ \tilde{C}_2 }

\def\G{ \tilde{G} }

\newcommand{ \eL         }{ {\tilde e}_{\scriptscriptstyle L} }
\newcommand{ \eR         }{ {\tilde e}_{\scriptscriptstyle R} }

\newcommand{ \lR         }{ { \tilde \ell}_R }

\newcommand{ \muL        }{ { \tilde \mu}_L }
\newcommand{ \muR        }{ { \tilde \mu}_R }
\newcommand{ \tauL       }{ { \tilde \tau}_L }
\newcommand{ \tauR       }{ { \tilde \tau}_R }
\newcommand{ \tauu       }{ { \tilde \tau}_1 }
\newcommand{ \taud       }{ { \tilde \tau}_2 }

\newcommand{\sss}{\scriptscriptstyle}

\newcommand{\Z}{Z^{\sss 0}}

\newcommand{\epem}{e^{\sss +}e^{\sss -}}

\newcommand{\stopu}{\tilde{t}_1}
\newcommand{\stopd}{\tilde{t}_2}

\newcommand{\sthw}{\sin\theta_{\sss W}}
\newcommand{\cthw}{\cos\theta_{\sss W}}

\newcommand{ \N }[1]{ {\tilde N}_{#1} }

\begin{document}
\pagestyle{myheadings} 
\markboth
{\it S.~Ambrosanio, G.~A.~Blair / Measuring GMSB Parameters...}
{\it S.~Ambrosanio, G.~A.~Blair / Measuring GMSB Parameters...}

\begin{titlepage}
\thispagestyle{empty} 
\null~\vspace{-3.0cm} 
\begin{flushright} 
\null                   \hfill  CERN-TH/99-109 \\ 
hep-ph/9905403          \hfill  DESY 98-199    \\ 
\end{flushright} 
\hrule\hfill

\vspace{-0.1cm}

\hrule\hfill

\vspace{0.2cm} 
                                        
\begin{center}
\hbox{\BrickRed{\Large \bf 
Measuring Gauge-Mediated SuperSymmetry Breaking}}
\vspace{0.2cm} 
{\BrickRed{\Large \bf 
Parameters at a 500 GeV $\epem$ Linear Collider}}$^{\: \star}$ \\ 
\end{center}

\begin{center}
{\large {\bf Sandro~Ambrosanio}~$^{a,c}$ \ and \ 
        {\bf Grahame~A.~Blair}~$^{b,c}$} \\
~\\
$^a$ CERN -- {\it Theory Division}, \\
     CH-1211 Geneva 23, Switzerland (current address) \\
     e-mail: {\sl ambros@mail.cern.ch} \\
~\\
$^b$ {\it Royal Holloway and Bedford New College}, \\
      University of London, Egham Hill, Egham, \\
      Surrey TW20 0EX, U.K. (current address) \\
      e-mail: {\sl g.blair@rhbnc.ac.uk} \\
~\\
$^c$ {\it Deutsches Elektronen-Synchrotron} DESY, \\
      Notkestra\ss e 85, D-22603 Hamburg, Germany \\
\end{center}

\vspace*{\fill}  

\begin{center} 
{\bf Abstract} \\
\end{center}
{\small 
We consider the phenomenology of a class of gauge-mediated supersymmetry 
(SUSY) breaking (GMSB) models at a $\epem$ Linear Collider (LC) with 
$E_{\rm c.o.m.}$ up to 500 GeV. In particular, we refer to a high-luminosity 
(${\cal L} \sim 3 \times 10^{34}$ cm$^{-2}$ s$^{-1}$) machine, and use 
detailed simulation tools for a proposed detector. 
Among the GMSB-model building options, we define a simple framework and 
outline its predictions at the LC, under the assumption that no SUSY signal 
is detected at LEP or Tevatron. 
We assess the potential of the LC to distinguish between the various 
SUSY model options and to measure the underlying parameters with high 
precision, including for those scenarios where a clear SUSY signal would 
have already been detected at the LHC before starting the LC operations. 
Our focus is on the case where a neutralino ($\NI$) is the next-to-lightest 
SUSY particle (NLSP), for which we determine the relevant regions of the GMSB 
parameter space. 
Many observables are calculated and discussed, including production 
cross sections, NLSP decay widths, branching ratios and distributions, for 
dominant and rare channels. We sketch how to extract the messenger and 
electroweak scale model parameters from a spectrum measured via, e.g. 
threshold-scanning techniques.
Several experimental methods to measure the NLSP mass and lifetime are 
proposed and simulated in detail. 
We show that these methods can cover most of the lifetime range allowed by 
perturbativity requirements and suggested by cosmology in GMSB models. 
Also, they are relevant for any general low-energy SUSY breaking scenario.  
Values of $c\tau_{\NI}$ as short as 10's of $\mu$m and as long as 
10's of m can be measured with errors at the level of 10\% or better 
after one year of LC running with high luminosity.   
We discuss how to determine a narrow range ($\ltap 5\%$) for the fundamental 
SUSY breaking scale $\sqrt{F}$, based on the measured $m_{\NI}$, 
$c\tau_{\NI}$. Finally, we suggest how to optimise the LC detector 
performance for this purpose. 
}

\vspace*{\fill}  

\noindent 
\parbox{0.4\textwidth}{\hrule\hfill} \\ 
{\small 
$^\star$\ To be published in  
{\it The European Physical Journal C}
}  

\end{titlepage}

\setcounter{page}{0} 

\thispagestyle{empty} 
~\\ 
\newpage 
\thispagestyle{plain} 

\section{Introduction}
\label{sec:intro}
\noindent 
If the world is supersymmetric at short distances, then the gauge hierarchy
problem can be naturally solved. The most compelling proof of this 
hypothesis would be direct detection of superpartners at colliders. 
This has not been achieved so far, which tells us that supersymmetry 
(SUSY) must be broken. In order for a SUSY theory to preserve 
its theoretically pleasant characteristics, supersymmetry breaking (SSB) can 
only occur in a ``soft'' way \cite{StevePrimer}. However, this constraint 
still allows a general phenomenological approach to SSB involving over a 
hundred new parameters in addition to the Standard Model (SM) ones. 
Strategies for searches at present and future colliders must then rely,
at least to start with, on theoretically well-motivated schemes for 
SSB, providing a more definite framework and living on a manageable 
parameter space. A related question is how this SSB is transmitted to 
the visible (light) sector of the theory, e.g. the particles of 
the Minimal SUSY extension of the SM (MSSM). Historically, the most
popular approach has been that SUSY is broken at very high energies
(HESB) of the order of the Planck mass or the scale of Grand-Unified 
Theories (GUT) and SSB is communicated to the MSSM sector through 
gravitational interactions.
Such an approach goes usually under the name of (minimal) Supergravity 
[(m)SUGRA] or, with some additional assumptions, Constrained MSSM (CMSSM) 
\cite{KKRW}. 
More recently, another equally attractive scenario has earned large consensus 
and recognition, both among theorists and experimentalists, the Low-Energy
Supersymmetry Breaking (LESB) option, and in particular, the Gauge-Mediated 
(GMSB) version of it \cite{GR-GMSB}. LESB, in itself, may already have 
striking phenomenological consequences, as it was shown in pioneering
works by Fayet \cite{Fayet}. Indeed, gravity enters the expression
for the gravitino mass, 
\be 
m_{3/2} = m_{\G} = \frac{F}{\sqrt{3}M'_P} \simeq 
\left(\frac{\sqrt{F}}{100 \; {\rm TeV}}\right)^2 2.37 \; {\rm eV},  
\label{eq:Gmass}
\ee
where $\sqrt{F}$ is the fundamental scale of SSB, 100 TeV is a typical 
value for it in LESB models, and $M'_P = 2.44 \times 10^{18}$ GeV is the 
reduced Planck mass. As a result, the gravitino is so light in LESB
models that it plays always the r\^ole of the lightest SUSY particle
(LSP) and can be treated as massless for all kinematics purposes at 
high energy colliders.
However, for $\sqrt{F} \ll M'_P$, the dominant gravitino
interactions come from its longitudinal, spin-1/2 components, namely
the goldstino components that the gravitino has acquired through 
the so-called SUSY-Higgs mechanism. Hence, gravity does not enter 
the strength of the gravitino couplings to matter, which in the relevant
approximation are proportional to the mass splitting between superpartner
masses and the ordinary SM particle masses and inversely proportional 
to $F$. The latter can be small enough to render the gravitino relevant
for collider phenomenology. (It should be noted here that a light gravitino
LSP can also be obtained within the framework of no-scale SUGRA models 
\cite{noscalemodels}.)

The phenomenological scenario in LESB with conserved $R$-parity 
(which we assume in the rest of the paper) can be summarised as follows: 
\begin{itemize} 
\item every produced SUSY particle has to decay to the $\G$, possibly 
through a cascade; 
\item since the goldstino interactions are still much weaker than 
the ordinary SM gauge and Yukawa interactions, every decay chain has 
to involve the next-to-lightest SUSY particle (NLSP), which in turn 
will finally decay to the gravitino; 
\item depending on $\sqrt{F}$, the production energy and details of the 
SUSY spectrum, the NLSP can decay close to the interaction point (i.p.), 
within or outside a collider detector, producing a plethora of new 
spectacular signatures. 
\end{itemize}
Among the possible mechanisms for transmitting LESB to the MSSM fields,
by far the most effective and theoretically satisfying is GMSB, where 
a so-called messenger sector is responsible for communication 
between the secluded sector where SSB takes place and the visible 
sector, via SM gauge interactions.  
Mainly motivated by a natural suppression of the SUSY contributions 
to flavour-changing neutral current (FCNC) and CP-violating processes,
such a scenario was first explored in various forms in several early
1980's works \cite{oldGMSB} and then recently revived in its
present version in the famous papers of Ref.~\cite{newGMSB}.
Remarkably, in addition to the appealing theoretical features, the 
minimal version of GMSB also provides a powerful tool for building very 
predictive models and calculating spectra from just a handful of parameters, 
as done e.g. in Refs.~\cite{GMSBmodels1,GMSBmodels2,AKM-LEP2}. 
An important boost to the popularity of LESB and GMSB models came a few years 
ago due to a possible explanation of the anomalous CDF 
$\epem\gamma\gamma\slashchar{E_T}$ event within this framework 
\cite{eegg-exp,eegg-th}. 
Today, such an explanation seems more unlikely, yet it worked fine in 
stimulating a considerable number of dedicated analyses and searches for  
GMSB-inspired new signals at LEP and Tevatron, which are of course of much 
broader interest \cite{eegg-exp,LEPsearches,D0searches}. 
Hence, it is now time to think about how similar searches could be 
pursued at next generation colliders and how the reach in the GMSB 
parameter space of such machines could be optimised. Some work in this respect
has already been carried out for the Tevatron Run II \cite{LESB-RunII} and 
for the LHC \cite{GMSB-LHC}. 
In this paper, we will be instead mainly concerned with GMSB phenomenology
at a first phase of operations of a $\epem$ Linear Collider (LC) with 
c.o.m. energy up to around 500 GeV, and will focus on the case where a 
neutralino is the NLSP.  
In particular, we will refer to a high-luminosity machine 
with ${\cal L} \sim 3 \times 10^{34}$ cm$^{-2}$ s$^{-1}$, such as being 
considered e.g. 
by the ECFA/DESY TESLA project, and the related proposed detector. 
Many of our results and experimental methods can be easily 
extended to more general LESB models and might even have an impact on other 
scenarios such as HESB models with $R$-parity violation, where delayed
NLSP or LSP decays can take place. 

The rest of this paper is organised as follows. In Sec.~\ref{sec:GMSBmodels},
we briefly describe our GMSB-model building framework and specify the region 
of the parameter space we are interested in here. In Sec.~\ref{sec:NLSPdecay}, 
we focus on the general phenomenology of models with a neutralino NLSP 
and discuss its possible (delayed) decays, including some new aspects 
of interest for the LC. In Sec.~\ref{sec:TESLA}, we introduce the 
main features of the proposed TESLA linear collider and give the expected 
machine parameters relevant to our study. In Sec.~\ref{sec:THRESCAN}, 
we discuss the general characteristics of the GMSB signal at the LC 
and show an example of how it is possible to extract a good amount of 
information about the GMSB parameters via a simple experimental technique in 
principle possible at such a machine. 
In Sec.~\ref{sec:BRAHMS}, we describe the relevant characteristics 
of the LC detector and the software we used for our simulations. 
In Sec.~\ref{sec:NLSPprop}, we discuss several methods for measuring the 
neutralino NLSP properties, and in particular its mass and lifetime, using 
different parts of the detector, and we show our results. Finally, in 
Sec.~\ref{sec:conc}, we draw our conclusions and comment on how the 
performance of the LC in measuring GMSB parameters depends on details of the 
machine and detector design. We also give a few suggestions to optimise such 
performance. 

\section{Models with Gauge-Mediated SUSY Breaking}
\label{sec:GMSBmodels}
\noindent
In addition to the automatic suppression of SUSY FCNC, GMSB models 
have many other interesting characteristics. For instance, the 
sparticles' masses have a transparent and common origin and all approximately 
scale with a single parameter $\Lambda$ which is the universal 
soft SUSY breaking scale for the visible sector. Also, the resulting 
spectrum is notably different from other SUSY scenarios. Further, 
it is possible to achieve radiative electroweak symmetry breaking 
(EWSB) nicely. There are however problems connected for instance with the lack
of a compelling dynamical mechanism for generating the SUSY parameter $\mu$,
but this is common to other SUSY frameworks. 

As far as GMSB-model building is concerned, we will follow closely 
the approach used in Ref.~\cite{AKM-LEP2} for LEP2 phenomenology, with some 
extensions of the parameter space to account for the wider kinematical 
reach of a LC. We will not repeat the technical details here, but in order to 
fix our framework and notations we remind that, after imposing EWSB, a 
minimal GMSB model can be constructed from the following parameters,
\be
M_{\rm mess}, \; N_{\rm mess}, \; \Lambda, \; \tan\beta, \; {\rm sign}(\mu),
\ee 
where $M_{\rm mess}$ is the overall messenger scale; $N_{\rm mess}$ is the 
so-called messenger index that parameterises the structure of the messenger
sector; $\Lambda$ is the universal soft SUSY breaking scale felt by the
low-energy sector; $\tan\beta$ is the ratio of the vacuum expectation 
values (VEVs) of the two Higgs doublets; sign($\mu$) is the ambiguity
left for the SUSY higgsino mass after EWSB conditions are imposed. 
The MSSM parameters and the sparticle spectrum are determined from 
renormalisation group equation evolution starting from boundary conditions 
at the $M_{\rm mess}$ scale, where 
$M_a = N_{\rm mess} \Lambda g(\Lambda/M_{\rm mess})\alpha_a$, ($a=1$, 2, 3) 
for the gaugino masses and 
$\tilde{m}^2 = 2 N_{\rm mess} \Lambda^2 f(\Lambda/M_{\rm mess}) 
\sum_a (\alpha_a/4\pi)^2 C_a$ for the scalar masses. Here $g$, $f$  
are the one, two-loop functions whose exact expression can be found e.g. 
in Ref.~\cite{AKM-LEP2}, and $C_a$ are the quadratic Casimir invariants 
for the scalar fields. The $A_f$ couplings are taken to be zero at the
messenger scale, since they are generated (first power) at the two-loop 
level. We use a phenomenological approach for $B\mu$, which is not 
assumed to vanish at $M_{\rm mess}$, but is instead determined together
with $|\mu|$ by requiring correct EWSB. 

For the purpose of exploring the GMSB parameter space of interest for 
the LC, we generated about 20,000 models, of which about 5,000 have  
a neutralino NLSP. The spectacular GMSB signatures, most of which
are free from SM-background, make it generally possible to exclude GMSB 
models at LEP2 with $m_{\rm NLSP} < \sqrt{s}/2 -$ few GeV \cite{AKM-LEP2}. 
We estimate that in a few years searches at LEP and Tevatron will only  
allow models where the whole MSSM spectrum is above about 100 GeV, at
least in most typical GMSB scenarios. 
(A remarkable exception is the case where the neutralino is the NLSP
and decays outside the detector, due to relatively large values of 
$\sqrt{F}$, but this is of no special interest here.) Hence, we limit
ourselves to models where 100 GeV $< m_{\rm NLSP} < 250$ GeV  = 
$\sqrt{s}_{\rm LC}/2$. As a result, the relevant range for $\Lambda$
is between about 60 TeV/$N_{\rm mess}$ and 200 TeV/$\sqrt{N_{\rm mess}}$.

For the sake of simplicity, at first we considered only models where 
$N_{\rm mess}$ is a positive integer between 1 and 10 (actually, we 
could not construct a model with $N_{\rm mess} > 8$ satisfying all constraints 
described above and below). As an example, if the messenger sector consists
of a ${\mbox{\boldmath$5$}} + ${\mbox{\boldmath$\overline{5}$}} 
of the global GUT group SU(5) $\supset$ 
SU(3)$_C \otimes$SU(2)$_L \otimes$U(1)$_Y$, then $N_{\rm mess} = 1$, 
while if it also includes a 
${\mbox{\boldmath$10$}} + ${\mbox{\boldmath$\overline{10}$}}, then 
$N_{\rm mess} = 1 + 3 = 4$.
Our messenger scale $M_{\rm mess}$ is bounded from below by several 
constraints. First, to avoid excessive fine-tuning of the messenger
masses, we impose $M_{\rm mess} > 1.01 \Lambda$. Second, we require 
that the mass of the lightest messenger scalar be much heavier than 
the MSSM particles (at least 10 TeV, that is 
$M_{\rm mess} > \frac{\Lambda + \sqrt{\Lambda^2+(20 \; {\rm TeV})^2}}{2}$).
Finally, to preserve  gauge-coupling unification, we also impose
$M_{\rm mess} > M_{\rm GUT} \; {\rm exp}(-125/N_{\rm mess})$. 
In this way, the lowest allowed value we obtained for $M_{\rm mess}$
is around 19 TeV. Further, to start with, we set a nominal upper bound on 
the messenger scale $M_{\rm mess} \ltap 10^{5} \Lambda \Longrightarrow 
M_{\rm mess} \ltap 2 \times 10^{10}$ GeV. We will see that this is overruled
by other constraints described below. As for $\tan\beta$, we require it 
to be larger than 1.2 (to avoid imminent bounds from SUSY Higgs searches
at LEP2 and non-perturbative blowing up of the top Yukawa coupling below 
the GUT scale) and we could not construct a coherent model with correct EWSB 
and $\tan\beta$ larger than about 55, with a mild dependence on $\Lambda$.  

In addition to these parameters, for each given GMSB model, a value
for the fundamental SUSY breaking scale $\sqrt{F}$ has to be specified
to complete the information needed for collider phenomenology.
The ratio $F_S/F = \Lambda M_{\rm mess} / F$, where $F_S$ is the scale
of SUSY breaking felt by the messenger particles, depends on details 
of the secluded sector and the communication between it and the messengers.
If this occurs, e.g., via a direct interaction and the goldstino 
superfield coincides with a single superfield entering the messenger 
superpotential (which we will assume in the following for simplicity), 
then one can infer from perturbativity arguments up to the GUT scale that 
the corresponding coupling has to be smaller than one \cite{AKM-LEP2}. 
In models with radiative secluded-messenger communication, the ratio can 
be even much smaller. In general, one can argue that
\be 
\sqrt{F} > \sqrt{\Lambda M_{\rm mess}} > \Lambda.
\ee
This allows the determination, for each given GMSB model, of a lower bound 
for the gravitino mass (and the NLSP lifetime, as we will see) and an upper 
bound for the strength of its interactions with matter $\sim 1/F$.
In our set of models of interest for the LC with 
100 GeV $< m_{\rm NLSP} < 250$ GeV, we find $m_{\G} \gtap 0.2$ eV and
$\sqrt{F} \gtap 30$ TeV. 
Unfortunately, there is no such compelling argument to put a strict upper
limit on $\sqrt{F}$ that can be of relevance to collider physics. 
In a simple cosmological scenario, one might invoke the argument that 
if the gravitino mass is too heavy 
($\gtap 1$ keV $\Longrightarrow \sqrt{F} \gtap$ few thousand TeV), 
then the gravitino relic density could over-close the universe \cite{Cosmo}.
This is of some use for our purposes and we will exploit this  
argument in the following. However, one has to keep in mind that 
a heavier gravitino can well be in agreement with cosmological scenarios
including an inflationary epoch. Barring the latter possibility, one
finds that the upper limit on the gravitino mass can only be satisfied
in our framework if $M_{\rm mess} \ltap 2 \times 10^8$ TeV, in models
of interest for the LC. This also implies that values of $N_{\rm mess}$ 
larger than 6 are highly disfavoured in this case. 

In this parameter space, we generated models by means of a private computer 
program called {\tt SUSYFIRE} \cite{SAsoft}, an updated, generalised and 
{\tt Fortran}-linked version of the program used in Ref.~\cite{AKM-LEP2}, 
which can produce minimal and non-minimal GMSB and SUGRA models.
For scanning, we used logarithmic steps for $\Lambda$, 
$\Lambda/M_{\rm mess}$ and $\tan\beta$. The program proceeds by iterating
the following: setting the masses and the gauge couplings at the weak scale;
evolving the RGE's to the messenger scale;
setting the messenger scale boundary conditions (see Eqs.~(23), (24) in 
Ref.~\cite{AKM-LEP2}) for the soft sparticle masses; evolving the RGE's 
back to the weak scale, taking care of decoupling each sparticle at
the proper threshold. We use two-loop RGE's for the gauge couplings, 
third generation Yukawa couplings and gaugino soft masses. The other 
RGE's are at the one-loop level. We require EWSB using the one-loop effective
potential approach (one-loop Higgs masses + consistent corrections from stops, 
sbottoms and staus) at the $\sqrt{m_{\stopu}m_{\stopd}}$ scale and we 
eliminate $|\mu|$ and $B\mu$ in favour of $\tan\beta$ and $M_Z$. 

The phenomenology of GMSB models is largely dependent on which particle is 
the NLSP or, better, on which sparticle(s) has (have) a large branching ratio 
(BR) for decaying to its SM partner and a gravitino.
Four main scenarios are possible: 

\begin{description} 

\item[Neutralino NLSP scenario:] Occurs whenever 
$m_{\NI} < (m_{\tauu} - m_{\tau})$. Here typically a decay of the $\NI$  
to $\G\gamma$ is the final step of decay chains following
any SUSY production process. As a consequence, the main inclusive signature 
at colliders is prompt or displaced photon pairs + X + missing energy. 
$\NI$ decays to $\G \Z$ and other minor channels are also important    
for this study, as we will see in the following. 
In the rest of this paper, we will focus on this possibility,
although we are well aware that the other scenarios are very 
relevant for LC phenomenology and we plan to devote further work
to them. 
A detailed discussion of the neutralino NLSP case will be carried out in 
Sec.~\ref{sec:NLSPdecay}. 

\item[Stau NLSP scenario:] Defined by 
$m_{\tauu} < {\rm Min}[m_{\NI}, m_{\lR}] - m_{\tau}$, 
features $\tauu \to \G \tau$ decays, producing $\tau$ pairs or 
charged semi-stable $\tauu$ tracks or decay kinks + X + missing energy.
Here $\ell$ stands for $e$ or $\mu$. 

\item[Slepton co-NLSP scenario:] When 
$m_{\lR} < {\rm Min}[m_{\NI}, m_{\tauu} + m_{\tau}]$, 
$\lR \to \G \ell$ decays are also open with large BR. In addition to the 
signatures of the stau NLSP scenario, one also gets $\ell^+\ell^-$
pairs or $\lR$ tracks or decay kinks. 

\item[Neutralino-stau co-NLSP scenario:] If 
$| m_{\tauu} - m_{\NI} | < m_{\tau}$ and $m_{\NI} < m_{\lR}$, 
both signatures of the neutralino NLSP and stau NLSP scenario are present
at the same time, since $\NI \leftrightarrow \tauu$ decays are not allowed
by phase space. 

\end{description}

Note that one always has $m_{\lR} > m_{\tauu}$ in the GMSB parameter
space we explored, and that the classification we give above is only 
valid in the limit $m_e$, $m_\mu \to 0$ and has to be intended as 
an indicative scheme.  Indeed, we did not take into account
very particular regions of the parameter space where, due to a fine-tuned
choice of $\sqrt{F}$ and the sparticle masses, one may achieve competition
between phase-space suppressed decay channels from one ordinary sparticle to 
another and sparticle decays to the gravitino \cite{AKM2}. Note also 
that we did not find in our sample any model with a sneutrino NLSP, 
since this is only possible in a corner of the parameter space where 
the lightest sparticle masses are well below 100 GeV.  

\begin{figure}
\centerline{
\epsfxsize=\textwidth
\epsffile{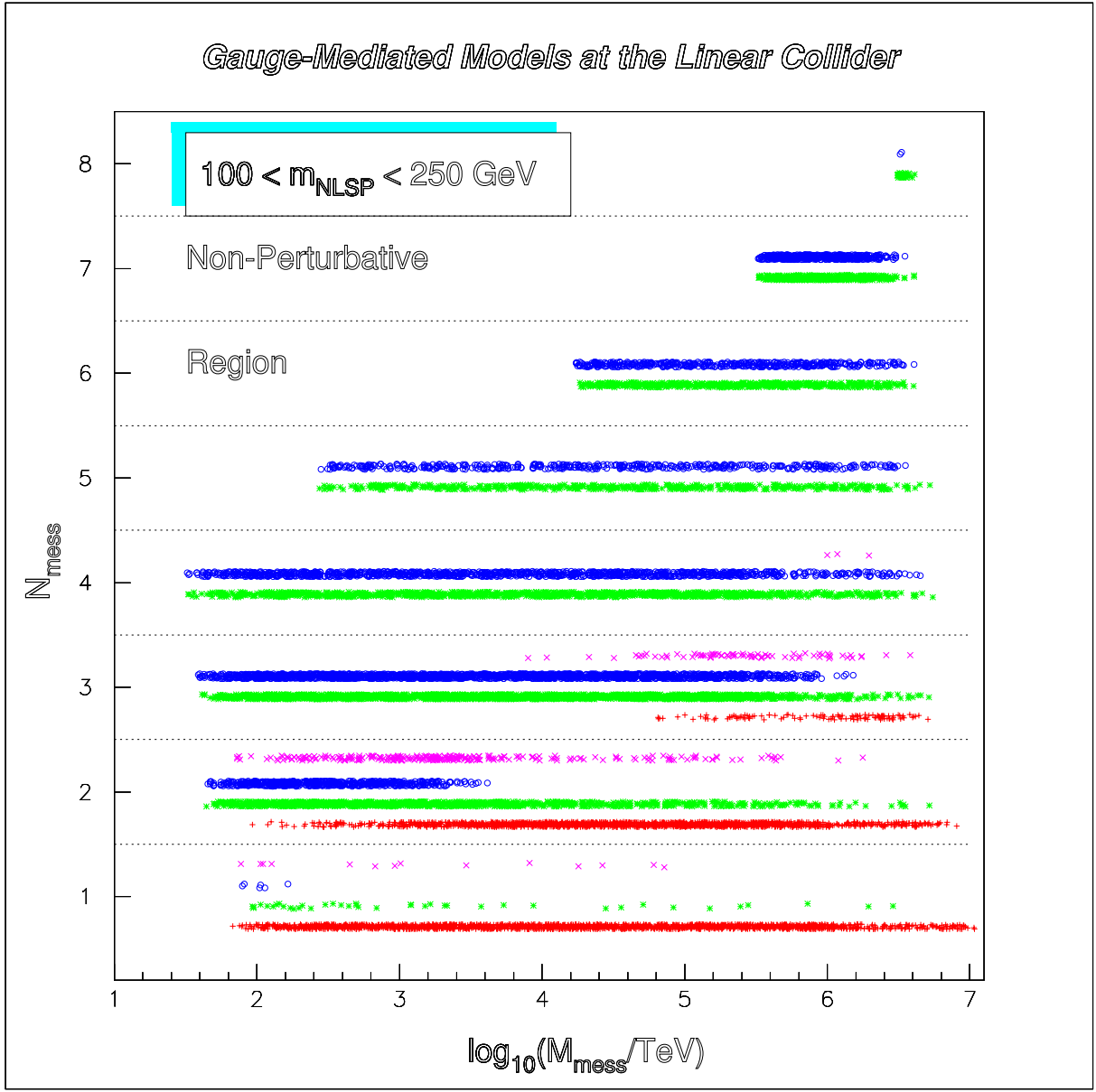}
}
\caption{\sl 
Scatter plot of GMSB models of interest for a 500 GeV $\epem$ linear collider 
(100 GeV $ < m_{\rm NLSP} < 250$ GeV) in the ($M_{\rm mess}$, $N_{\rm mess}$) 
plane. The region in the upper left corner is excluded by requiring 
perturbativity up to the GUT scale.
For each integer value of $N_{\rm mess}$ 
(fractions along the y axis have here no meaning and are for display 
purposes only), from bottom to top and in different grey scale, 
we display neutralino-NLSP models (+), stau-NLSP models (*), 
slepton co-NLSP models (o), and neutralino-stau-coNLSP models (x). 
}
\label{fig:NmessvsMmess}
\end{figure}

In Fig.~\ref{fig:NmessvsMmess}, 
we show where in the ($M_{\rm mess}, N_{\rm mess})$ 
plane the scenarios described above are of relevance, for a GMSB spectrum 
of interest for the LC. 
One can see that the neutralino-NLSP scenario can only occur (but does not 
need to) for $N_{\rm mess} =$ 1, 2 or 3.
For $N_{\rm mess} = 3$, it is also necessary to have a messenger scale
as high as $10^8$ GeV or more. Neutralino-stau co-NLSP models exist 
also for $N_{\rm mess} = 4$, but only for very high $M_{\rm mess}$.
Stau NLSP and slepton co-NLSP models are instead possible for all allowed 
values of $N_{\rm mess}$, but slepton co-NLSP models need 
$M_{\rm mess}$ to be lower than $10^6$ ($10^7$) GeV, if $N_{\rm mess} = 1$
(2). Perturbativity requirements up to the GUT scale start to be effective
in excluding relatively low values of $M_{\rm mess}$ for 
$N_{\rm mess} \ge 5$, while models with $N_{\rm mess} = 7$ or 8 are not
possible if one imposes the simple cosmology-inspired condition 
$m_{\G} \ltap 1$ keV.
 
Within a given scenario, the specific topology of the signatures is 
determined by the value of $\sqrt{F}$. We discuss this in detail in the 
next section, for the specific case where a neutralino is the NLSP. 
We now analyse a few important characteristics of neutralino NLSP 
models with 100 GeV $< m_{\NI} < 250$ GeV in our sample. 

\begin{figure}
\centerline{
\epsfxsize=0.5\textwidth
\epsffile{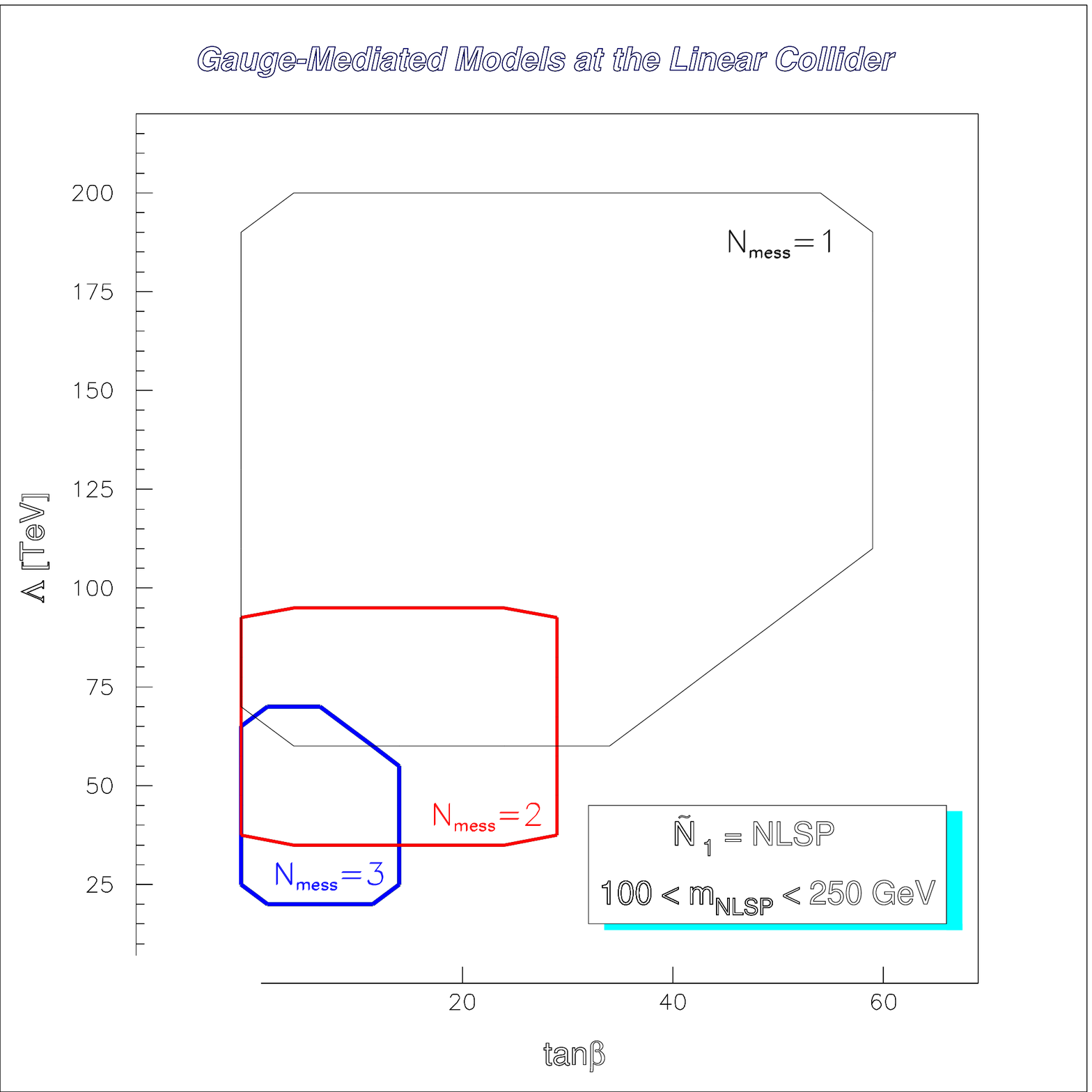}
}
\caption{\sl 
Allowed regions in the ($\tan\beta$, $\Lambda$) plane
for neutralino-NLSP models of interest for the 
linear collider, for different values of $N_{\rm mess}$
(necessary, but not sufficient conditions). 
}
\label{fig:Lambdavstanb}
\end{figure}

First, in Fig.~\ref{fig:Lambdavstanb}, we show the allowed regions 
in the ($\tan\beta$, $\Lambda$) plane for such a scenario to be realized. 
All the neutralino NLSP models we generated fall within the regions shown 
in Fig.~\ref{fig:Lambdavstanb} for a given value of $N_{\rm mess}$, but note that 
it is generally possible to construct models giving rise to different NLSP 
scenarios that also fall in the same regions of this plane. 
The regions in figure are sketched with a regular form to give a more 
intuitive feeling and are a bit wider than those
actually populated by the relevant models in our sample. 
Also note that, due to the stau L--R mixing producing lower mass eigenvalues 
for large $\tan\beta$, it is impossible to build a neutralino NLSP model 
of interest here for $\tan\beta \gtap 30$ (15), when the messenger sector
is not the simplest possible one, namely $N_{\rm mess} = 2$ (3). 

\begin{figure}
\centerline{
\epsfxsize=0.5\textwidth
\epsffile{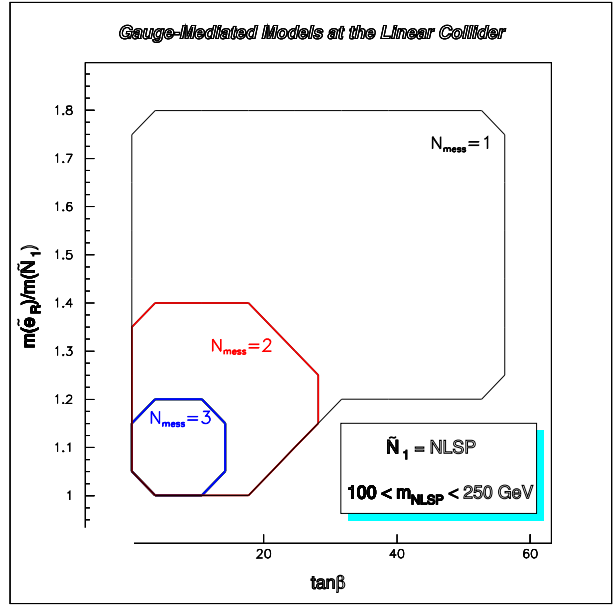}
}
\caption{\sl 
Bounds on the ratio between $R$-selectron and the $\NI$ masses 
in neutralino NLSP models of interest for the LC. Contours of the 
populated regions are shown for different values of $N_{\rm mess}$.
}
\label{fig:Rselvstanb}
\end{figure}

Second, it is important to determine how much heavier the other sparticles
can be compared to the $\NI$. This tells us what the likelihood is that once 
$\NI$-pairs are produced as an isolated signal at the LC, one can turn 
other SUSY processes on by just slightly raising the available c.o.m. energy. 
In all neutralino-NLSP models,
the next-to-NLSP particles are the $R$-sleptons, and in particular the 
$\tauu$ (which turns always out to be dominated -- 85\% or more -- by the 
$R$ component). The $\eR$ mass is particularly relevant, since it largely 
determines the $\epem\to\NI\NI$ cross section, 
together with the $\NI$ physical composition, due to the large contribution
from $t$-channel $\eR$-exchange graphs. Indeed, we will see that in GMSB 
models, the $\NI$ is dominated by the bino  component, more strongly coupled
to $R$-particles, and the $\eL$ is always much heavier than the $\eR$. 
The contribution from $t$-channel $\eL$-exchange graphs is hence generally
negligible. In Fig.~\ref{fig:Rselvstanb}, we show the ratio $m_{\eR}/m_{\NI}$ 
as a function of $\tan\beta$ and for different messenger multiplicity.
We chose $\tan\beta$ here as the independent variable mainly for visual 
purposes. The main information that can be extracted from 
Fig.~\ref{fig:Rselvstanb} is that there are no neutralino-NLSP models where 
the $\eR$ is more than 1.8 (1.4, 1.2) times heavier than the $\NI$ for 
$N_{\rm mess} = 1$ (2,3). 

\begin{figure}
\centerline{
\epsfxsize=0.7\textwidth
\epsffile{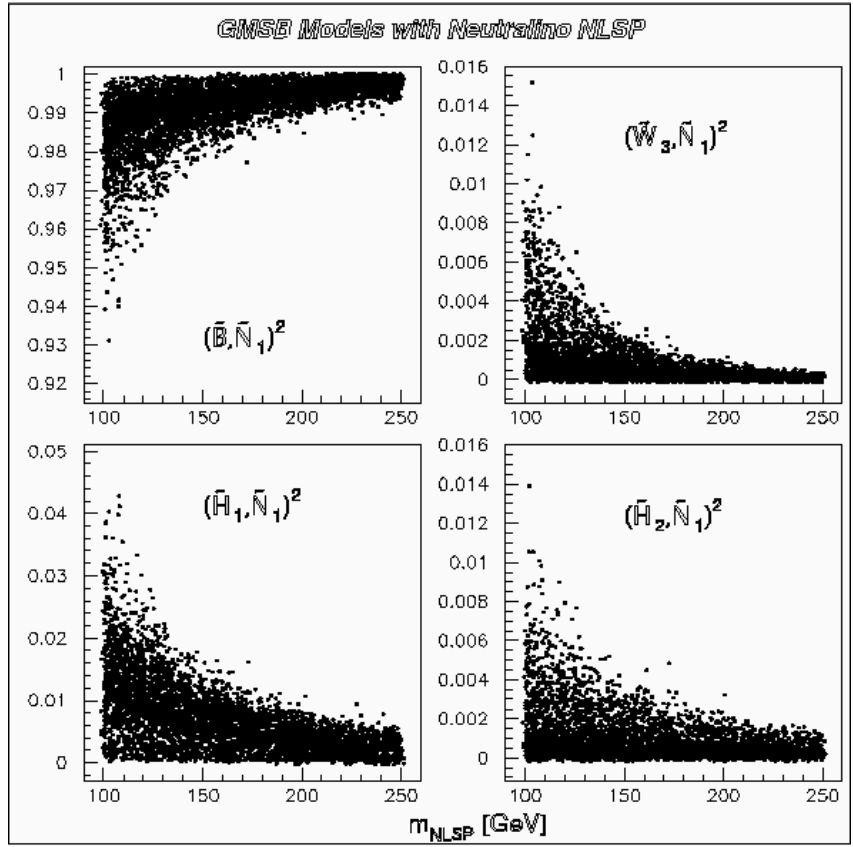}
}
\caption{\sl 
$\NI$ composition in GMSB models with neutralino NLSP of interest
for the LC. From top to bottom and left to right, we show the bino, wino, 
higgsino$_1$ and higgsino$_2$ components. Our basis for the neutralino 
mass matrix is the same as in Ref.~\protect\cite{Haber-Kane}.
}
\label{fig:N1comp} 
\end{figure}

Both in connection with the production cross section and with the
decay properties to be discussed in the next section, it is essential 
to specify the possible physical composition of the $\NI$ in GMSB
models with neutralino NLSP. Fig.~\ref{fig:N1comp} shows clearly that 
the bino component is always well above 90\%, while the wino component 
never reaches the 2\% level. The total higgsino component in the $\NI$ 
can only rarely reach the 5\% level, hence in some cases we will just
neglect it in the following and assume that the $\NI$ is a pure gaugino.
Note also that in the EW-diagonalised basis, the photino component 
$|\langle\NI|\tilde{\gamma}\rangle|^2$ of the NLSP is always included in 
the 0.60--0.85 range, while the zino component is in the 0.15--0.35 range.

The reader must be warned, however, that this is only true in the simple
GMSB framework we use in this paper. Here, we chose the phenomenological
approach where one just assumes the existence of the $\mu$ and $B$ terms 
at the messenger scale and determines them through EWSB conditions. As a 
consequence of this and the particular characteristics of the stop 
spectrum in GMSB, $|\mu|$ turns always out to be $\gtap 300 \; {\rm GeV}
\gg M_Z$ in neutralino GMSB models with 100 GeV $< m_{\NI} < 250$ GeV. 
Further, the relation $|\mu| \gtap 2 M_1$ always holds at the EW scale 
(here $M_1$ is the bino soft mass). Such a circumstance produces decoupling
between the gaugino and higgsino blocks in the neutralino and chargino
mass matrices and the characteristic relations $m_{\CI} \simeq m_{\NII} 
\simeq M_2 \simeq 2 m_{\NI} \simeq 2 M_1$ approximately hold, while the 
heavier neutralino and chargino mass eigenvalues are always of order $|\mu|$.

However, there are many possible sources of more complex scenarios. 
For instance, if one attempts to put together a radiative mechanism 
to generate $|\mu|$ and $B$, one may find extra corrections to the 
Higgs soft (mass)$^2$ parameters, which in turn can modify the value of 
$|\mu|$, often lowering it 
\cite{GMSBmodels1,GMSBmodels2,AKM-LEP2,higgsinoNLSP}.
Further, it is possible to build coherent GMSB models that have 
unequal messenger multiplicity relative to the three SM gauge groups. 
Models in this class exist where the higgsino component of the $\NI$
NLSP is large, with remarkable phenomenological consequences, both 
for the $\NI\NI$ production cross section and $\NI$ decay BR's to be 
discussed below. Many other variations in the messenger sector are 
possible \cite{Steve-gen}, but the associated phenomenology
is beyond the scope of this paper. In the following, we will always
assume that the higgsino components of the $\NI$ NLSP are small and
possibly negligible. 

We list here three reference GMSB models with neutralino NLSP 
in our sample that we will use in the following. 
We chose these particular models because they are qualitatively different 
for our experimental studies of Sec.~\ref{sec:NLSPprop}. They cover 
a good spectrum of possibilities and provide a feeling of the various 
problems that the experimenters could face if nature had chosen GMSB 
and the neutralino as the NLSP. 
 
Model \#~1 features are summarised in Tab.~\ref{tab:Model1}, where 
spectrum, production cross sections at a 500 GeV LC, and other relevant 
details such as sparticle physical composition and dominant decay channels 
are given (details about the NLSP decay are deferred to the next section). 
The precise values reported for masses and cross sections
(which include ISR and running $\alpha_{\rm em}$ effects),  
depend slightly on details of the spectrum calculation, higher-order 
corrections etc. They should be considered as an approximation at the level 
of a few percent. (Since here we are not particularly interested in the Higgs 
sector, we give only indicative information about it.) 

\begin{table}
\renewcommand{\arraystretch}{1.2}
\begin{center}
\begin{tabular}{|c||c|c|c|} \hline
\multicolumn{4}{|c|}{{\bf Model \#~1. INPUT:} 
$M_{\rm mess} = $ 161 TeV; $N_{\rm mess} = 1$; $\Lambda = 76$ TeV; 
$\tan\beta = 3.5$; $\mu > 0$} \\ \hline
Particle       & Mass              & Production @ 500 GeV LC
& Comments    \\ \hline \hline
$\G =$ LSP     & $\gtap 2.9$ eV    & indirect only 
& stable             \\ \hline
$\NI=$ NLSP    & 100.0 GeV         & $\sigma(\NI\NI)     = 256$ fb 
& $|\langle\NI|\tilde{B}\rangle|^2 = 0.97$; \ \  decays to $\G$  \\ \hline
$\tauu$        & 136.6 GeV         & $\sigma(\tauu\tauu) = 56.9$ fb
& $\simeq \tauR$; \ \  decays to $\NI$ \\ \hline
$\eR$, $\muR$  & 137.1 GeV         & 
                          $\sigma(\eR\eR, \muR\muR) = 274, 56.6 $ fb 
& decay to $\NI$     \\ \hline 
$\CI$          & 183.3 GeV         & $\sigma(\CI\CI) = 137$ fb        
& $|U_{11}|^2 = 0.87$; \ \ $|V_{11}|^2 = 0.94$  \\ \hline 
$\NII$         & 184.6 GeV         & 
                   $\sigma(\NI\NII,\NII\NII) = 39.1, 38.3 $ fb
& $|\langle\NII|\tilde{W}_3\rangle|^2 = 0.9$              \\ \hline
$\tilde{\nu}_{e,\mu,\tau}$ 
               & 264.4 GeV         & -- -- 
& $\tilde{\nu}_{\tau}$ slightly lighter                  \\ \hline
$\eL$, $\muL$  & 274.3 GeV         & $\sigma(\eL\eR)     = 101$ fb
&                   \\ \hline
$\taud$        & 274.5 GeV         & $\sigma(\tauu\taud) < 0.1 $ fb 
& $\simeq \tauL$                   \\ \hline\hline
$h^0$          & $\sim$ 105 GeV    & $\sigma(h Z) \sim 70$ fb 
&                   \\ \hline\hline
$\NIII$, $\NIIII$, $\CII$
               & $> 400$ GeV       & -- -- 
&                   \\ \hline
$H^0$, $A^0$, $H^{\pm}$
               & $> 500$ GeV       & -- -- 
&                   \\ \hline 
$\tilde{g}$    & $\sim 650$ GeV    & -- -- 
&                   \\ \hline
$\tilde{q}$    & $> 700$ GeV       & -- -- 
&                   \\ \hline
\end{tabular}
\end{center}
\caption{\sl
Input parameters, output spectrum and basic characteristics of a 
typical GMSB model with a 100 GeV neutralino NLSP: Model \#~1. 
}
\label{tab:Model1}
\end{table}

Model \#~1 is a model with a rather light spectrum, 
in particular the NLSP mass is right at our assumed LEP2/Tevatron bound 
of 100 GeV. The next-to-NLSP, the $R$-sleptons, are in the middle of their 
allowed mass range for such a NLSP mass. If this GMSB scenario were to be 
realized, the sparticles that could be produced with appreciable cross 
section at a 500 GeV LC would be gauginos, $R$-sleptons and $L$-selectron 
(in association with $\eR$). 
The total GMSB signal would be in this case quite ``generous.'' 
Heavy interacting sparticles are definitely out of reach, even for a possible
second phase of LC operations with c.o.m. energy at or slightly above 1 TeV. 
These large mass splittings are a well-known characteristics of GMSB models 
(cfr. e.g. Refs.~\cite{GR-GMSB,LCRep}) and are due to the fact that gaugino 
and scalar masses are proportional to the relevant gauge couplings. 
(The light Higgs is close to the edge of detectability at LEP2/Tevatron, 
depending on fine details and higher-order corrections that we do not take 
into account. In any case, models with a slightly heavier $h^0$ and no
significant differences in the other sectors can easily be constructed
with small changes to the input parameters. The rest of the Higgs sector 
is very heavy.) 
Notice that such a model would not be expected to produce a large signal 
at the LHC, due to the heaviness of gluino and squarks, hence a careful 
search and study at the LC would be most likely necessary, if not for initial 
SUSY discovery, then at least for a confirmation and for determining with 
good accuracy the source of the anomalous signal and the underlying SUSY-model
parameters. 

\begin{table}
\renewcommand{\arraystretch}{1.2}
\begin{center}
\begin{tabular}{|c||c|c|c|} \hline
\multicolumn{4}{|c|}{{\bf Model \#~2. INPUT:} 
$M_{\rm mess} = $ 309 TeV; $N_{\rm mess} = 1$; $\Lambda = 146$ TeV; 
$\tan\beta = 3.5$; $\mu > 0$} \\ \hline
Particle       & Mass              & Production @ 500 GeV LC
& Comments    \\ \hline \hline
$\G =$ LSP     & $\gtap 11$ eV     & indirect only 
& stable             \\ \hline
$\NI=$ NLSP    & 200.0 GeV         & $\sigma(\NI\NI)     = 42.3$ fb 
& $|\langle\NI|\tilde{B}\rangle|^2 = 0.99$; \ \  decays to $\G$  
                                                          \\ \hline\hline
$h^0$          & $\sim$ 115 GeV    & $\sigma(h Z) \sim 63$ fb 
&                   \\ \hline\hline
$\tauu$        & 256.4 GeV         & -- -- 
& $\simeq \tauR$; \ \  decays to $\NI$ \\ \hline
$\eR$, $\muR$  & 256.8 GeV         & -- -- 
& decay to $\NI$     \\ \hline 
$\CI$          & 374.1 GeV         & -- -- 
& $|U_{11}|^2 = 0.95$; \ \ $|V_{11}|^2 = 0.98$  \\ \hline 
$\NII$         & 374.4 GeV         & -- -- 
& $|\langle\NII|\tilde{W}_3\rangle|^2 = 0.96$             \\ \hline
$\tilde{\nu}_{e,\mu,\tau}$ 
               & 511.5 GeV         & -- -- 
& $\tilde{\nu}_{\tau}$ slightly lighter                   \\ \hline
$\eL$, $\muL$  & 516.7 GeV         & -- -- 
&                   \\ \hline
$\taud$        & 516.7 GeV         & -- -- 
& $\simeq \tauL$                  \\ \hline
$\NIII$, $\NIIII$, $\CII$
               & $> 700$ GeV       & -- -- 
&                   \\ \hline
$H^0$, $A^0$, $H^{\pm}$
               & $> 900$ GeV       & -- -- 
&                   \\ \hline 
$\tilde{g}$    & $\sim 1150$ GeV   & -- -- 
&                   \\ \hline
$\tilde{q}$    & $> 1300$ GeV      & -- -- 
&                   \\ \hline
\end{tabular}
\end{center}
\caption{\sl
Input parameters, output spectrum and basic characteristics of a 
typical GMSB model with a 200 GeV neutralino NLSP: Model \#~2. 
}
\label{tab:Model2}
\end{table}

Model \#~2 (see Tab.~\ref{tab:Model2}) is much more of an ``avaricious'' 
model, with a 200 GeV NLSP mass. 
It is obtained from Model \#~1 by just raising the input value of 
$\Lambda$, leaving the $M_{\rm mess}/\Lambda$ ratio and the other 
parameters untouched. The only GMSB signal present at a 500 GeV LC
would be NLSP pair production in this case. Note that changing the 
$\NI$ mass with respect to Model \#~1 does not only result in a drastic 
reduction of the cross section for $\NI\NI$ production, but also in an 
important change in the $\NI$ decay BR's, as described in detailed in 
Sec.~\ref{sec:NLSPdecay}.
As a consequence, even focussing on $\NI\NI$ production only, this model
would produce a considerably different signal at the LC compared to 
Model \#~1, both quantitatively and qualitatively.   
In this case, gluino and squarks are very heavy, possibly close to a
reasonable bound from naturalness arguments and the GMSB signal at the 
LHC would be rather scarce. 

\begin{table}
\renewcommand{\arraystretch}{1.2}
\begin{center}
\begin{tabular}{|c||c|c|c|} \hline
\multicolumn{4}{|c|}{{\bf Model \#~3. INPUT:} 
$M_{\rm mess} = $ 110 TeV; $N_{\rm mess} = 1$; $\Lambda = 100$ TeV; 
$\tan\beta = 3$; $\mu < 0$} \\ \hline
Particle       & Mass              & Production @ 500 GeV LC
& Comments    \\ \hline \hline
$\G =$ LSP     & $\gtap 2.6$ eV     & indirect only 
& stable             \\ \hline
$\NI=$ NLSP    & 165.0 GeV         & $\sigma(\NI\NI)     = 136$ fb 
& $|\langle\NI|\tilde{B}\rangle|^2 = 0.99$; \ \  decays to $\G$  
                                                          \\ \hline
$\tauu$        & 171.5 GeV         & $\sigma(\tauu\tauu) = 34.6$ fb 
& $\simeq \tauR$; \ \  decays to $\NI$ \\ \hline
$\eR$, $\muR$  & 171.8 GeV         & $\sigma(\eR\eR,\muR\muR) = 78.7,34.5$ fb 
& decay to $\NI$     \\ \hline 
$\NII$         & 315.0 GeV         & $\sigma(\NI\NII) = 1.07$ fb 
& $|\langle\NII|\tilde{W}_3\rangle|^2 = 0.96$             \\ \hline\hline
$h^0$          & $\sim$ 105 GeV    & $\sigma(h Z) \sim 70$ fb 
&                   \\ \hline\hline
$\CI$          & 315.1 GeV         & -- -- 
& $|U_{11}|^2 = 0.93$; \ \ $|V_{11}|^2 = 0.99$  \\ \hline 
$\tilde{\nu}_{e,\mu,\tau}$ 
               & 342.5 GeV         & -- -- 
& $\tilde{\nu}_{\tau}$ slightly lighter                   \\ \hline
$\eL$, $\muL$  & 349.8 GeV         & -- -- 
&                   \\ \hline
$\taud$        & 349.8 GeV         & -- -- 
& $\simeq \tauL$                  \\ \hline
$\NIII$, $\NIIII$, $\CII$
               & $> 500$ GeV       & -- -- 
&                   \\ \hline
$H^0$, $A^0$, $H^{\pm}$
               & $> 650$ GeV       & -- -- 
&                   \\ \hline 
$\tilde{g}$    & $\sim 950$ GeV    & -- -- 
&                   \\ \hline
$\tilde{q}$    & $> 950$ GeV       & -- -- 
&                   \\ \hline
\end{tabular}
\end{center}
\caption{\sl
Input parameters, output spectrum and basic characteristics of a 
typical GMSB model with a very-short lived neutralino NLSP and nearly 
degenerate light sparticles: Model \#~3. 
}
\label{tab:Model3}
\end{table}

Model \#~3 (see Tab.~\ref{tab:Model3}) is a special model presenting  
some unusual and challenging characteristics. 
First of all, the $R$-slepton masses are very close
to the neutralino NLSP mass of 165 GeV. (Note that when the difference 
between the $\tauu$ mass and the NLSP mass approaches the tau mass, 
one falls in the neutralino-stau co-NLSP scenario that we are not 
treating here.) 
As we will see, this poses the problem
of separating the various GMSB signals from each other at the LC in order 
to perform specific measurements. Another experimental challenge follows
from the fact that the relatively low minimum gravitino 
mass combined with a quite large $\NI$ mass makes it possible for the 
neutralino to decay very close to the interaction region at the LC 
(see Sec.~\ref{sec:NLSPdecay}) in this case. 

\begin{figure}
\centerline{
\epsfxsize=0.6\textwidth
\epsffile{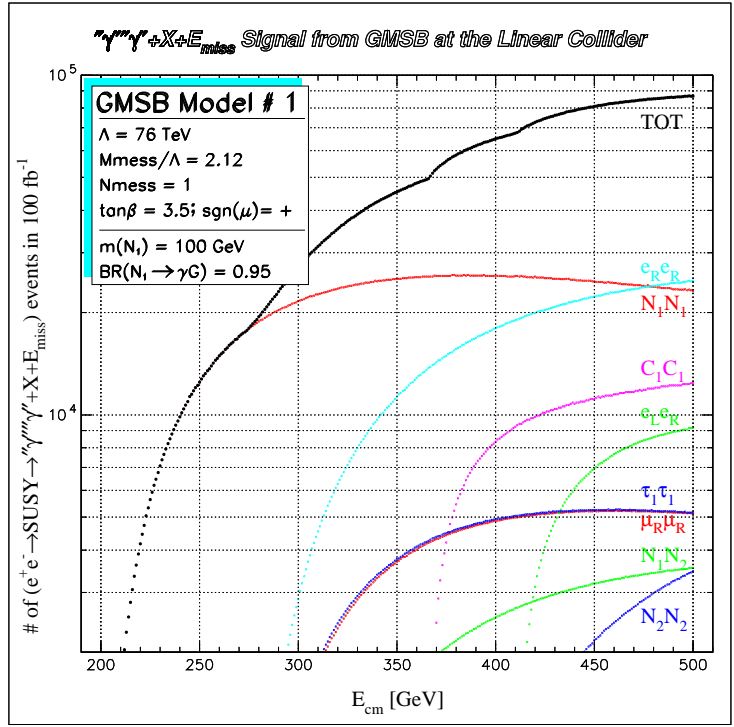}
}
\caption{\sl 
Inclusive signal ``$\gamma$''``$\gamma$'' + X + $\slashchar{E}$ from 
GMSB Model \#~1 as a function of $\sqrt{s}$ in the range of interest
for a 500 GeV LC. The total as well as all contributions to the signal 
from each sparticle-pair production process are shown.  
The normalisation is based on an integrated luminosity of 100 fb$^{-1}$.
}
\label{fig:xsec}
\end{figure} 

In Fig.~\ref{fig:xsec}, we plot the cross-section for the various SUSY
production processes as a function of $\sqrt{s}$ in the range of interest 
for a 500 GeV LC, for Model \#~1. 
Actually, Fig.~\ref{fig:xsec} contains some more 
information, since the normalisation of the y axis takes into account 
the inclusive nature of the GMSB signal and the typical luminosities 
of the LC. As we will see in Sec.~\ref{sec:NLSPdecay}, for all neutralino
NLSP models $\NI \to \G \gamma$ is the dominant NLSP decay channel,
with BR's always greater than about 85\%. (For Model \#~1 this is 
actually 95\%.) As a consequence, each time
a sparticle pair is produced, there is a large probability of getting
a final state with 2 photons, some other particle resulting from cascade
decays and large missing energy. In Fig.~\ref{fig:xsec}, the quotation 
marks for $\gamma$ mean that (one of) the photons might escape detection 
if the $\NI$ lifetime is very large (cfr. Sec.~\ref{sec:NLSPdecay}).
The number of ``$\gamma$''``$\gamma$'' + X  + $\slashchar{E}$ events is 
normalised to an integrated luminosity of 100 fb$^{-1}$, which is a typical 
value for a few-month run at the LC (cfr. Sec.~\ref{sec:TESLA}). 
Notice here that for the case of Model \#~1, running at c.o.m.~energies 
of order 270 GeV would still allow for order 20,000 GMSB events, while 
selecting pure $\gamma\gamma\slashchar{E}$ events only. This circumstance
will be exploited for our experimental studies in Sec.~\ref{sec:NLSPprop}.
We will also use Fig.~\ref{fig:xsec} as a basis for the study of  
Sec.~\ref{sec:THRESCAN}. 

\section{Neutralino NLSP Decays} 
\label{sec:NLSPdecay}

\noindent 
In this section, we analyse the properties of the NLSP decay in GMSB 
models, with focus on the case where $\NI =$ NLSP. 

In Ref.~\cite{AKKMM2}, all the formulas for 2-body decays involving
the gravitino can be found, in the limit where the gravitino interactions
can be approximated by those of the goldstino and its mass can be kinematically
neglected, which is always the case in GMSB at collider energies. 
For a generic decay $\tilde{S} \to S \G$, where S is a SM particle and 
$\tilde{S}$ its MSSM superpartner, one has for the corresponding width, 
\be
\Gamma = \frac{{\cal A}_S} {48 \pi} 
\frac{m_{\tilde{S}}^5}{{M'_P}^2 m_{\G}^2} (\beta_S^*)^8 = 
         \frac{{\cal A}_S} {16 \pi}
\frac{m_{\tilde{S}}^5}{\sqrt{F}^2} (\beta_S^*)^8 ,
\label{eq:NLSPwidth}
\ee
where the gravitino mass is given by (\ref{eq:Gmass}), $\beta_S^*$ is the 
relativistic factor $\sqrt{1- (m_S/m_{\tilde{S}})^2}$
if $S$ is a vector or scalar boson. If $S$ is a massless fermion, 
$\beta_S^* \to 1$. 
${\cal A}_S$ is a constant depending on the $S$, $\tilde{S}$ spin 
and possibly a mixing matrix element. For example, if $S$ is a SM lepton or 
quark and $\tilde{S}$ a slepton or squark, then simply ${\cal A}_S = 1$. 

We are here interested in the $\tilde{S} = \NI$ case, since in our neutralino
NLSP models the only particle that can undergo a 2-body decay to a gravitino
with a non-negligible width is the lightest neutralino. The relevant 
expressions for ${\cal A}_S$ can be found in Tab.~\ref{tab:N1decfacs}, where 
we use the notation of Ref.~\cite{Haber-Kane} for the neutralino mixing 
matrix and $\alpha$ is the mixing angle in the MSSM neutral Higgs sector. 

\begin{table}
\renewcommand{\arraystretch}{1.2}
\begin{center}
\begin{tabular}{|r||c|c|} \hline
Decay Channel     & ${\cal A}_S =$      & where                 \\ 
$\N{i} \to S \G$  &                     &                       \\ \hline\hline
$S = \gamma$      & $\kappa_{i\gamma}$  & 
  $\kappa_{i\gamma} = |N_{i1} \cthw + N_{i2} \sthw |^2$         \\ \hline
         $\Z$      & $\kappa_{iZ_T} + \frac{1}{2}\kappa_{iZ_L}$ & 
  $\kappa_{iZ_T}    = |N_{i1} \sthw - N_{i2} \cthw |^2$         \\ 
                  &                                            &
  $\kappa_{iZ_L}    = |N_{i3} \cos\beta - N_{i4} \sin\beta|^2$  \\ \hline
         $h^0$    & $\kappa_{ih^0}/2$   &
  $\kappa_{ih^0}    = |N_{i3} \sin\alpha - N_{i4} \cos\alpha|^2$ \\ \hline 
         $H^0$    & $\kappa_{iH^0}/2$   &
  $\kappa_{iH^0}    = |N_{i3} \cos\alpha + N_{i4} \sin\alpha|^2$ \\ \hline 
         $A^0$    & $\kappa_{iA^0}/2$   &
  $\kappa_{iA^0}    = |N_{i3} \sin\beta + N_{i4} \cos\beta|^2$   \\ \hline 
\end{tabular}
\end{center}
\caption{\sl 
Constant factors entering the expressions for the widths of neutralino 2-body
decays to gravitino.}
\label{tab:N1decfacs}
\end{table}

Due to the absence of the $(\beta^*)^8$ kinematic suppression and the 
$\NI$ physical composition in the models of interest here (cfr. 
Fig.~\ref{fig:N1comp}), the $\NI$ decay is always dominated by the photon
channel. However, in the context of this paper where $m_{\NI} > 100$ GeV
and fine details of the neutralino decay will be used in the following,
it is important to note that the BR for decaying to a $\Z$ can be sizeable
and also to address the problem of the 3-body decay channels
$\NI \to f \bar{f} \G$, where $f$ is a SM lepton or quark \cite{GMSBmodels2}.  

\begin{figure}
\centerline{
\epsfxsize=\textwidth
\epsffile{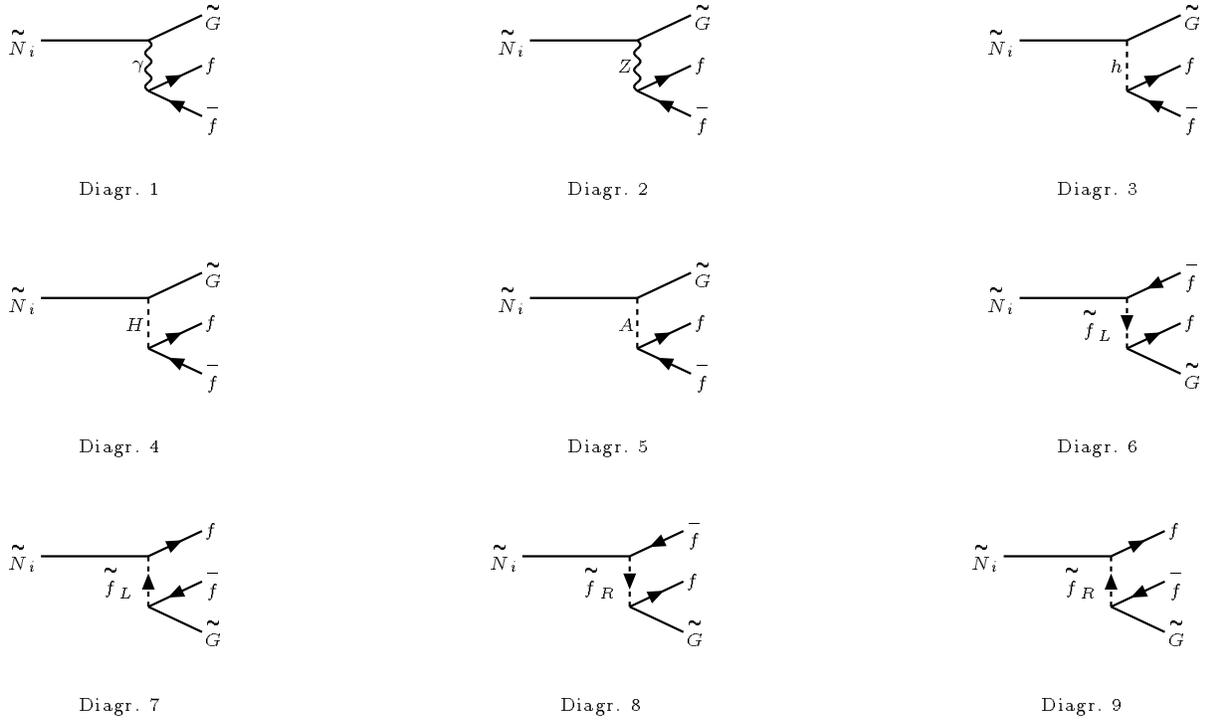}
}
\caption{\sl 
Feynman diagrams contributing to 3-body neutralino decays 
$\N{i} \to f \bar{f} \G$, where $f$ is a charged SM fermion. 
The channels $\N{i}\to\nu_j\bar{\nu}_j\G$ receive contribution from 
Diags.~2, 6 and 7 only.
}
\label{fig:fey3body}
\end{figure} 

The Feynman diagrams contributing to the 3-body processes are shown
in Fig.~\ref{fig:fey3body}. An analytical formula for the sum over
final-state SM-fermions of the total widths for these decays via real
or virtual boson exchange has appeared in Ref.~\cite{GMSBmodels2}. 
However, that summed formula could not take into account Diags.~6--9 where 
a $L$- or $R$-sfermion is exchanged, which are a-priori not less relevant
than Diags.~2--5, where other heavy intermediate particles are involved. 
Also, in Ref.~\cite{GMSBmodels2} the virtual photon contribution (Diagr.~1)
was calculated including an overall detector-dependent cutoff on the fermion 
pair invariant mass, chosen to be of order 1 GeV. In the context of this 
paper, however, we have at our disposal a full detector simulator 
(cfr. Sec.~\ref{sec:BRAHMS}) and we will be interested in the individual
BR's for each $f\bar{f}$ pair (cfr. Sec.~\ref{sec:NLSPprop}). 
Also, the kinematical distributions of these decays are relevant to 
our following studies. In the rest of the paper, we will often
use the name ``neutralino charged decays'' when referring to the 
$\NI\to f\bar{f}\G$ channels and in particular to the case where 
the subsequent final state includes either a charged lepton or a 
charged ``stable'' hadron. 

To account for these channels, we proceeded as follows. 
Using the general lagrangian for the goldstino interactions with matter 
without assuming on-shell conditions (cfr., e.g., Refs.~\cite{GR-GMSB,AKKMM2}),
we input all the relevant vertices\footnote{For some of our gravitino 
vertices and using the goldstino lagrangian as an input, we checked that there 
is agreement with the output of {\tt LanHEP 1.5.06} 
\cite{LanHEP}} involving the gravitino in {\tt CompHEP 3.3.18} \cite{CompHEP} 
in a limit suitable for collider physics. For the other vertices involving 
MSSM particles, we used the home-made lagrangian\footnote{For the relevant 
MSSM vertices and in the relevant limit, we checked that there is numerical 
agreement with our results when using the {\tt CompHEP 3.3}-compatible MSSM 
lagrangian of Ref.~\cite{MSSMlag} instead.} that was first checked
against analytical calculations and then used for numerical evaluations
in the work of Ref.~\cite{AKM2}

 We named the resulting software {\tt Gravi-CompHEP} \cite{SAsoft}. 
Using {\tt Gravi-CompHEP}, we found that the contribution to the 
total width from virtual photons is in very good numerical agreement 
(for the electron and muon case up to 4 digits) with the analytical 
formula \cite{Steve-priv}, valid to lowest order in $m_f/m_{\NI}$, 
\be
\Gamma(\NI\to f\bar{f}\G) = \Gamma(\NI\to\gamma\G) 
\frac{\alpha_{\rm em}}{3\pi} N_f^c Q_f^2 
\left[{\rm ln}\left(\frac{m_{\NI}}{m_f}\right)^2 - \frac{15}{4} \right]
\label{eq:Steve-for}
\ee
where $Q_f$ is the final fermion electric charge in units of $e$, 
$N_f^c = 1 (3)$ for leptons (quarks) and the cutoff is naturally 
provided by the $f$ mass. Note that, e.g. for a 100 GeV neutralino mass
as in Model \#~1, Eq.~(\ref{eq:Steve-for}) gives for the case of electrons
in the final state numbers about twice (five times) as large as for the 
case of muons (taus). For hadronic final states, a realistic evaluation
must take hadronization effects and higher-order corrections into account. 
However, since hadrons will not be our main focus in the analyses of 
Sec.~\ref{sec:NLSPprop} and we will be most interested in the BR's for the 
leptonic channels, we chose a reasonable approximation using  
a rough cutoff for the invariant mass of the final fermion pair at 
$2\Lambda_{\rm QCD} \sim 300$ MeV for light quarks and 
Eq.~(\ref{eq:Steve-for}) for heavy quarks. 

 The contribution from $\Z$-exchange to $\Gamma(\NI\to f\bar{f}\G)$ 
(Diagr.~2) is obtained from Eq.~(\ref{eq:NLSPwidth}) by replacing 
\[ 
{\cal A}_Z (\beta_Z^*)^8 \to 
\left[\kappa_{1Z_T} I_1(Z) + \frac{\kappa_{1Z_L}}{2} I_0(Z)\right] 
{\rm BR}(Z\to f\bar{f}), 
\] 
where $I_0$ and $I_1$ are kinematical factors taking finite $\Z$-width 
effects into account\footnote{Analytical expressions can be found in 
Refs.~\cite{GMSBmodels2,AKKMM2}}. In our model sample for the LC, one 
finds that as long as $m_{\NI} \gtap 120$ GeV the on-shell $\Z$ approximation 
is accurate at the level of 10\% or better. For lighter neutralinos, 
the full calculation is required, since e.g. for $m_{\NI} = 100$ GeV one
has $I_0(Z) = 0.0052$ and $I_1(Z) = 0.0023$, whereas $(\beta_Z^*)^8 = 0.00081$,
and the on-shell $\Z$ approximation underestimates the Diagr.~2 contribution 
by a factor 2.5--3. On the other hand, for models with $m_{\NI} \ltap 120$ 
GeV in our sample, the $\Z$-exchange contribution is always 
$\ltap 3 \; (15)\%$ of the virtual photon contribution, 
e.g. for the $f=e \; (\tau)$ case.  

The interference between the $\gamma$- and $\Z$-exchange
diagrams turns out to be always small, generally at the level of a few
\% or less of the pure Diagr.~1 contribution (cfr. also 
Ref.~\cite{GMSBmodels2}).  

Diagr.~3 is always negligible for the models of our interest here.  
Indeed, one has both a dynamical suppression due to the lack of higgsino
components in the $\NI$ (cfr. Fig.~\ref{fig:N1comp}) and a kinematical
$\sim (\beta^*)^8$ suppression, for $m_{\NI} \gtap 150$ GeV 
(in this range, $m_h/m_{\NI} \ltap 0.85$ always in our model sample
and the on-shell approximation applies).  
Taking off-shell effects into account for $m_{\NI} \ltap 150$ GeV (the  
formulas are similar to those for the $\Z$ case described above)
does not help either, since the $h^0$ width is typically very small 
in the MSSM. Also, Diagr.~3 contributes to the channels with
heavy fermions in the final state only, which have typically lower BR's. 
Diags.~4 and 5 are even more strongly suppressed, because the masses 
of the CP-odd and heavy CP-even Higgses only rarely drop below 300 GeV 
in our model sample. Interferences involving Diags.~3--5 are basically
zero. In the rest of this section, we will often assume for simplicity 
that the $\NI$ is pure bino; this makes all the contributions from 
Diags.~3--5 zero and is justified by Fig.~\ref{fig:N1comp}. 

Finally, as far as Diags.~6--9 are concerned, the $\tilde{f}$
exchanged is necessarily heavier than the initial $\NI$ by definition
of neutralino NLSP model. For the case of hadronic final states,
these diagrams do not count, since the squarks are too heavy in our models.
However, it turns out that limited to the case 
of $\lR$ exchange, the contribution to the width is often 
non-negligible and at the level of several to 10\% of the total, especially 
for those models where $m_{\NI}/m_{\lR}$ is close to 1 and for the case of 
heavier leptons in the final state where Diagr.~1 is less dominant. 
This is again due to the relatively
large $\tilde{B}-\lR$ coupling and the fact that the $\lR$ can never
be much heavier than the NLSP (cfr. Fig.~\ref{fig:Rselvstanb}). Diags.~6--7
are always negligible, even in the leptonic case, due to the relative 
heaviness of the $L$-sleptons. 

\begin{figure}[ht]
\centerline{
\epsfxsize=0.7\textwidth
\epsffile{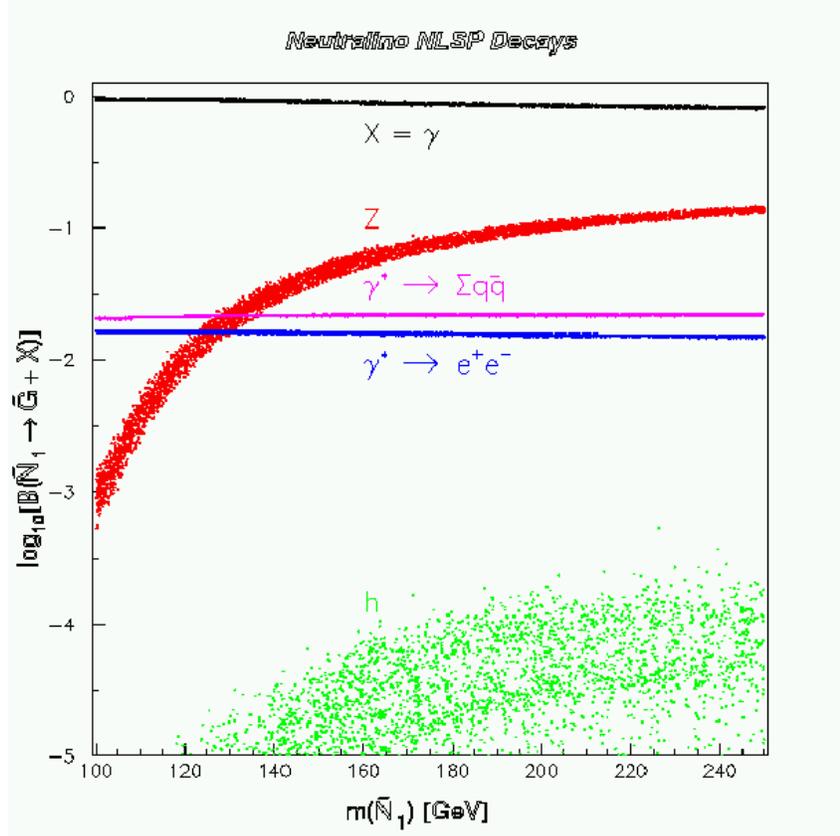}
}
\caption{\sl
Scatter plot for the BR's of various neutralino NLSP decay channels as 
a function of the $\NI$ mass. Dots in different grey scale (colours) refer 
to the decays $\NI \to \gamma\G$, $\NI\to Z \G$ (including off-shell effects), 
and to hadrons or $\epem$ via virtual photon, as labelled. 
For reference, we also report results for the two body $\NI\to h^0\G$ decay 
in the on-shell approximation, whose BR is always negligible. 
}
\label{fig:N1BRs}
\end{figure}

 In Fig.~\ref{fig:N1BRs}, we give a general idea of the behaviour of the 
BR's for the main neutralino NLSP decay channels as a function of $m_{\NI}$
for all the models in our sample of interest for the LC. 
From top to bottom, we show the BR's for the dominant two-body channel to 
a photon, the two-body channel to a $\Z$ including off-shell effects 
(so that the contribution from Diagr.~2 to 3-body channels can be readily 
extracted by multiplying by the appropriate $\Z$ BR), the hadronic and 
$\epem$ 3-body channels from Diagr.~1. For comparison, we also report 
our results for the BR of the 2-body $\NI\to h^0\G$ decay in the on-shell
approximation. The logarithmic scale does not allow inspection of fine 
effects. However, it is evident that the $\Z$ channel can be important 
with BR's up to about 15\% for heavy neutralinos, while the main 
3-body channels via virtual photon are always relevant with BR's at the 
level of a few \%. The Higgs channel has always BR's less than 0.1\%. 
Note also that, due to the homogeneous physical composition of the $\NI$
in our sample, the important BR's are basically dependent only on the 
neutralino mass. 

\begin{figure}[ht]
\centerline{
\epsfxsize=0.7\textwidth
\epsffile{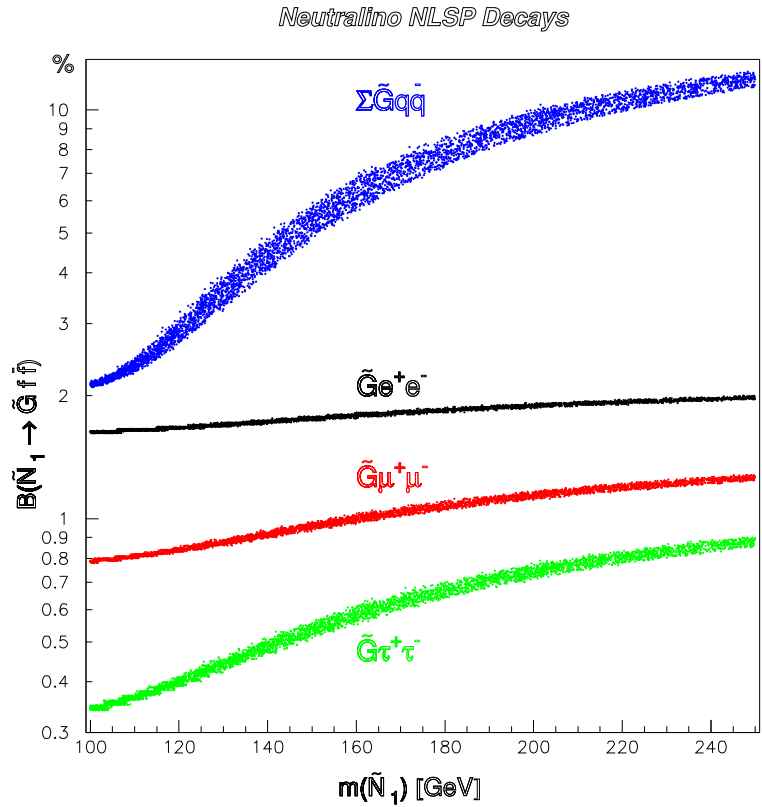}
}
\caption{\sl
Scatter plot for the BR's of neutralino NLSP decays to several 3-body \
$f \bar{f} \G$ channels as a function of the $\NI$ mass.
Dots in different grey scale (colours) refer to decays to quarks, electrons
muons and taus, from top to bottom. Contributions from all Diags.~1--9 in 
Fig.~\ref{fig:fey3body} are included here. 
}
\label{fig:3bodyBRs}
\end{figure}

 Fig.~\ref{fig:3bodyBRs} is a scatter plot for our model sample 
showing in detail the BR's for all the 

\noindent 
3-body $\NI$ decay channels
(excluding $\NI\to\nu\bar{\nu}\G$) as a function of the neutralino mass.
From top to bottom, hadronic + $\G$, $\epem\G$, $\mu^+\mu^-\G$ and 
$\tau^+\tau^-\G$ final states are calculated including all contributions
from Diags.~1--9 in Fig.~\ref{fig:fey3body}. 
The BR's for all 3-body channels all increase for heavier neutralinos.
The hadronic channel occurs about 2\% to 15\% of the times, the electron 
channel 1\%--1.5\%, the muon channel 0.8\%--1.1\%, the tau channel 
0.3\%--0.7\%. Again, fixing the $\NI$ mass basically determines these
BR's, with some more uncertainty for the hadron and $\tau$ channels that
receive relatively larger contributions from $\Z$-exchange. 

\begin{figure}
\centerline{
\epsfxsize=0.8\textwidth
\epsffile{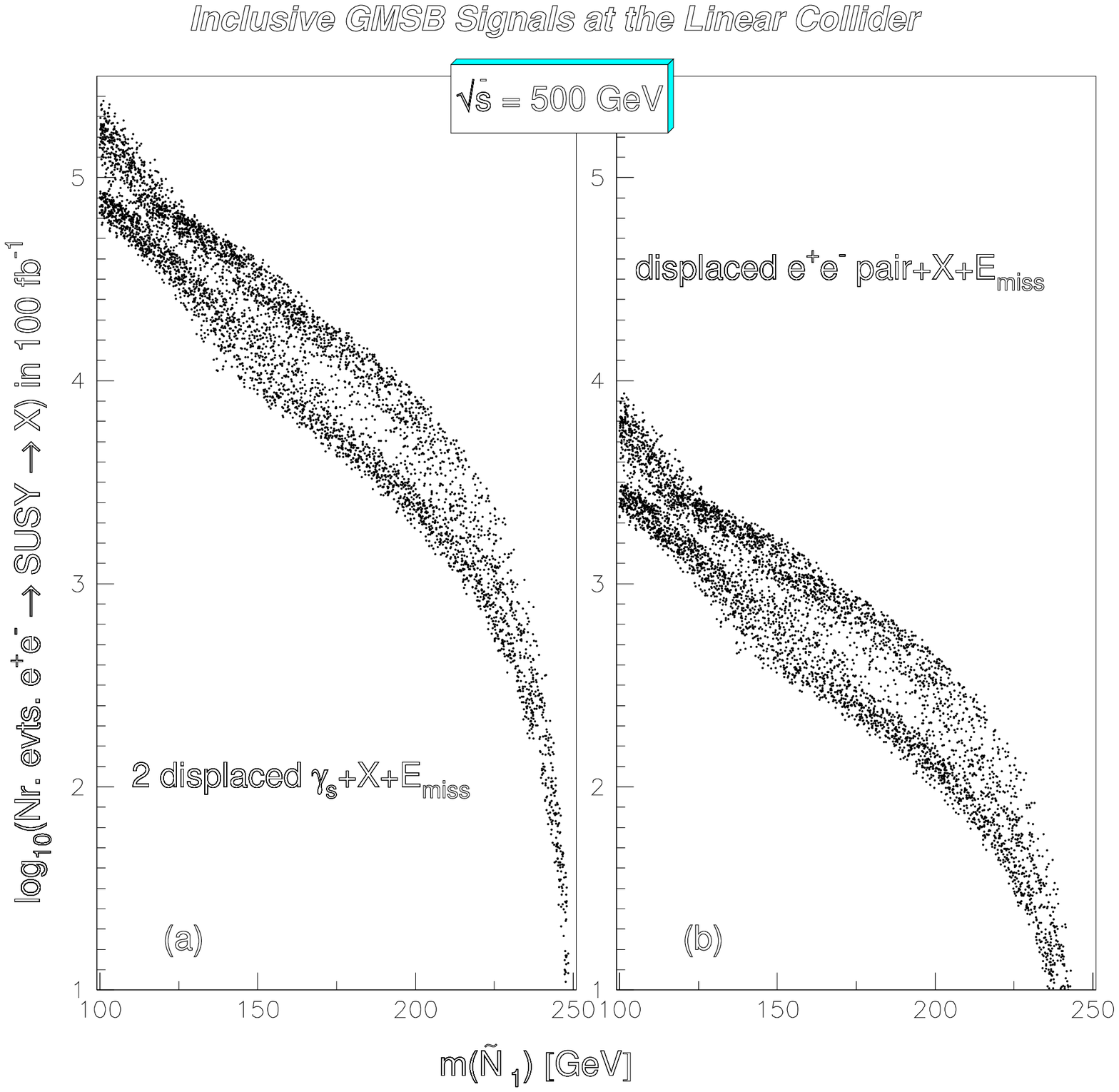}
}
\caption{\sl
Scatter plot for some inclusive GMSB signals in neutralino NLSP models
of interest for a LC. We report the number of events expected after 
a nominal 100 fb$^{-1}$ run at 500 GeV c.o.m. energy. 
(a) Events including two displaced photons and missing energy. 
(b) Events including a displaced $\epem$ pair and missing energy. 
Here we assume that the delayed decays of the $\NI$ NLSP all occur within 
the detector. 
}
\label{fig:inclusive_500}
\end{figure}

 In spite of the fact that the BR for the leptonic 3-body channels is 
often quite low, due to the large integrated luminosity that might
be available at a LC, the possible number of events featuring a (displaced)
$\ell^+\ell^-$ pair is still large in most cases. In 
Fig.~\ref{fig:inclusive_500}, we show scatter plots for our neutralino NLSP 
model sample referring to inclusive GMSB signals at a $\sqrt{s} = 500$ GeV 
LC as functions of the neutralino mass. We refer to a nominal 100 fb$^{-1}$ 
run and sum over all SUSY production processes. In 
Fig.~\ref{fig:inclusive_500}(a) we report
the number of events including two (displaced) photons and missing energy 
coming from two long-lived neutralino decays. 
In Fig.~\ref{fig:inclusive_500}(b), we consider all 
events including at least a displaced $\epem$ pair and missing energy. 
In the models we are interested in, one gets at least 100 such events if 
$m_{\NI} < 200$ GeV, and up to about 10,000 events for lighter neutralinos. 
Notice that the meaning of ``displaced'' here is that the tagged particles 
are produced at some distance from the interaction region, where the distance 
depends on the neutralino lifetime and the specific processes considered. 
In some cases, the displacement might be so large that the particles are 
actually produced outside the detector, as we will see in the following. 

\begin{figure}
\centerline{
\epsfxsize=\textwidth
\epsffile{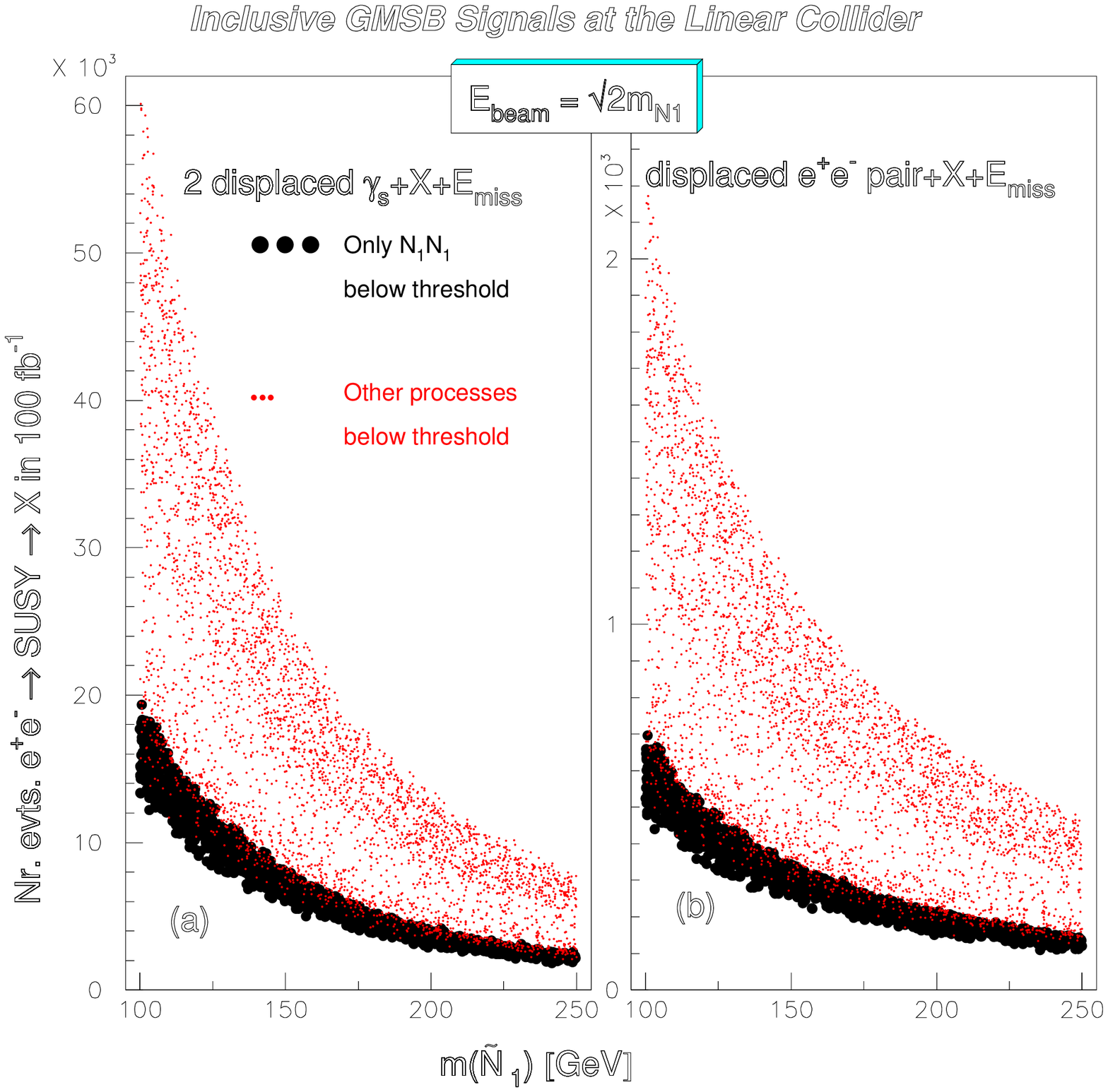}
}
\caption{\sl
Scatter plot for some inclusive GMSB signals in neutralino NLSP models
of interest for a LC, as in Fig.~\ref{fig:inclusive_500}, but for 
$\sqrt{s} = \sqrt{2} m_{\NI}$. Big black dots are for models where 
only $\NI\NI$ pairs can be produced at such an energy, while little
grey dots refer to models where other processes are also below threshold.
}
\label{fig:inclusive_sqrt2mn1}
\end{figure}

 In Sec.~\ref{sec:NLSPprop}, we will see that it is
sometimes useful to run a LC at a c.o.m. energy not far from the $\NI\NI$
threshold to isolate the signal from neutralino pair production. To
give a feeling about this problem, in Fig.~\ref{fig:inclusive_sqrt2mn1}
we show a scatter plot similar to Fig.~\ref{fig:inclusive_500}, but
for $\sqrt{s} = \sqrt{2} m_{\NI}$. The big black dots refer to models 
for which only $\NI\NI$ pairs can indeed be produced at such an energy,
while small grey dots are for models where other processes (typically
pair production of $R$-sleptons) are also below threshold. Note that the 
number of events including a displaced $\epem$ pair is always larger than 
about 100 for an integrated luminosity of 100 fb$^{-1}$. 
The BR's for all the main $\NI$ decay channels for the three reference
models introduced in Sec.~\ref{sec:GMSBmodels} are shown in 
Tab.~\ref{tab:BRsModels1-3} and will be referred to in the analyses 
of Sec.~\ref{sec:NLSPprop}. 

\begin{table}
\begin{tabular}{|c||c|c|c|} \hline
Decay Channel    & BR in Model \#~1 & BR in Model \#~2 & BR in Model \#~3 
\\ \hline\hline
$\NI\to\gamma\G$                  & 0.9507 & 0.8395     & 0.8913 \\ \hline
$\NI\to Z\G \; ^{\rm (a)}$        & 0.0003 & 0.1115     & 0.0585 \\ \hline 
$\NI\to\epem\G \; ^{\rm (b)}$     & 0.0164 & 0.0191     & 0.0179 \\ \hline
$\NI\to\mu^+\mu^-\G \; ^{\rm (b)} $   
                                  & 0.0079 & 0.0117     & 0.0101 \\ \hline
$\NI\to\tau^+\tau^-\G \; ^{\rm (b)}$  
                                  & 0.0034 & 0.0076     & 0.0059 \\ \hline
$\NI\to\sum_q q\bar{q}\G\; ^{\rm (b)} $ 
                                  & 0.0213 & 0.0999     & 0.0631 \\ \hline
$\NI\to\sum_i \nu_i\bar{\nu_i}\G\; ^{\rm (b)}  $ 
                                  & 0.0002 & 0.0223     & 0.0117 \\ \hline
$\NI\to h^0\G \; ^{\rm (a)}$      & -- --  & $< 0.0001$ & $< 0.0001$ \\ \hline 
\end{tabular}
\caption{BR's for the main decay $\NI$ decay channels in our three reference
GMSB models. $^{\rm (a)}$ Entries include the on-shell $\Z$ ($h^0$) 
contribution only. 
$^{\rm (b)}$ Entries include all contributions. As a consequence, due 
to double counting of the on-shell $\Z$ ($h^0$) contributions, the BR's in
each column do not sum up to 1.} 
\label{tab:BRsModels1-3}
\end{table}

 In Sec.~\ref{sec:NLSPprop}, we will heavily use the characteristics 
of the three-body decays of the neutralino, including their kinematical
distributions. Using {\tt Gravi-CompHEP}, we calculated such distributions
for our reference models and we show here our results. We then implemented
numerically the corresponding differential widths into our event generator
to perform the Monte Carlo simulation. 

\begin{figure}
\centerline{
\epsfxsize=0.6\textwidth
\epsffile{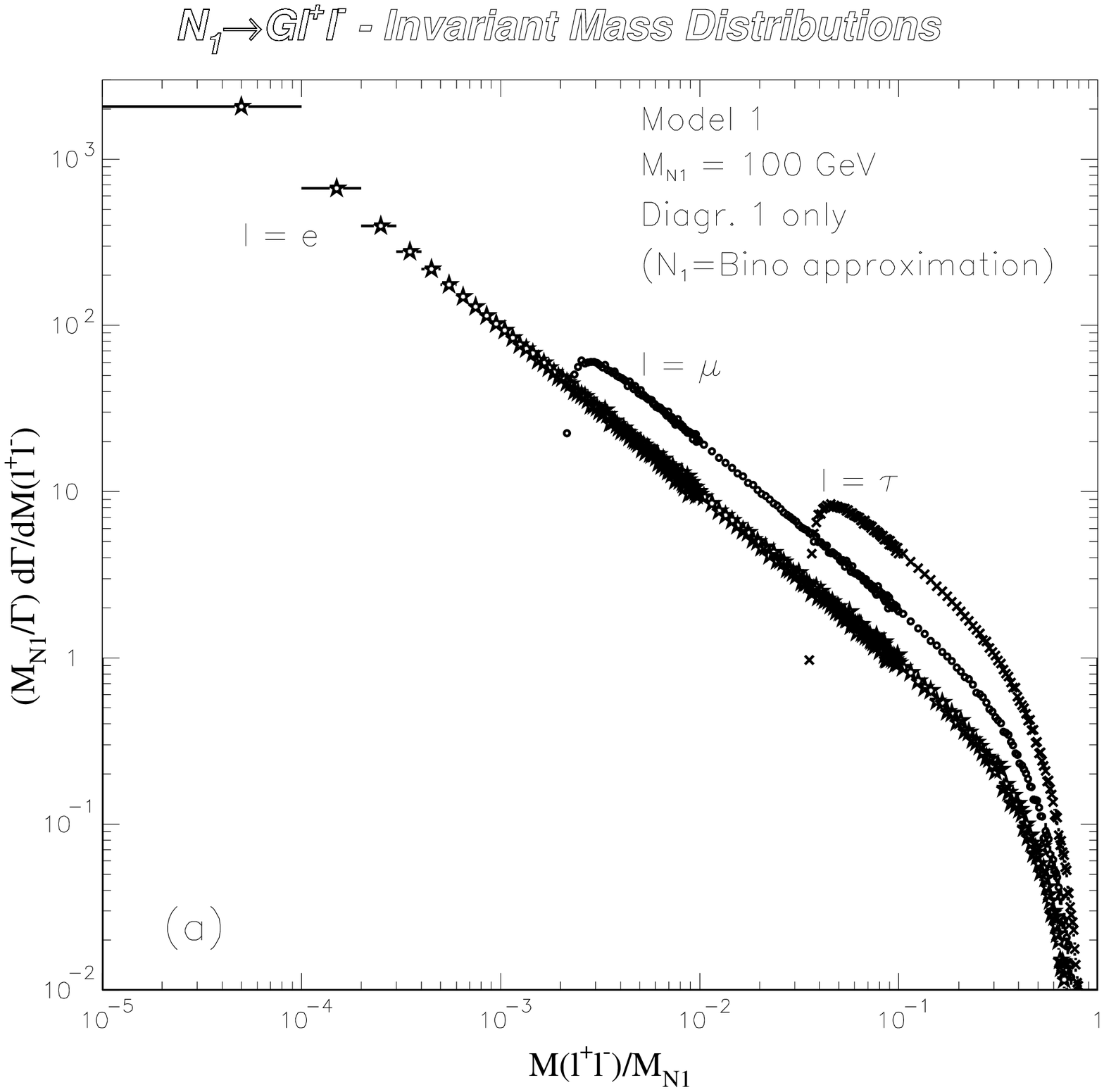}
\epsfxsize=0.6\textwidth
\epsffile{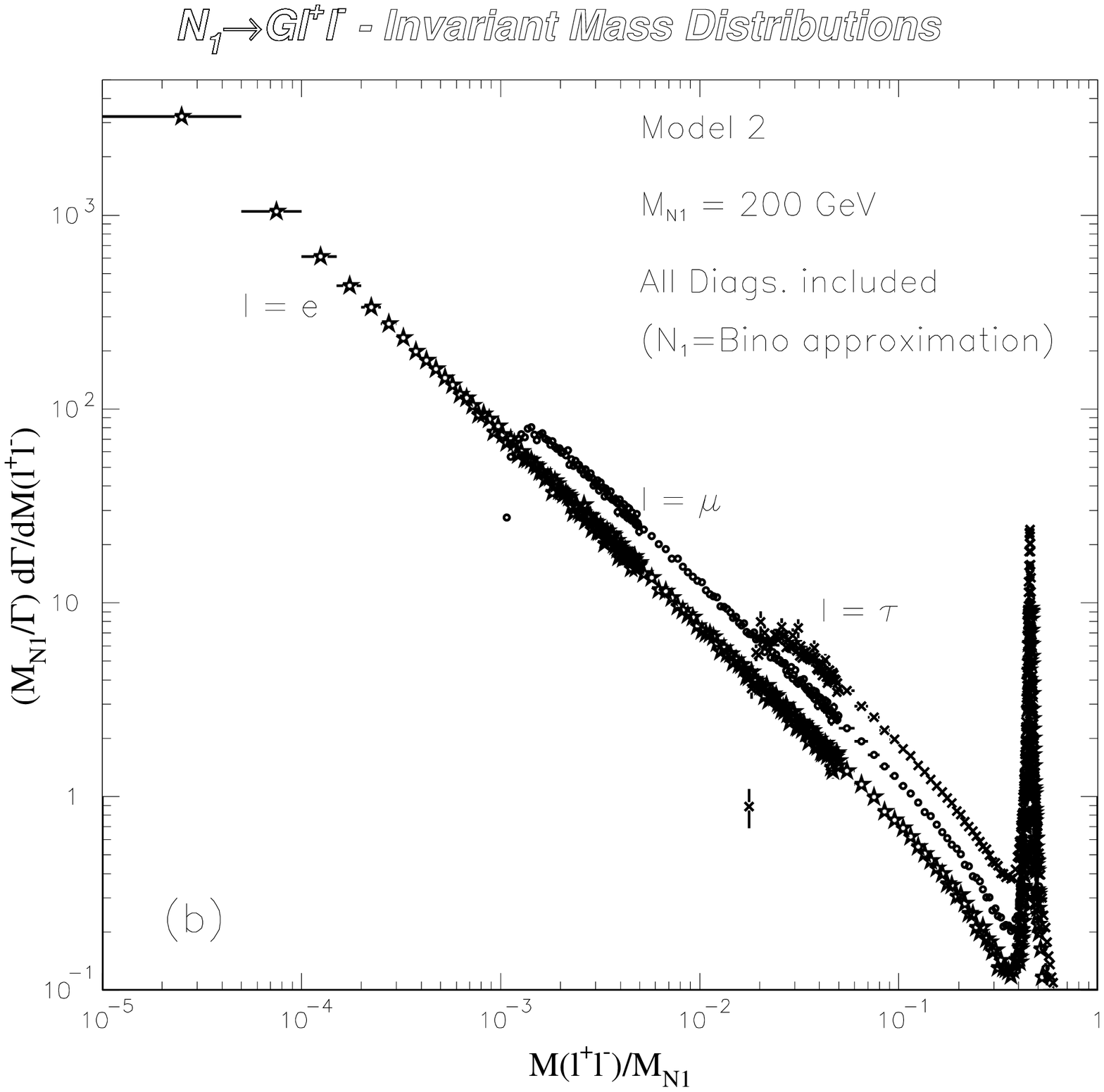}
}
\caption{\sl
Normalised $\ell^+\ell^-$ invariant mass distributions for the leptonic
three-body decays of the $\NI$ in Model \#~1 (a) and Model \#~2 (b).
Stars, circles, crosses refer to the electron, muon, and tau case, 
respectively.
}
\label{fig:mlldistmodel1-2}
\end{figure}

 In Fig.~\ref{fig:mlldistmodel1-2}, the normalised invariant $\ell^+\ell^-$ 
mass distribution for all the leptonic channels is plotted. The stars, circles,
crosses are the central points of our results for the electron, muon, tau 
cases respectively in Model \#~1 (a) and Model \#~2 (b). The horizontal 
bars show the $M_{\ell\ell}/m_{\NI}$ binning we used, while the errors coming 
from numerical phase space integration on the distributions are too small 
to be visible in logarithmic scale in most cases. Note that the distributions
are sharply peaked for low invariant masses close to $2M_{\ell\ell}$ and 
this is more and more true for lighter leptons. As a consequence, e.g. the
$\epem$ pairs coming from $\NI\NI$ production at the LC and subsequent 
three-body decay of (one of) the neutralinos tend to be generated with 
small separation angles, which introduces some experimental challenges
(cfr. Sec.~\ref{sec:NLSPprop}). In Fig.~\ref{fig:mlldistmodel1-2}(b),
the peak corresponding to the $\Z$-exchange contribution is also evident. 
To allow a better inspection of the scaling properties of the distributions 
with $m_{\NI}$ and $m_f$, we used here the (very good) $\NI = \tilde{B}$ 
approximation for both models, so to keep the $\NI$ physical composition 
constant when going from Model \#~1 to Model \#~2. 
For the case of Model \#~1, we included Diag.~1 only of 
Fig.~\ref{fig:fey3body}, since the other contributions would hardly be 
visible in the plot anyway.  

\begin{figure}
\centerline{
\epsfxsize=0.6\textwidth
\epsffile{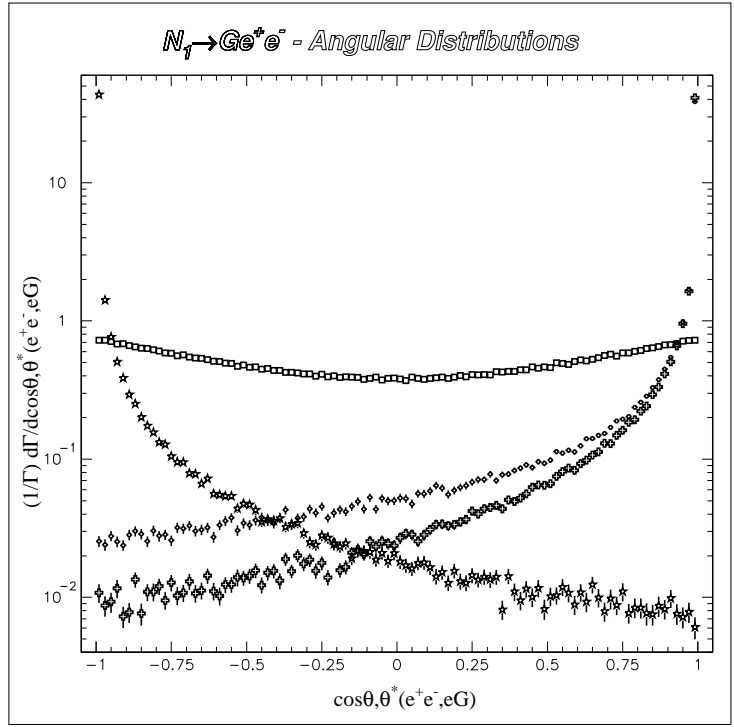}
}
\caption{\sl
Normalised angular distributions for the $\NI\to\epem$ decay in 
Model \#~1. Circles, squares, crosses and stars refer to different
angles, as defined in the text. 
}
\label{fig:angdistmodel1ee}
\end{figure}

 Limited to the case of Model \#~1, in Fig.~\ref{fig:angdistmodel1ee}, 
we show some relevant angular distributions for the $\NI\to\epem\G$ decay, 
the three-body channel we will be most interested in. 
The circles refer to the normalised $\cos\theta(\epem)$ distribution, where 
$\theta(\epem)$ is the angle between the electron and the positron momenta 
in the decaying $\NI$ rest frame. As expected, the $e^+$ and the $e^-$
prefer to proceed along the same direction, due to the dominance of the 
virtual photon contribution. The stars correspond to the normalised 
$\cos\theta(e^\pm \G)$ distribution, where $\theta(e^\pm \G)$ is the angle 
between the electron (or positron) and the gravitino momenta in the $\NI$ 
rest frame. The $\epem$ pair is produced in the direction opposite to the
$\G$ in the great majority of cases. The squares refer to the normalised
$\cos\theta^*(\epem)$ distribution, where $\theta^*(\epem)$ is the angle
between the electron (or positron) momentum and the direction of the boost
of the $\epem$ system with respect to the $\NI$ rest frame, calculated 
in the $\epem$ rest frame. In our case, this is basically the angle between 
the electron (or positron) and the virtual photon momenta and the almost
constant behaviour is then expected. Finally, the crosses show the 
$\cos\theta^*(e^\pm \G)$ normalised distribution, where 
$\theta^*(e^\pm \G)$ is defined as above and refers to the $e^\pm G$ system 
instead of the $\epem$ one. The general behaviour of these angular 
distributions is better understood in the light of the fact that 
the two-body $\NI\to\gamma\G$ decay is isotropic. Some instability of 
our results due to numerical phase space integration is visible, but 
it is well within the shown vertical error bars. 

\begin{figure}
\centerline{
\epsfxsize=0.55\textwidth
\epsffile{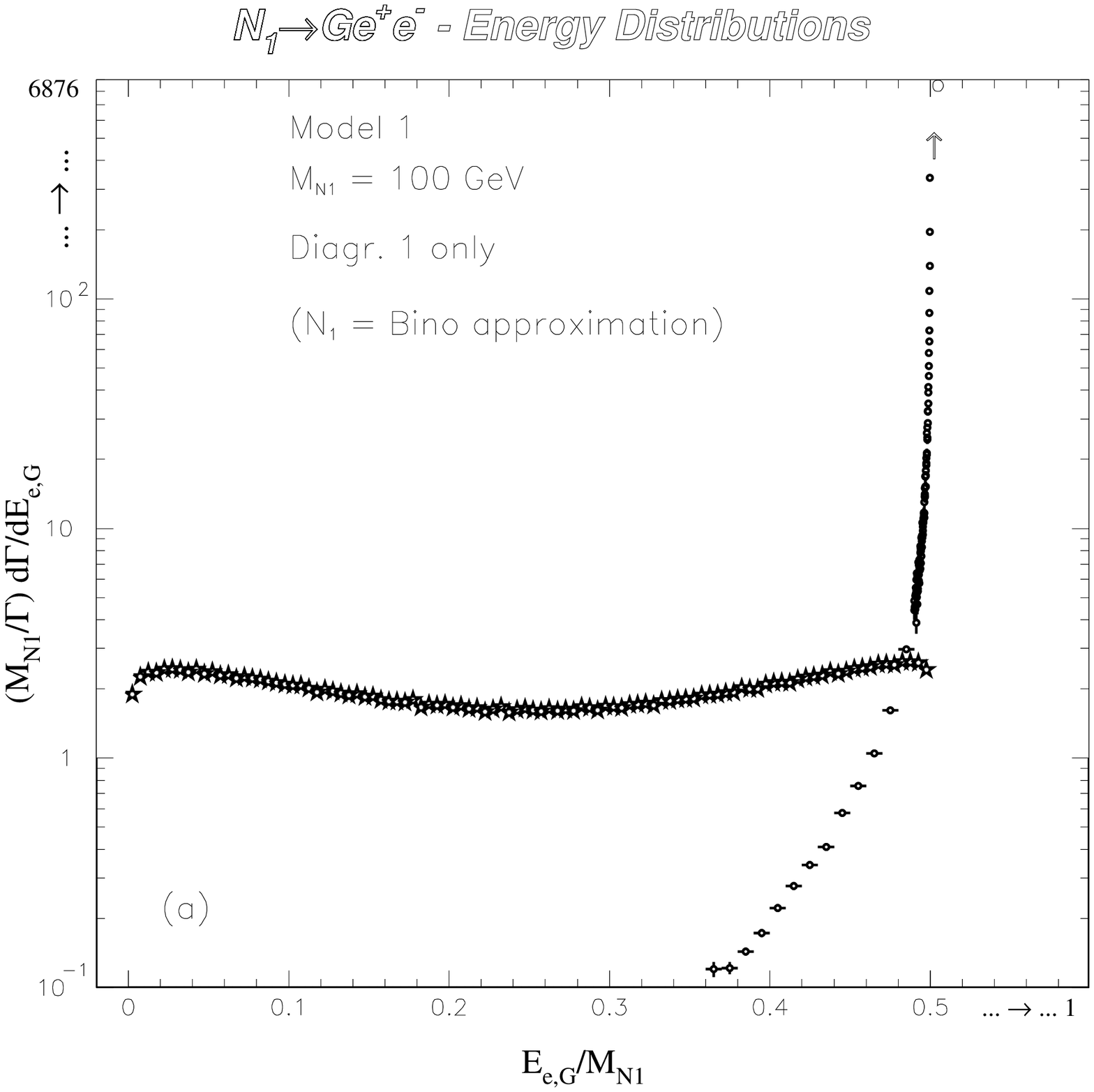}
\epsfxsize=0.55\textwidth
\epsffile{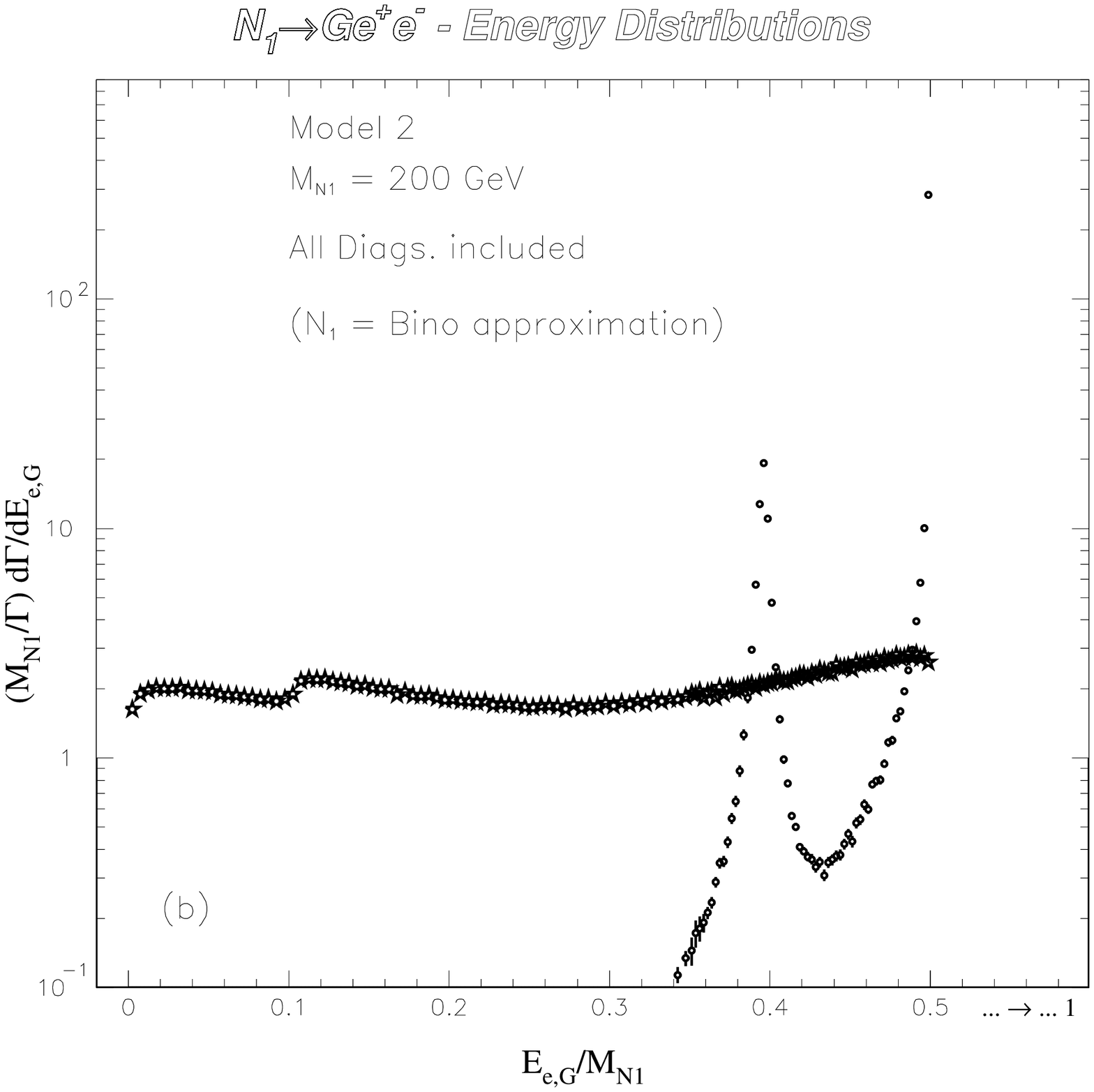}
}
\caption{\sl
Normalised electron or positron (stars) and gravitino (circles) energy 
distributions for the $\NI\to\epem$ decay in Model \#~1 (a) and Model 
\#~2 (b). 
}
\label{fig:enedistmodel1-2ee}
\end{figure}

 In Fig.~\ref{fig:enedistmodel1-2ee}, we show the normalised energy 
distributions for the $\NI\to\epem\G$ decay in Model \#~1 (a) and 
Model \#~2 (b). The stars refer to the electron (or positron) energy,
while the circles are for the gravitino energy, in the decaying $\NI$
rest frame. For the case of Model \#~1, the $\G$ tends to take half of the 
available energy, while the electron and positron share the rest with 
an almost uniform distribution between $m_e$ and about $m_{\NI}/2$. 
Model \#~2 features evident effects of the $\Z$-exchange contribution that 
add to a behaviour similar to the one for Model \#~1 coming from the still 
dominant virtual photon contribution. Again, in order to allow a cleaner
comparison between the two models, we used here the bino approximation. 
Note that both the scales on the x and y axes in 
Fig.~\ref{fig:enedistmodel1-2ee} are interrupted for display convenience. 

 After having inspected the various possible decay channels, we turn now to 
the discussion of the total width and the lifetime of the neutralino.
As anticipated above, this determines the topology of the signatures of 
neutralino GMSB models at colliders. A single neutralino 
produced with energy $E_{\NI}$ will decay before travelling a distance
$\lambda$ with a probability given by 
\bea 
P(\lambda) & = & 1 - {\rm exp}(-\lambda/L) \; \; {\rm where} \\ 
\label{eq:NLSPprob}
L & =  & c \tau_{\NI} (\beta\gamma)_{\NI}.
\label{eq:bigell}
\eea 
$L$ is the $\NI$ ``average'' decay length and $(\beta\gamma)_{\NI}$
is the kinematical factor $(E_{\NI}^2/m_{\NI}^2 - 1)^{1/2}$. Note that
for $\NI$ pairs directly produced at the LC with $\sqrt{s} = 500$ GeV, 
$(\beta\gamma)_{\NI} = 2.29$ (0.75) if $m_{\NI} =  100$ (200) GeV.  

Using Eqs.~(\ref{eq:Gmass}), (\ref{eq:NLSPwidth}), and (\ref{eq:Steve-for}), 
the neutralino lifetime can be conveniently expressed in the suggestive form,  
\bea
c \tau_{\NI} & = & \frac{16\pi}{{\cal B}} \frac{\sqrt{F}^4}{m_{\NI}^5}
\nonumber \\ 
& \simeq & \frac{1}{100 {\cal B}} 
\left(\frac{\sqrt{F}}{100 \; {\rm TeV}}\right)^4 
\left(\frac{m_{\NI}} {100 \; {\rm GeV}}\right)^{-5},
\label{eq:NLSPtau}
\eea 
which stresses the scaling properties with the 5$^{\rm th}$ 
inverse power of the neutralino mass and the 4$^{\rm th}$ power of the 
fundamental SSB scale. ${\cal B}$ is a number of order unity that can be well 
approximated by ${\cal B} \simeq {\cal A}_{\gamma} = 
\kappa_{1\gamma}$ when the two-body $\NI\to\gamma\G$ channel widely 
dominates, as e.g. in Model \#~1, or by simple expressions in most cases. 
In general, however, it is a complicated function of the neutralino 
composition, the GMSB model spectrum etc., when the full contributions to 
the three-body channels are taken into account. 

\begin{figure}
\centerline{
\epsfxsize=0.65\textwidth
\epsffile{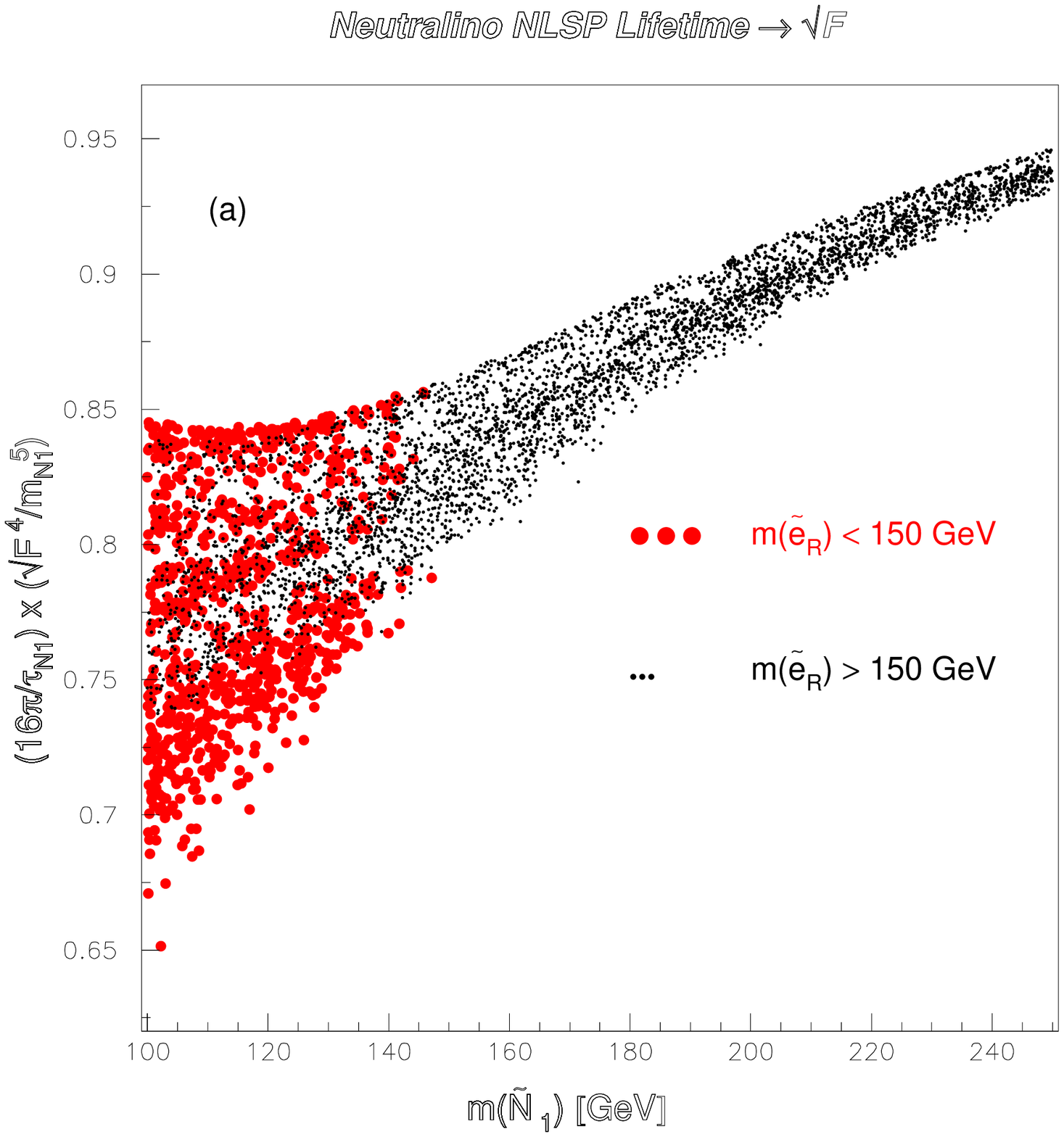}
\hspace{-1.0cm}
\epsfxsize=0.65\textwidth
\epsffile{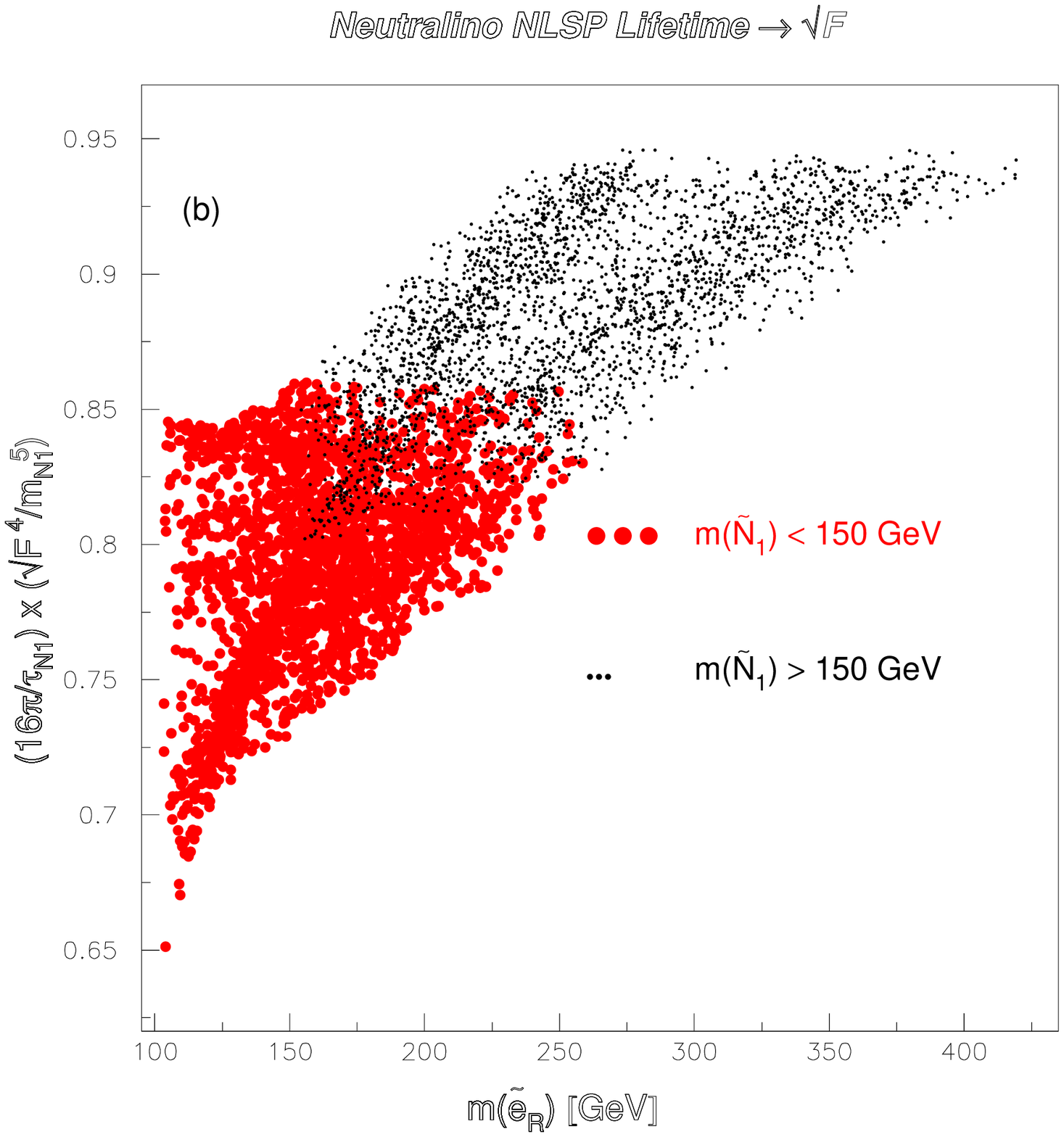}
}
\caption{\sl
Scatter plot showing the relation between the neutralino NLSP
lifetime and the fundamental scale of SUSY breaking $\sqrt{F}$, i.e. the 
factor ${\cal B}$ in Eq.(\ref{eq:NLSPtau}), as a function of the 
neutralino mass (a) or the $R$-selectron mass (b) in GMSB models for the LC. 
Big grey dots in (a) represent neutralino NLSP models with a light 
$R$-selectron (102--150 GeV), small black dots are for the heavier selectron 
case (150--430 GeV). In (b), big grey (small black) dots are for models 
with $m_{\NI} < (>) 150$ GeV. 
}
\label{fig:lifetosqrtf}
\end{figure}

 Once the neutralino mass and lifetime are measured (cfr. 
Sec.~\ref{sec:NLSPprop}), one can get striking information on $\sqrt{F}$ 
from Eq.~(\ref{eq:NLSPtau}). The uncertainty is then only due to the 
factor ${\cal B}$. If the BR's for the various $\NI$ decay channels 
are also measured with good precision, or the neutralino
composition and the (light) GMSB spectrum is extracted by measuring 
other observables (production cross sections, distributions), the 
fundamental SUSY breaking scale can be determined precisely. However, it
is remarkable that even without collecting additional information, the 
knowledge of $m_{\NI}$ and $c\tau_{\NI}$ is sufficient to constrain the value 
of $\sqrt{F}$ in a narrow range, based on the well defined characteristics of 
GMSB models. 
In Fig.~\ref{fig:lifetosqrtf}, we report scatter plots of our neutralino
NLSP model sample for the LC showing ${\cal B}$ in Eq.~(\ref{eq:NLSPtau})
as a function of the neutralino mass (a) and the right selectron mass (b). 
To stress the existence of some correlation between ${\cal B}$ and 
the right selectron mass for models with a fixed light neutralino mass, 
in Fig.~\ref{fig:lifetosqrtf}(a), the big grey (small black) dots refer to 
models where 102 (150) $< m_{\eR} <$ 150 (430) GeV. This can be compared to 
Fig.~\ref{fig:lifetosqrtf}(b), where big grey (small black) dots correspond
to models with $m_{\NI} < (>) 150$ GeV. From this, one can e.g. infer that
if $m_{\NI} \simeq 120$ GeV and the neutralino lifetime is measured to be
about 1 cm, then $360 \ltap \sqrt{F} \ltap 385$ TeV. If, in addition, 
$m_{\eR}$ is measured to be heavier than about 150 GeV (for instance, from
$\eR\eR$ threshold scanning, see Sec.~\ref{sec:THRESCAN}, or using its  
impact on the $\NI\NI$ cross section), then the allowed range is further
reduced to 370--385 TeV. For a 200 GeV neutralino, a 1 cm lifetime gives
$725 \ltap \sqrt{F} \ltap 740$ TeV. 

 To summarise, we note that in the absence of further information, 
the theoretical error on determining $\sqrt{F}$ from given values 
of $\NI$ mass and lifetime amounts to about 3\% in the worst case, 
helped by the 4$^{\rm th}$ power dependence in Eq.~(\ref{eq:NLSPtau}). 

\begin{figure}
\centerline{
\epsfxsize=0.7\textwidth
\epsffile{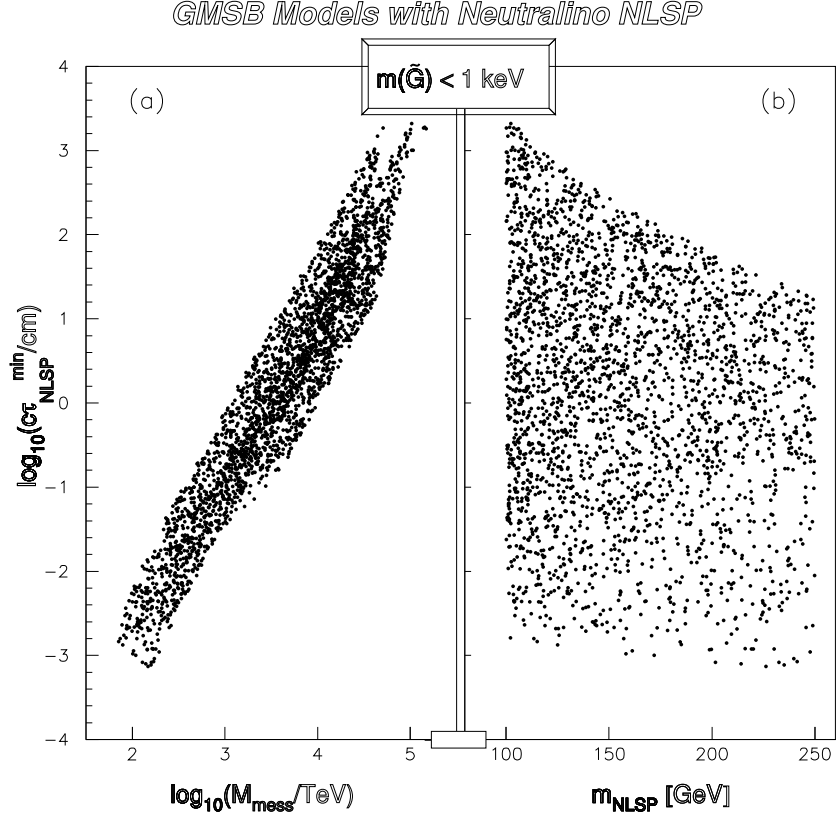}
}
\caption{\sl
Scatter plot of the neutralino NLSP lifetime as a function of the messenger
scale $M_{\rm mess}$ (a) and $m_{\NI}$ (b). For each set of GMSB 
model input parameters ($\Lambda$, $M_{\rm mess}$, etc.), the lower limit of 
the NLSP lifetime is plotted, corresponding to 
$\sqrt{F} \simeq \sqrt{F}_{\rm mess} = \sqrt{\Lambda M_{\rm mess}}$. 
Only models that fulfil the limit on the gravitino mass 
($m_{\G} \ltap 1 \; {\rm keV} \Rightarrow \sqrt{F}_{\rm mess} 
\ltap \sqrt{F} \ltap 2000$ TeV) suggested by simple cosmology 
are used. 
}
\label{fig:NLSPtau}
\end{figure}

 It is of primary importance for collider phenomenology to assess the range 
of variation for $c\tau_{\NI}$. As anticipated in Sec.~\ref{sec:GMSBmodels},
it is possible to use a lower limit from theory on $\sqrt{F}$, while 
significant upper limits can only come from weak cosmological arguments
suggesting $m_{\G} \ltap 1$ keV. The lower limit defines a minimum value 
$c \tau^{\rm min}_{\NI}$ for the neutralino lifetime as well as 
$m_{\G}^{\rm min}$ for the gravitino mass on a GMSB model-by-model basis. 
In Fig.~\ref{fig:NLSPtau}, we plot this limit as a function of 
$M_{\rm mess}$ (a) and $m_{\NI}$ (b) for our model sample of 
interest for the LC. In this plot, we also use the cosmological 
upper limit on $m_{\G}$, in the sense that those models where
$m_{\G}^{\rm min} > 1$ keV are not plotted. As a result, one can 
see that the neutralino lifetime can be anywhere between about 5 microns
and about 25 metres (or more if no cosmological arguments are used). 
Models with a high messenger scale produce longer neutralino
lifetimes. For instance, if $M_{\rm mess} \gtap 10^4$ TeV, then
$c \tau_{\NI}$ is always larger than about 1 cm. Also note that
shorter lifetimes are obtained for heavier neutralinos. A 100 GeV neutralino
will always live more than about 15 microns. On the other hand, a 250
GeV neutralino will tend to decay well within a typical detector size,  
with lifetimes always smaller than about 20 cm, if cosmological arguments 
are used.

\begin{figure}
\centerline{
\epsfxsize=0.8\textwidth
\epsffile{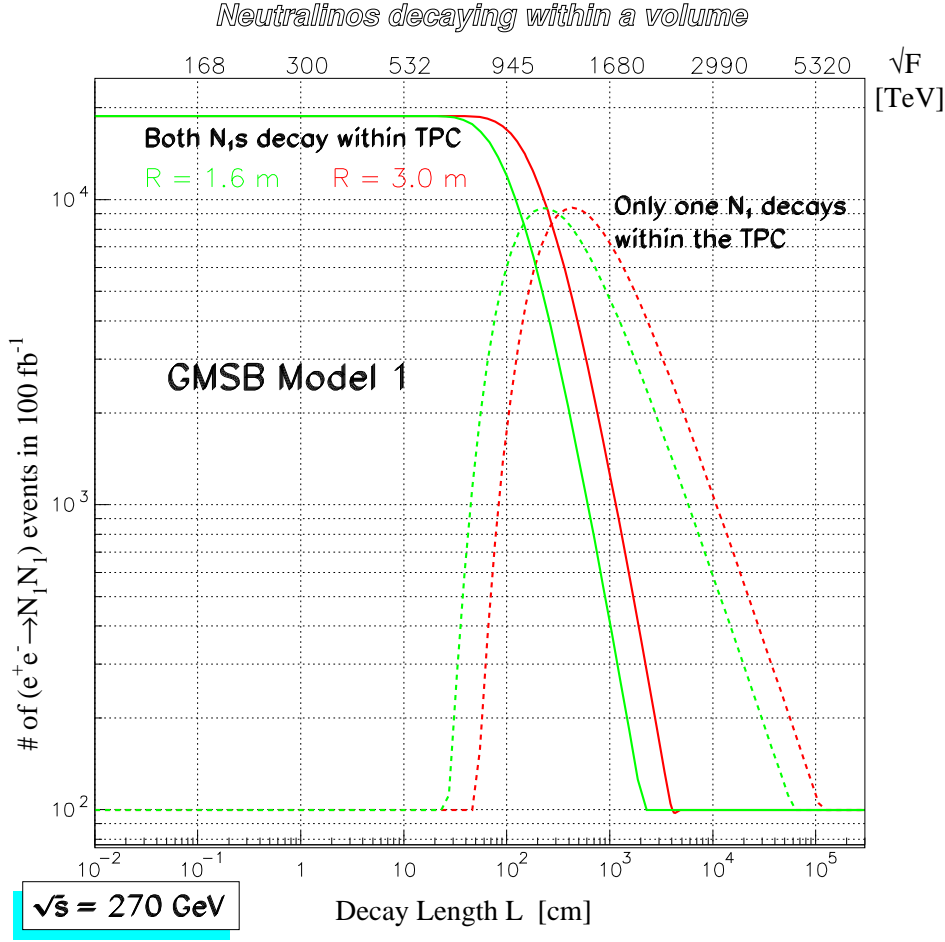}
}
\caption{\sl
Number of events in 100 fb$^{-1}$ featuring one or both neutralinos decaying 
within two reference spheres of radius 1.6 and 3.0 m (different grey scale 
or colours) after $\NI\NI$ production at $\sqrt{s} = 270$ GeV as a function of 
$L$ or $\sqrt{F}$ for Model \#~1. 
}
\label{fig:decays_in_tpc}
\end{figure}

 When SUSY pairs are produced in a neutralino NLSP scenario, the resulting 
final states always include two neutralinos, which in turn decay to
a gravitino + X. The probabilities of both, one or zero neutralinos
decaying within a given volume of the detector depend on the
neutralino's decay length $L$ of Eq.~(\ref{eq:bigell}), which in turn 
depends on the specific SUSY process, model and collider c.o.m. energy
one is considering.   
If we define a spherical volume of radius $R$, then the probabilities 
associated with these circumstances are of course given by \ \ $P(R)^2$, 
\ \ $2P(R)[1-P(R)]$, \ \ and \ \ $[1-P(R)]^2$. 
In Fig.~\ref{fig:decays_in_tpc}, we show how many events are 
expected with two (solid line) or one (dashed line) neutralino decays as a 
function of $L$ (or $\sqrt{F}$) for the case of direct neutralino-pair 
production in Model \#~1.  
We show curves for two reference spheres with 
a radius of 160 and 300 cm (we will see in Sec.~\ref{sec:BRAHMS} that the outer
cylinder of a typical proposed TPC for a LC detector is included between such 
two spheres). Our numbers refer to a 100 fb$^{-1}$ run at the LC with 
$\sqrt{s} = 270$ GeV, where only $\NI\NI$ pairs can be produced with a 
cross section of 188 fb. This choice of parameters will be of relevance for 
the studies to be presented in Sec.~\ref{sec:NLSPprop}. Note that for
neutralino decay lengths as large as 1 km (and $\sqrt{F}$ larger than 
5000 TeV), there still are about 100 events where one neutralino decays  
within the reference volume. We warn however that this is only true if 
a sum over all possible final states coming from neutralino decays is performed
and no angular or other detector acceptance cuts are taken into account. 
Based on this, a refined statistical study based on a realistic 
cylindrical detector and experimental framework for the LC is performed 
in Sec.~\ref{subsec:stat}. 

\section{The Linear Collider and the TESLA Project}
\label{sec:TESLA}
\noindent 
The LHC will explore the next high energy frontier and can be expected 
to be among the prime sources of new physics discoveries into the next 
decade and beyond.  However, ongoing studies of the physics potential 
of a $e^+e^-$ LC operating at c.o.m. energies ranging up to 500 GeV, 
or higher (1--2 TeV), are revealing many complementary
measurements that could be made at such a machine on a similar timescale
to that of the LHC. Additional options available at a LC are a considerable 
electron (and possibly also positron) polarisation, 
$e\gamma$ and $\gamma\gamma$ options, as well as the potential 
for $e^-e^-$ collisions, making a LC a very flexible and relevant 
facility, with particular application to detailed studies of new 
physics signals. 

Several linear collider designs are presently under discussion
\cite{LCWorld} and much of the discussion in this paper is applicable
to any machine. However, in order to relate our study to a specific
case, we explore the machine parameters of the high-luminosity
TESLA option.  The TESLA machine proposal is described in some detail 
in vol.~2 of the ECFA/DESY ``Conceptual Design Report'' (CDR) \cite{CDR}.  
The most recent proposals involve two phases of operation; an earlier phase 
operating at $\sqrt{s} = 500$ GeV or less and a later phase operating at 
$\sqrt{s} = 800$ GeV or less, with luminosities of 
$3.1\times 10^{34}$ cm$^{-2}$s$^{-1}$ and 
$5.0\times 10^{34}$ cm$^{-2}$s$^{-1}$ respectively \cite{tesla_parameters}.  
In this study, we mostly limit ourselves to the first 
phase foreseen for such a collider, with $\sqrt{s}$ varying between 
approximately 200 and 500 GeV.
In most cases, we will consider results that can be obtained after
collecting an integrated luminosity of 200 fb$^{-1}$, corresponding 
to approximately one year of running at TESLA and to a few years 
of running if parameters proposed for other linear collider options 
(such as JLC or NLC \cite{LCWorld}) are used. 

We should stress that one of the highly desirable features of a
LC is the ability to tune the c.o.m. energy to explore
thresholds with precision. In this way, specific signals, e.g. from
SUSY, can be enhanced from among others, unless the production 
thresholds are too closely degenerate. For instance, we use this
property below for our GMSB models \#~1--2 to isolate the neutralino 
pair production process for individual study.  
In addition, the energy can be tuned to alter appreciably 
the Lorentz ($\beta\gamma$) factors of the produced neutralinos and hence 
extend the range of NLSP lifetime measurements.  
Neither of these options will be available at the LHC.

A further advantage to this study of a LC over the LHC is the fact that 
the effective c.o.m.~energy is known precisely, up to effects of initial 
state radiation (ISR) and beamstrahlung.  The values of 
$\frac{\delta E_{\rm beam}}{E_{\rm beam}}$ due to beamstrahlung are 
estimated to be
2.8\% and 4.7\% for the first and second phase of TESLA respectively
\cite{tesla_parameters}.
For the processes we study in this paper, it is of prime importance to 
know the energy of the pair-produced neutralinos in order to be able to
reconstruct the neutralino decay length, as described further
in Sec.~\ref{sec:NLSPprop}.  
We wish to stress here the complementary nature of the LC with
respect to the LHC and we envision the pleasing scenario where the 
LHC provides a wealth of interesting data, which is subsequently
investigated at the LC with high precision. This may indeed 
be necessary in order to distinguish conclusively between 
GMSB and other possible SUSY realizations (e.g., no-scale SUGRA models) 
and to measure the fundamental parameters with the precision needed to 
extract striking conclusions concerning physics at the messenger and 
higher scales. 
Because of this, the present LC detector design proposals 
should be flexible enough to benefit at a late stage from LHC new physics 
data. Our study attempts to address this issue and we will further comment
on this in Sec.~\ref{sec:conc}. 
Moreover, as we already pointed out in Sec.~\ref{sec:GMSBmodels},
GMSB models often feature very heavy strongly interacting sparticles, 
which could not provide a large signal at the LHC. In this case, 
the r\^ole of the LC in determining the origin of the new physics signal
would be even more important. 

\section{Disentangling a Signal from GMSB at the Linear Collider}
\label{sec:THRESCAN}

\noindent 
In this section, we provide an example of how it might be possible to 
extract a good amount of information about the parameter values of 
the underlying model from the observation of an abundant GMSB signal 
at the LC and a simple threshold scannning technique. 
Let's assume that Nature has chosen GMSB and that Model \#~1 is realized.
Let's also assume for simplicity that $\sqrt{F}$ is not too large, so 
that (most of) the produced NLSP's decay within the detector.  
If this is the case, just a few weeks of LC running at some initial 
c.o.m. energy between 200 and 500 GeV would be enough to recognize the 
presence of an evident GMSB-like scenario. Indeed, a copious number of
events with two $\gamma$'s and large missing energy would show up,
due to the inclusive characteristics of the GMSB signal. Further, 
we will see in Sec.~\ref{sec:NLSPprop} that in most cases it will be
possible to show that these photons do not point to the interaction 
region, and hence are likely to come from a delayed neutralino decay,
since the SM background is essentially zero. 
(Of course, at least in the case of Model \#~1, it is very reasonable that 
at the moment of starting the LC operations clear indications for GMSB 
would have already come from the LHC.) Among the two-photon events, 
there will be many coming from $\NI\NI$ production featuring no other 
particles and, if $\sqrt{s} \gtap 300$ GeV, many others including (soft) 
$\epem$ pairs from selectron-pair production as well.
If the c.o.m. energy is even larger, then more complex events, many with 
hadronic activity, would also appear from, e.g., $\CI\CI$ production.  

A feeling of the situation can be obtained from inspection of 
Fig.~\ref{fig:xsec}. In Sec.~\ref{sec:TESLA}, we stressed the importance 
of the ability of a LC of tuning the c.o.m. energy to explore thresholds
with precision. 
First, one could vary $\sqrt{s}$ in big steps and just inclusively count 
two-photon events to get the rough location of the thresholds for the various 
SUSY-production processes (cfr. thick line labeled ``TOT'' in 
Fig.~\ref{fig:xsec}). Then, one could focus on the individual thresholds, 
observe more exclusive characteristics of the signal (for instance, 
$\lR\lR$ production gives $\ell^+\ell^-\gamma\gamma\slashchar{E}$ events
only, with well-defined lepton energy spectra, since $R$-sleptons will 
always decay to $\NI$, and so forth), and vary $\sqrt{s}$ in finer steps 
to get a precise value of the sparticle masses involved in the corresponding 
production process. For the case of a GMSB model like Model \#~1 where about 
10 thresholds are present below $\sqrt{s} = 500$ GeV, it seems reasonable
to assume that a 200 fb$^{-1}$ run (less than 1 year, based on the TESLA 
expected performance) would allow extraction of the light masses with errors 
at the level of fractions of a GeV. Of course, the fine details depend 
on the slope and the magnitude in the vicinity of the thresholds of the curves
for the various individual cross-sections as functions of $\sqrt{s}$ 
(cfr. Fig.~\ref{fig:xsec}). For instance, in absence of important $t$-channel 
contributions, one would expect a steeper $\sim \beta^3$ behaviour for 
gaugino-pair (fermion) production compared to $\sim \beta$ for slepton-pair
(scalar) production, so that gaugino masses could generally be determined with 
higher precision \cite{threshold_slope}. 
On the other hand, in our case the $t$-channel contributions are important 
and, in addition, one can always imagine to spend more machine 
time running close to the ``harder'' thresholds and also use other 
observables (e.g. distributions) to get additional information on the 
spectrum. 

Our intent here is not to simulate fully such a complex study, but to 
evaluate what could be the sensitivity in determining the 
GMSB parameters from the knowledge of the light spectrum that could 
come from a roughly uniform threshold-scanning. Based on Model \#~1 
and a total of 200 fb$^{-1}$ collected between 200 and 500 GeV c.o.m. 
energies, we estimated the following approximate precisions for the 
sparticle masses:
\bea 
\Delta(m_{\NI})   \sim 0.2 \; {\rm GeV}; & 
\Delta(m_{\NII})  \sim 0.8 \; {\rm GeV}; & 
\Delta(m_{\CI})   \sim 0.1 \; {\rm GeV}; \nonumber \\
\Delta(m_{\eL})   \sim 0.2 \; {\rm GeV}; &
\Delta(m_{\eR})   \sim 0.2 \; {\rm GeV}; &
\Delta(m_{\muR})  \sim 0.8 \; {\rm GeV}; \label{eq:errmass} \\
\Delta(m_{\tauu}) \sim 0.8 \; {\rm GeV}; &
\Delta(m_{\taud}) \sim 2.0 \; {\rm GeV}; &
\Delta(m_{h^0})   \sim 0.1 \; {\rm GeV}. \nonumber
\eea 
The assumption on $\Delta(m_{h^0})$ is based on the fact that many 
$\epem\to h^0\Z$ events
would be observed at the LC if Model \#~1 is realized and many other 
Higgs events would have already seen and studied at the LHC\footnote{
Here, we assume that the theoretical error on determining the lightest 
Higgs mass from any SUSY-model input parameters, currently at the level 
of at least a few GeV \cite{Sven} will be reduced by that time by more 
detailed calculations. Similarly, we imagine that the theoretical error 
on the sparticle masses will also be brought at a level comparable to 
the numbers quoted above.} 

 Also, in Sec.~\ref{subsec:N1mass}, we will see that a measurement 
of the $\NI$ mass with a precision at the level of a few tenths GeV 
can be easily achieved by looking at the $\gamma$ energy spectrum from 
$\NI$ decays. 

\begin{figure}
\centerline{
\epsfxsize=0.55\textwidth
\epsffile{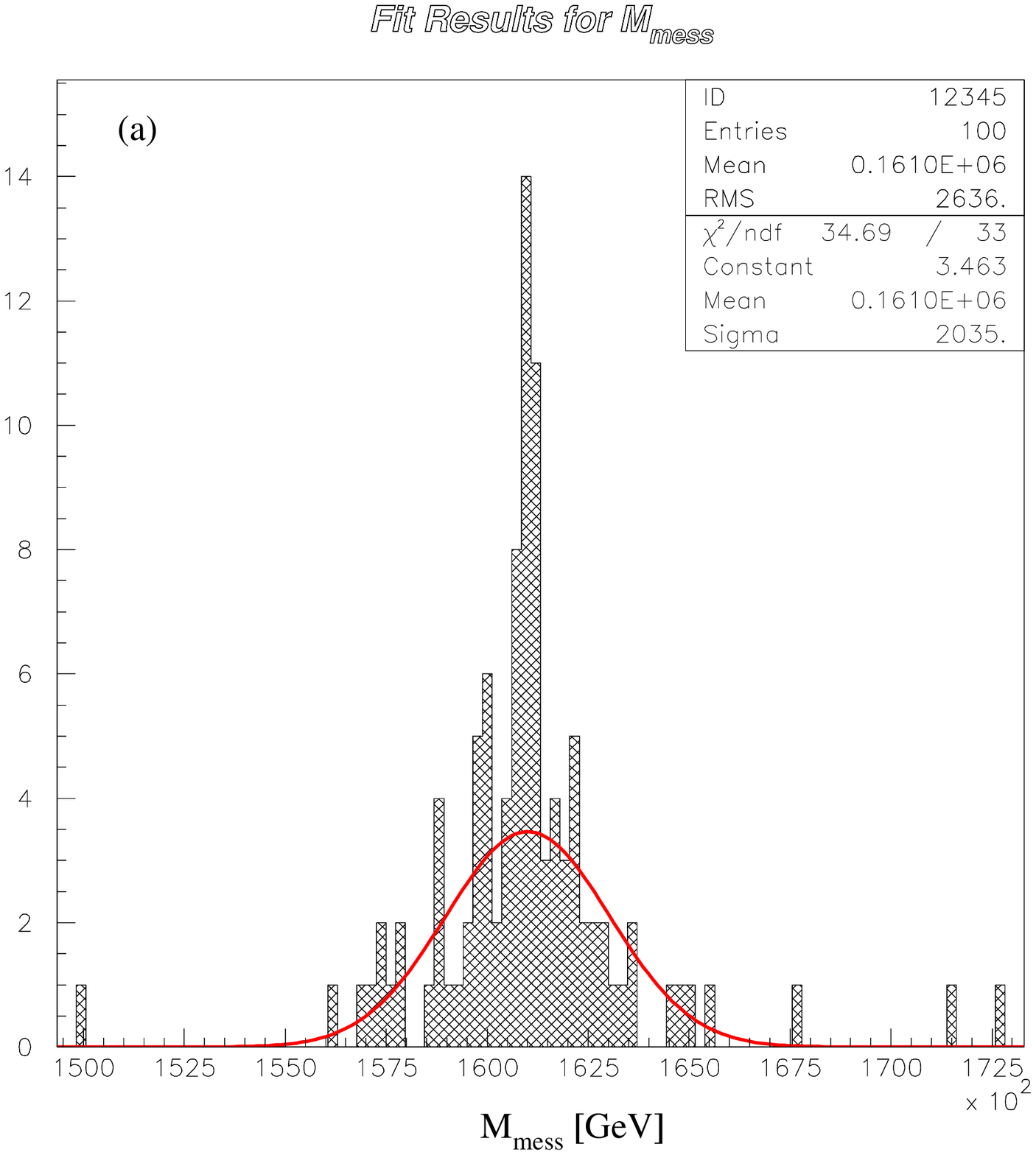}
\epsfxsize=0.55\textwidth
\epsffile{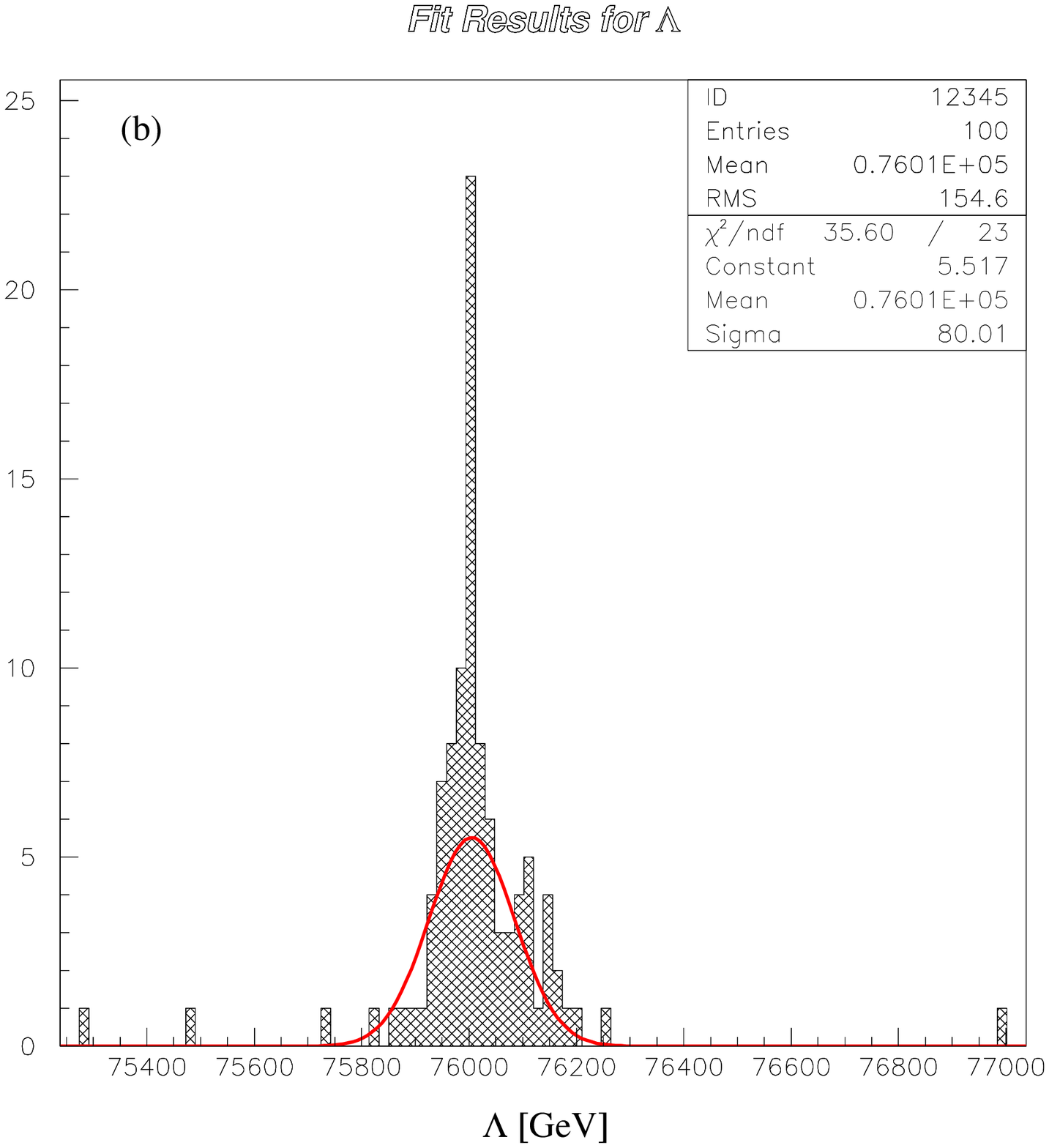}
}
\caption{\sl
Reconstructed values and fit to $M_{\rm mess}$ (a) and $\Lambda$ (b) for 
Model \#~1 as a result of 100 possible sparticle spectrum measurements 
from threshold scanning, as described in the text.   
}
\label{fig:mmesslambdafit}
\end{figure}

\begin{figure}
\centerline{
\epsfxsize=0.55\textwidth
\epsffile{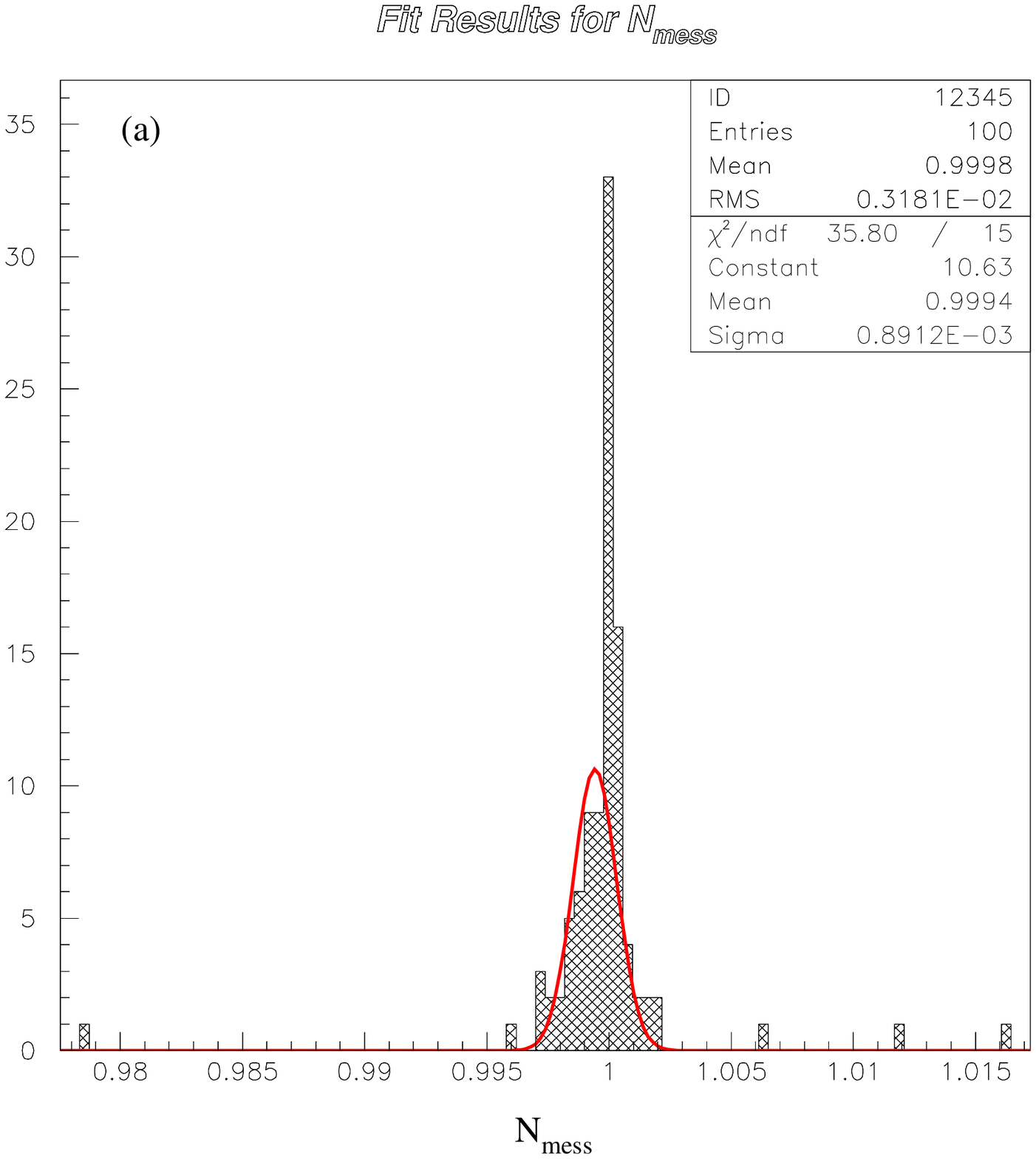}
\epsfxsize=0.55\textwidth
\epsffile{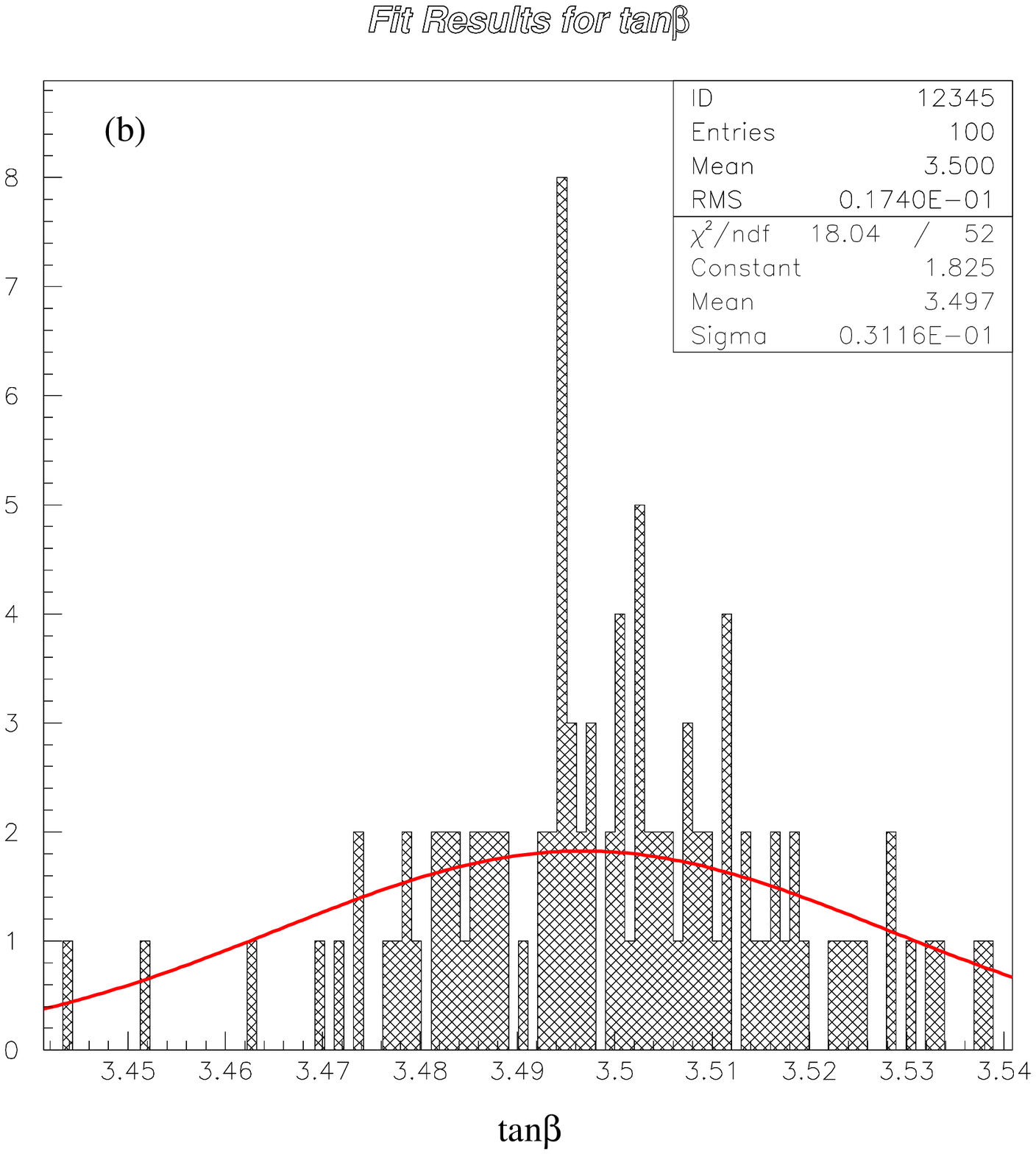}
}
\caption{\sl
As in Fig.~\ref{fig:mmesslambdafit}, but for $N_{\rm mess}$ (a)
and $\tan\beta$ (b).
}
\label{fig:nmesstanbetafit}
\end{figure}

 We used a home-made computer program called 
{\tt MinuSUSY} \cite{SAsoft}, interfaced to {\tt SUSYFIRE} and {\tt Minuit}
\cite{Minuit}, to perform fits to SUSY-model basic parameters starting 
from information on the sparticle spectrum\footnote{{\tt MinuSUSY} does
not take higher-order corrections to the sparticle masses into account,
but these can typically be reabsorbed in a redefinition of the basic 
model parameters and a shift of the starting values needed to generate a 
given spectrum. For our purpose here, however, the precise values of the 
SUSY parameters are not the point, since we are only interested in 
evaluating the level of sensitivity one could reach by using these 
techniques. We believe our indications in this respect to be found below 
are still valid without taking fine effects into account.}. 
The program works both with (m)GMSB and (m)SUGRA models and in ``global'' 
or ``local'' mode. 
The ``global'' mode is intended to determine which class of SUSY models 
and which approximate values of the basic parameters best recover the 
input spectrum. We used this run mode starting from the light spectrum 
of Model \#~1 (cfr. Tab.~\ref{tab:Model1}) and found that indeed there is 
no mSUGRA model that can reasonably fit it. Such a spectrum could be 
recovered with good precision only by releasing one or more of the unification 
assumptions at the GUT scale. In contrast to mSUGRA, the minimal GMSB
framework allowed us to single out a successful region of the parameter
space including Model \#~1.    
Once a rough knowledge of the basic parameter values is 
obtained, it is possible to run {\tt MinuSUSY} in ``local'' mode around these 
values and get optimised values and errors on them based on the input errors 
on the sparticle masses. Basically, one simulates a large number of possible 
sets of mass measurements using a gaussian distribution for the masses around 
the central values that one would get from the chosen underlying SUSY model.  
Starting from the errors on the masses for Model \#~1 quoted in 
Eq.~(\ref{eq:errmass}), we simulated 100 sets of measurements of the light 
sparticle spectrum.
Our results for the 100 subsequent reconstructions of the GMSB parameter set 
performed with {\tt MinuSUSY} are summarized in 
Figs.~\ref{fig:mmesslambdafit}, \ref{fig:nmesstanbetafit}, for 
$M_{\rm mess}$ and $\Lambda$, $N_{\rm mess}$ and $\tan\beta$, respectively. 
Here we did not require $N_{\rm mess}$ to be an integer and considered it
as a real variable to perform the fits. (Notice that non-integer 
values of $N_{\rm mess}$ are possible in some non-minimal classes of 
GMSB models, cfr. e.g. Ref.~\cite{Steve-gen}.) 
A gaussian (+ constant) fit to the distributions gives the results
shown in Tab.~\ref{tab:GMSBfit}. As for the sign of $\mu$, we found 
that the best fits are obtained for $\mu > 0$, as expected.   

\begin{table}
\renewcommand{\arraystretch}{1.2}
\begin{center}
\begin{tabular}{|c||c|}                          \hline
Parameter      & Fitted value                 \\ \hline \hline  
$M_{\rm mess}$ & ($161    \pm 2$)      TeV    \\ \hline
$\Lambda$      & ($76.01  \pm 0.08$)   TeV    \\ \hline
$N_{\rm mess}$ & $0.9994 \pm 0.0009$          \\ \hline
$\tan\beta$    & $3.50   \pm 0.03$            \\ \hline
\end{tabular}
\end{center}
\caption{\sl
Results of fits to the parameters of GMSB Model \#~1 starting from a 
possible set of light sparticle masses measurements via threshold scanning 
technique, as described in the text. A 200 fb$^{-1}$ run at the LC is
assumed. 
}
\label{tab:GMSBfit}
\end{table}

Tab.~\ref{tab:GMSBfit} indicates that it seems possible to determine 
$\Lambda$ and $N_{\rm mess}$ with a precision of about 1 part in $10^3$ 
and $\tan\beta$ with 1 part in $10^2$, by just using threshold scanning
and sparticle masses as observables and running for less than 1 year at a LC
with $\sqrt{s} \le 500$ GeV. Of course, to achieve such an impressive goal, 
it is crucial that one can count on the high-luminosity, such as the option 
proposed for TESLA. 
The only parameter that could not be determined at a level of 1\% 
or better is $M_{\rm mess}$, but this is well understandable since the 
sparticle masses depend only logarithmically on it. Better precision could
well be reached if other observables (total cross sections, distributions, 
branching ratios etc.) were added to the global fits. On the other hand, 
it must be said that Model \#~1 is a particularly ``easy'' model, in the
sense that it yields a light spectrum and the various thresholds are well
separated. It would be much more difficult if the scenario of Model \#~2 
(for which LC energies well above 500 GeV would be needed to extract the 
GMSB parameters) or Model \#~3 (with many sparticles almost degenerate) 
were realized. 

\section{Event and Detector Simulation} 
\label{sec:BRAHMS}

\noindent
To generate GMSB events we used a modified version of {\tt SUSYGEN 2.20/03}
\cite{SUSYGEN}, where the 3-body neutralino decays were added and the 
corresponding kinematical distributions were input numerically, according to 
the discussion of Sec.~\ref{sec:NLSPdecay} and the results obtained with 
{\tt Gravi-CompHEP}, for our reference Models \# 1--3. 
For each GMSB model, the relevant input cards to {\tt SUSYGEN} were calculated 
with {\tt SUSYFIRE}, keeping $\sqrt{F}$ (and hence the $\NI$ lifetime) as a 
free parameter, subject to the bounds discussed in Secs.~\ref{sec:intro} and 
\ref{sec:NLSPdecay}. The generated events were then passed to our detector 
simulation software, {\tt BRAHMS} \cite{BRAHMS}. 
This is a {\tt GEANT}~3.21 \cite{Geant321} code including material and 
tracking detectors, as motivated by the ECFA/DESY CDR \cite{CDR}. 
The relevant detector components are simulated as follows.  

The beampipe is taken as a tube of beryllium of radius 1.0 cm and thickness 
$0.14\% X_0$, where $X_0$ is the radiation length. 
Five layers of vertex detectors (VXD), each of thickness $0.12\% X_0$
and point resolution of 3.5 $\mu$m are located at radial positions
of 1.2, 2.4, 3.6, 4.8 and 6.0 cm, with respective half $z$-lengths
of 2.5, 5.0, 7.5, 10.0 and 12.5 cm. 

We include an intermediate tracking chamber (ITC) as material with a total 
of  $0.23\% X_0$  and dimensions 12 cm $< r < 30$ cm and $|z| < 100$ cm,
but we do not consider this detector for the track fit.
In the forward and rear directions, we include a forward tracking detector 
(FTD) made of disks of silicon strip detectors each of thickness 
300 $\mu$m at $z$-positions of 40, 50, 120, 140 and 160 cm, with outer 
radii of 10, 10, 30, 30, 30 cm respectively and inner radii of
2.5, 2.5, 10.1, 11.7, 13.3 cm. All elements have $r\phi$ resolution
of 25 $\mu$m.

We use a time projection chamber (TPC) as central tracker with
inner active radius of 38.6 cm, outer active radius of 162.6 cm
and active longitudinal half-length of 250 cm. The active volume
is filled with gas (which we take to be argon) and provides a maximum of
118 hit points along a track, each with point resolution of 160 $\mu$m
in $r\phi$ and 0.1 cm in $z$. The inner wall to the TPC
consists of a total of $3\% X_0$ of aluminium.  This is an important
source of conversions which we discuss further in our analysis below.

For the calorimeter part of our simulator, we use an electromagnetic
calorimeter (ECAL) and assume simple gaussian smearing 
with resolutions motivated by those in the CDR and
given in Tab.~\ref{tab:calparms}.

\begin{table}
\renewcommand{\arraystretch}{1.4}
\begin{center}
\begin{tabular}{|l||c|} \hline
Angular Coverage  & $|\cos \theta | < 0.95$ \\ \hline
Barrel $r$- Dimensions (cm) & $ 172 < r < 210$ \\ \hline
Endcap $z$-Dimensions (cm) & $280 < |z| < 330$\\ \hline
Energy Resolution (\%) &  $10.3/\sqrt{E \; [{\rm GeV}]} + 0.6$ \\ \hline
Spatial Resolution (cm)&  $4/\sqrt{E \; [{\rm GeV}]} + 2 $\\ \hline
Angular Pointing Resolution (mrad) & 
$50/\sqrt{E \; [{\rm GeV}]}$ \\ \hline
Time Resolution (ns) & $ 2/\sqrt{E \; [{\rm GeV}]} + 0.5 $\\ 
\hline
\end{tabular}
\end{center}
\caption{\sl 
Calorimeter parameters used in Monte Carlo smearing.} 
\label{tab:calparms}
\end{table}

In addition to the detector resolutions, there is an additional
uncertainty in the position of the i.p. due to the
beam spot size. In the following, we take the beam spot dimensions
as given by the TESLA machine design parameters at 500 GeV c.o.m. energy
\cite{tesla_parameters} and thus apply gaussian smearing to the 
production vertex of the neutralinos with
$\sigma_x = 553$ nm , $\sigma_y = 5$ nm and $\sigma_z=400$ $\mu$m.

\section{Measuring the NLSP Properties and the \\ 
         Fundamental SUSY Breaking Scale at the LC}
\label{sec:NLSPprop}
\noindent 
In this section, we focus on practical methods to measure the NLSP 
properties and, in particular, its mass and lifetime. As discussed in 
Sec.~\ref{sec:NLSPdecay}, in GMSB (or in more general LESB) models, 
the NLSP lifetime can be macroscopic and this opens a very important
window for inspecting SUSY breaking physics, which is not available 
in HESB models, like mSUGRA. Indeed, as Eq.~(\ref{eq:NLSPtau}) shows, 
measuring $m_{\rm NLSP}$ and $c\tau_{\rm NLSP}$ determines the fundamental
scale of SUSY breaking $\sqrt{F}$ up to the factor ${\cal B}$ 
(cfr. Fig.~\ref{fig:lifetosqrtf}), that can also be measured in principle. 
We refer to the specific case of neutralino NLSP scenarios and, in particular, 
to the typical Models \#~1--3 discussed in Sec.~\ref{sec:GMSBmodels}. 

\subsection{Overview of Experimental Techniques}
\label{subsec:overview}
\noindent 
In the following, we will first describe several techniques that we propose 
for performing such measurements. 
Depending on the $\NI$ lifetime, these methods involve (and test) different 
parts of the detector, requiring high and 
somewhat unusual performances. Hence, this study could be an important 
benchmark in the process of designing a LC detector and the related simulation 
software and should not only be seen as limited to SUSY searches.

We assume here that at least one SUSY production process (i.e. 
$\epem\to\NI\NI$) is accessible at the LC, so that two neutralinos (plus 
possibly some cascade decay products) appear for each SUSY pair produced.   
The neutralinos must then decay to a $\G$, based on the discussion of 
Sec.~\ref{sec:NLSPdecay}. 

\begin{figure}
\centerline{
\epsfxsize=0.55\textwidth
\epsffile{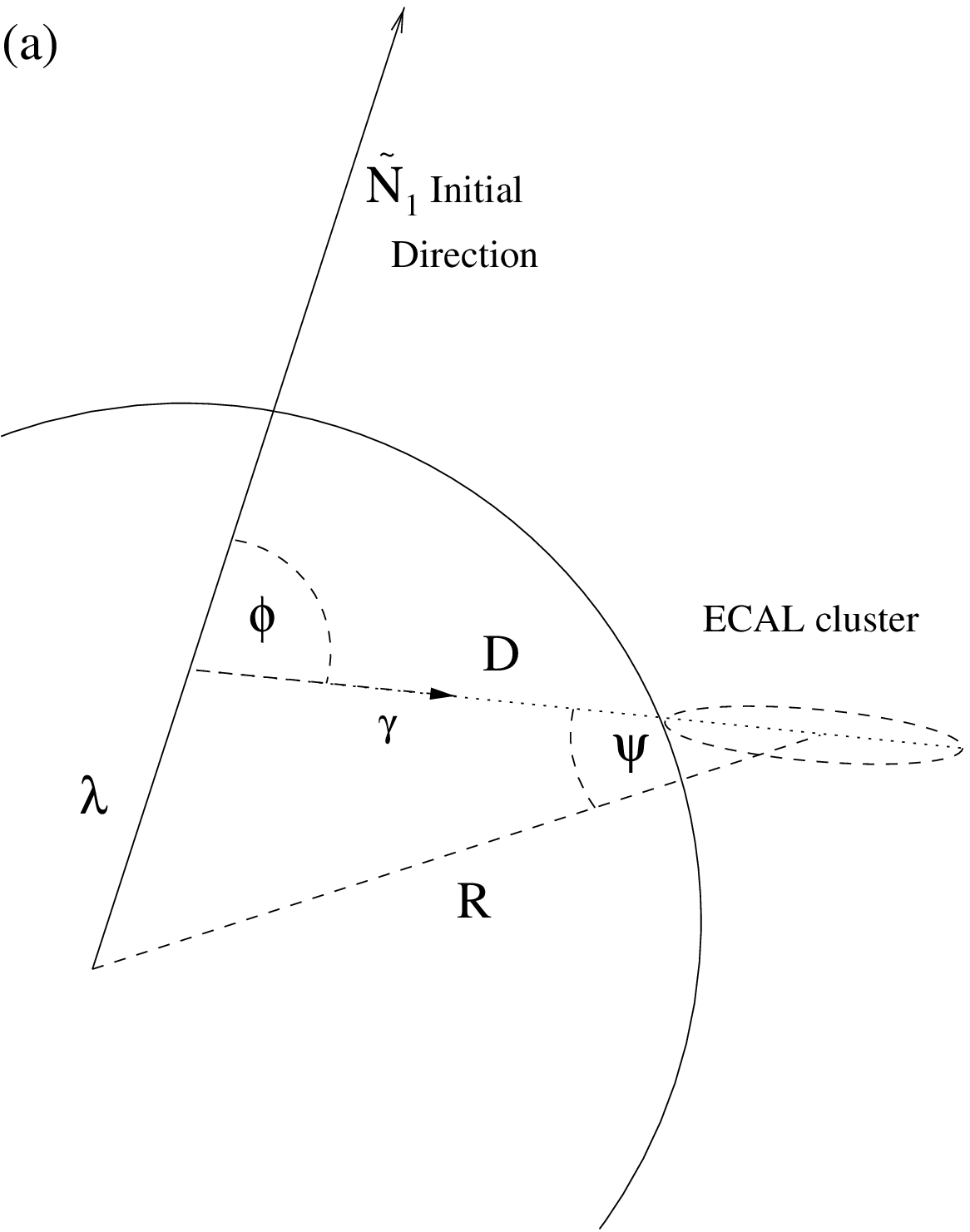}
\epsfxsize=0.40\textwidth
\epsffile{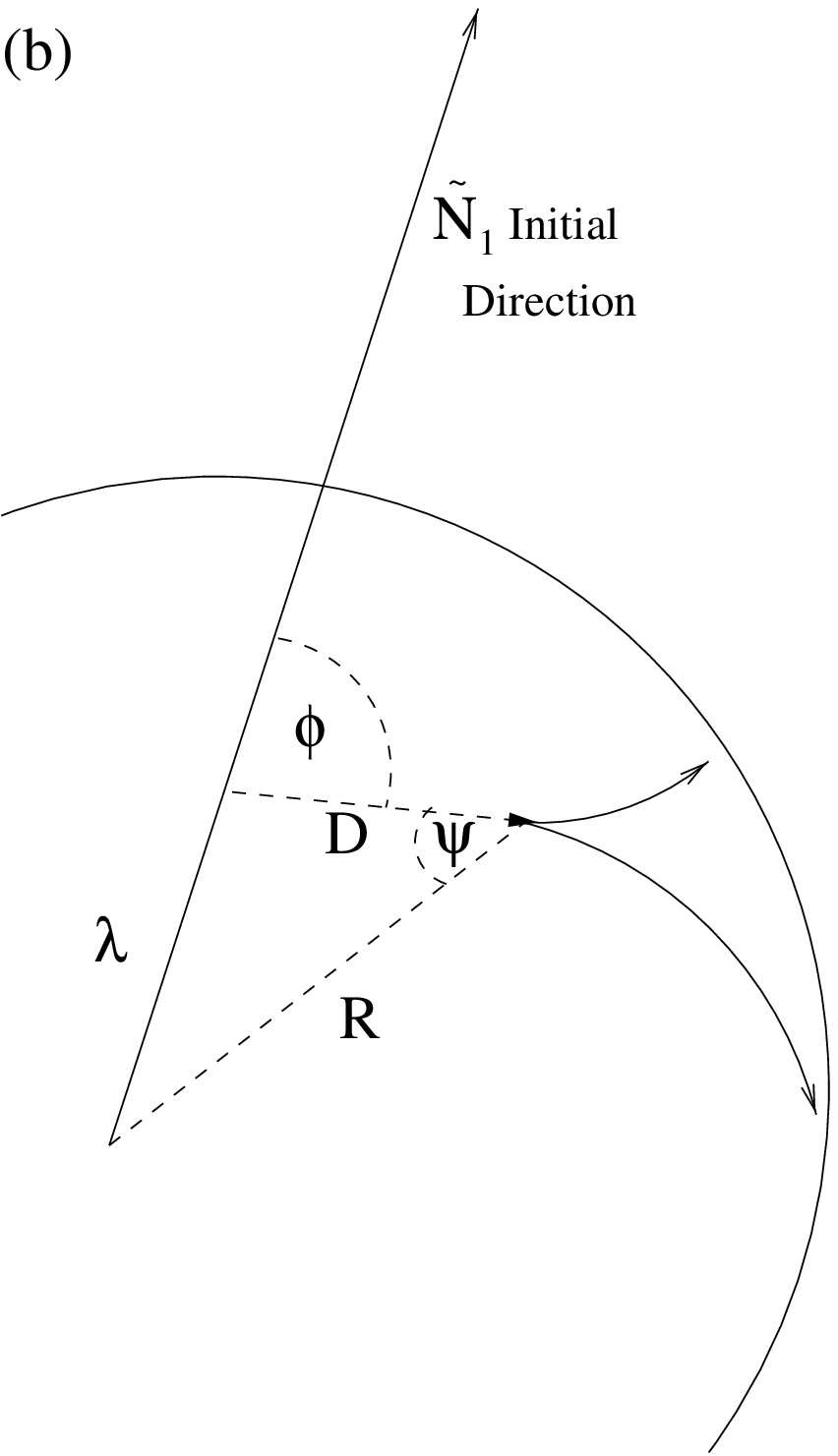}
}
\caption{\sl 
Topologies of neutralino ($\NI$) decays. (a) Pure photonic decay;
(b) charged decay (where $D=0$) or photonic decay plus $\gamma$ 
conversion ($D\neq 0$)
}
\label{fig:neutralino_decay}
\end{figure}

The topology of the $\NI$ decay is sketched in 
Fig.~\ref{fig:neutralino_decay} where Fig.~\ref{fig:neutralino_decay}(a)
shows the case for a purely photonic $\NI\to\gamma\G$ decay and 
Fig.~\ref{fig:neutralino_decay}(b) shows the cases where a ``charged 
decay'' (e.g. $\NI\to\mu^+\mu^-\G$) has occurred ($D=0$) or where the 
$\gamma$ has converted subsequent to a photonic decay ($D\neq 0$). 
In each event, another $\NI$ is present. If the process generating the 
neutralinos is simply NLSP-pair production, then the other $\NI$ moves 
approximately colinearly with the one shown, where the acolinearity angle 
depends only on the ISR and beamstrahlung. 
This information could be used to constrain the event reconstruction, but
it requires detailed knowledge of the acolinearity angle distribution. 
For simplicity, we treat each neutralino decay as independent in
the following analyses.

We have seen in Sec.~\ref{sec:NLSPdecay} that in most cases the $\NI$ decays 
to a $\G$ (which escapes the detector) and an observable component. 
The latter can be a photon or a visible $f \bar{f}$ pair coming from a 
``charged decay''. In the first case, the photon can be observed via 
its shower in the ECAL. However, in a real detector there is always material 
between the i.p. and the tracking volumes, so a fraction of the 
photons coming from SUSY production will undergo conversion in the material.  
Using the tracking detectors to reconstruct the resulting $\epem$ pairs, it 
is possible to obtain a very accurate determination of the original photon
energy and direction.

In the following, we list the observable final states (for each decayed
neutralino) and introduce the concepts and methods that we will use
later for $\NI$ lifetime measurements.    
\begin{description}
\item[a) Photon.]  We use this final state in 
the calorimeter pointing technique (see Sec.~\ref{subsec:pointing}) 
for laboratory decay lengths $L = \beta\gamma c \tau_{\rm NLSP}$ between 
approximately 5 cm and 200 cm. 
Calorimeter timing provides additional information for decay 
lengths from about 20 cm to about 120 cm (see Sec.~\ref{subsec:timing}).
Also, the statistical method of Sec.~\ref{subsec:stat} will be based
on counting events with photonic final states.   
Furthermore, photon conversions in detector material can also be used 
to measure $L$ in the short range, however we will show below that greater 
precision can generally be obtained by using ``charged decays'' only, after 
eliminating conversions by using appropriate experimental cuts. 

\item[b) $e^+e^-$ pairs.] ``Charged decays'' to $\epem$ occur relatively 
abundantly with BR's of order 1--2\% percent (cfr Sec.~\ref{sec:NLSPdecay}) 
and we will use them to measure values of $L \ltap 20$ cm.
In addition, this final state also occurs at the level of a few percent 
when photons convert in detector material. We will show in 
Sec.~\ref{subsec:3d} how to differentiate between conversions and 
``charged decays". For the latter case, 
the reconstructed $\epem$ vertex corresponds to the decay vertex of the 
$\NI$. For the former case, an extrapolation is required 
in order to obtain the neutralino decay point,  
see Fig.~\ref{fig:neutralino_decay}(b). 
This procedure is discussed in detail in Sec.~\ref{subsec:3d} below.

\item[c) $\mu^+ \mu^-$ or $h^+h^-$ pairs.] (Here $h^\pm$ is a charged 
``stable'' hadron.) To improve statistics slightly, we will use these 
final states together with b) in the tracking methods and consider an 
inclusive general ``two-track" topology. Note that in these cases, the 
events are always a result of $\NI$ ``charged decays'' and the relevant
BR's are typically at the level of several percent 
(cfr Sec.~\ref{sec:NLSPdecay}). 
\end{description}

In order to reconstruct the decay parameters from observation of 
$\epem\to\NI\NI$ events, we use the formula
\be
\cos \phi = \frac{E_0}{p_0 } 
- \frac{m_{\NI}^2}{2 p_0 E_\gamma }\; , 
\ee 
where $E_0$ is the nominal beam energy corrected for average
losses due to ISR and beamstrahlung, and 
$p_0 = (E_0^2-m_{\NI}^2)^\frac{1}{2}$. This formula allows the 
angle $\phi$ to be determined directly from the measured photon energy.
The explicit occurrence of $m_{\NI}$ in this equation 
emphasises the necessity of a good neutralino mass measurement.

For the cases where the photon converts to an $e^+e^-$ pair,
or on the case of a ``charged decay'', the line of the photon flight
is determined from the reconstructed vertex and momentum of the pair.
This line combined with the value of $\phi$ obtained from the energy
measurement gives an unambiguous value for $\lambda$ 
[cfr. Fig.~\ref{fig:neutralino_decay}(b)]. 

In the calorimeter pointing method (discussed in detail in 
Sec.~\ref{subsec:pointing} below), the angle $\psi$ and the distance 
$R$ are determined directly from the calorimeter shower reconstruction.  
In this way, the decay length $\lambda$ is determined on an event by event 
basis.

In the calorimeter timing method (discussed further in 
Sec.~\ref{subsec:timing} below), the time measurement gives the
quantity $D+\lambda-R$, the shower position reconstruction gives the 
value of $R$ and the energy measurement gives the value of $\phi$.
Closure of the triangle allows a solution for $\lambda$ to be obtained
in most cases (up to a quadratic ambiguity which is resolved by
the requirement that the decay should take place within the dimensions of
the ECAL).  

\subsection{$m_{\NI}$ Measurement by $E_\gamma$ Spectrum End-Point} 
\label{subsec:N1mass}
\noindent
As we have seen, a good neutralino mass measurement is a central 
requirement to most of the lifetime measurement techniques discussed below.  
Further, precise knowledge of $m_{\NI}$ is essential to extract the parameter
$\sqrt{F}$ from $c\tau_{\NI}$ (cfr Sec.~\ref{sec:NLSPdecay}). 
Here we discuss how this measurement could be made at a LC.

\begin{figure}
\centerline{
\epsfxsize=10.0cm
\epsffile{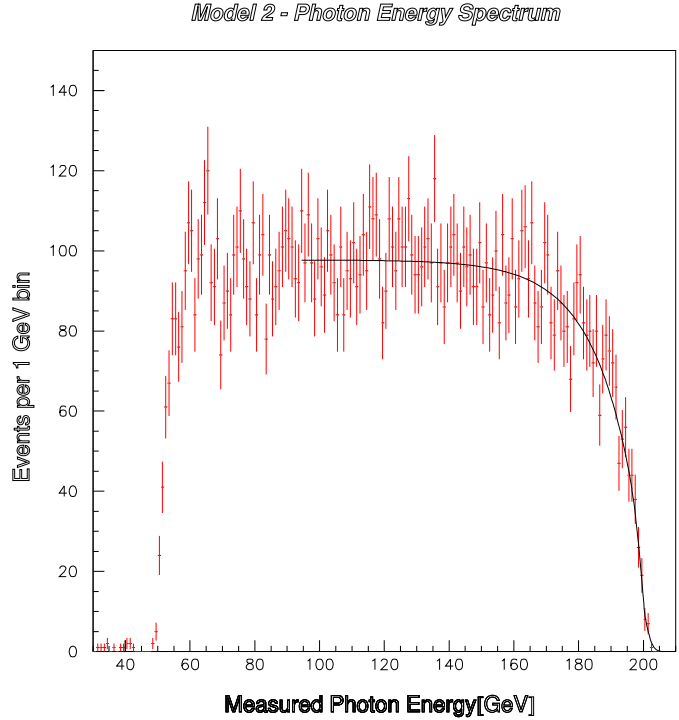} 
}
\caption{\sl 
Photon spectrum resulting from 200 GeV neutralino pair
production with 200 fb$^{-1}$ at $\sqrt{s} = 500$ GeV. ISR,
beamstrahlung and detector effects have been included. Also shown
is the result of the fit described in the text.
}
\label{fig:gamma_spectrum}
\end{figure}

One way to measure $m_{\NI}$ is by determining the end points
of the photon energy spectrum from the $\NI$ decay. When many SUSY 
production channels are open in a neutralino NLSP scenario, 
one gets photons from $\NI\to\G\gamma$ decays with a complicated energy 
spectrum. 
If the process $\epem\to\NI\NI$ is the only one allowed by kinematics,
then before radiative corrections the lower and upper ends of the 
$E_\gamma$ spectrum are always given by
\be 
E_{\gamma}^{\rm min,max} = \frac{1}{4}
\left( \sqrt{s} \mp \sqrt{s-4m_{\NI}^2} \right). 
\label{eq:eminmax}
\ee
In general, while the most energetic photons will always come from those 
neutralinos that are directly pair-produced, the lower end of the spectrum 
will be degraded by the presence of softer photons coming from other SUSY 
processes, in addition to the SM background (cfr. Sec.~\ref{subsec:bkgd}).     
For this reason, we concentrate here on the upper end of the spectrum to 
extract the $\NI$ mass.
The spectrum which would be obtained after detector effects for 
200 GeV neutralino pair production at $\sqrt{s} = 500$ GeV is shown
in Fig.~\ref{fig:gamma_spectrum} for 200 fb$^{-1}$ integrated luminosity
(corresponding to a run of less than 1 ``year'' of $10^7$ s).  
Here we have simulated the SUSY signal that one would detect if Model \#~2
was realized. In this case, we have seen in Sec.~\ref{sec:GMSBmodels} 
that $\sigma(\NI\NI) = 42.3$ fb and 
all the other SUSY-production processes would be below threshold at
$\sqrt{s} = 500$ GeV. 

In order to extract the functional form of the high edge of the spectrum 
after ISR, beamstrahlung and detector effects, a much larger number of Monte 
Carlo events was used to obtain a fit function using the known $\NI$ mass
as an input. The functional form includes two exponentials
to allow for ISR and beamstrahlung together with a cumulative
normal distribution, Freq, to account for the
calorimeter resolution integrated over the sharp edge of the spectrum.  
The function thus obtained was \\

\bea
\frac{1}{n_0} \frac{dn}{dE} & = &
0.788 - 0.716 \exp \left(\frac{E-E_\gamma^{\rm max}}{13.5}\right)
-0.0722 \exp \left( \frac{E-E_\gamma^{\rm max}}{6.06} \right) \nonumber \\ 
&  &  \mbox{}
+ 0.212 \: {\rm Freq}\left( \frac{E-E_\gamma^{\rm max}}{1.87} \right), 
\label{eq:egamma_fit}
\eea
where $n_0$ and $E_\gamma^{\rm max}$ are now free parameters.
This functional form was then used to fit to the 
200 fb$^{-1}$ worth of simulated data as shown in 
Fig.~\ref{fig:gamma_spectrum} to give the result
$E_\gamma^{\rm max} = (200.21 \pm 0.21)$ GeV.  
To extract the neutralino mass from this measurement we use 
Eq.~(\ref{eq:eminmax}) to obtain $m_{\NI}= (199.7 \pm 0.3$) GeV.  
In principle, the $\NI$ mass can also be obtained to high precision 
at a LC by scanning over the threshold region and previous studies \cite{CDR} 
suggest that a precision of order 100 MeV could be obtained from
such scans (cfr. also Sec.~\ref{sec:THRESCAN}).  
However, our point here is that even from an early run with a few months 
worth of data collected at a nominal, fixed c.o.m. energy, a neutralino mass 
measurement with precision at the level of 2\% would be possible.  

\subsection{Measuring the Neutralino NLSP Decay Length Using 2D Projective 
            Tracking}
\label{subsec:tracking}

\noindent 
In this section, we concentrate on determining the $\NI$ average decay
length $L = \beta\gamma c \tau$ from the distribution of reconstructed 
two-track events.
To this purpose, we require events with at least one photon in the ECAL 
(coming from one of the two neutralinos produced) and in addition two charged 
tracks which can be reconstructed to form a vertex (coming from the 
other neutralino decaying through ``charged channels'' or to a photon
that converts).   

\begin{figure}
\centerline{
\epsfxsize=0.55\textwidth 
\epsffile{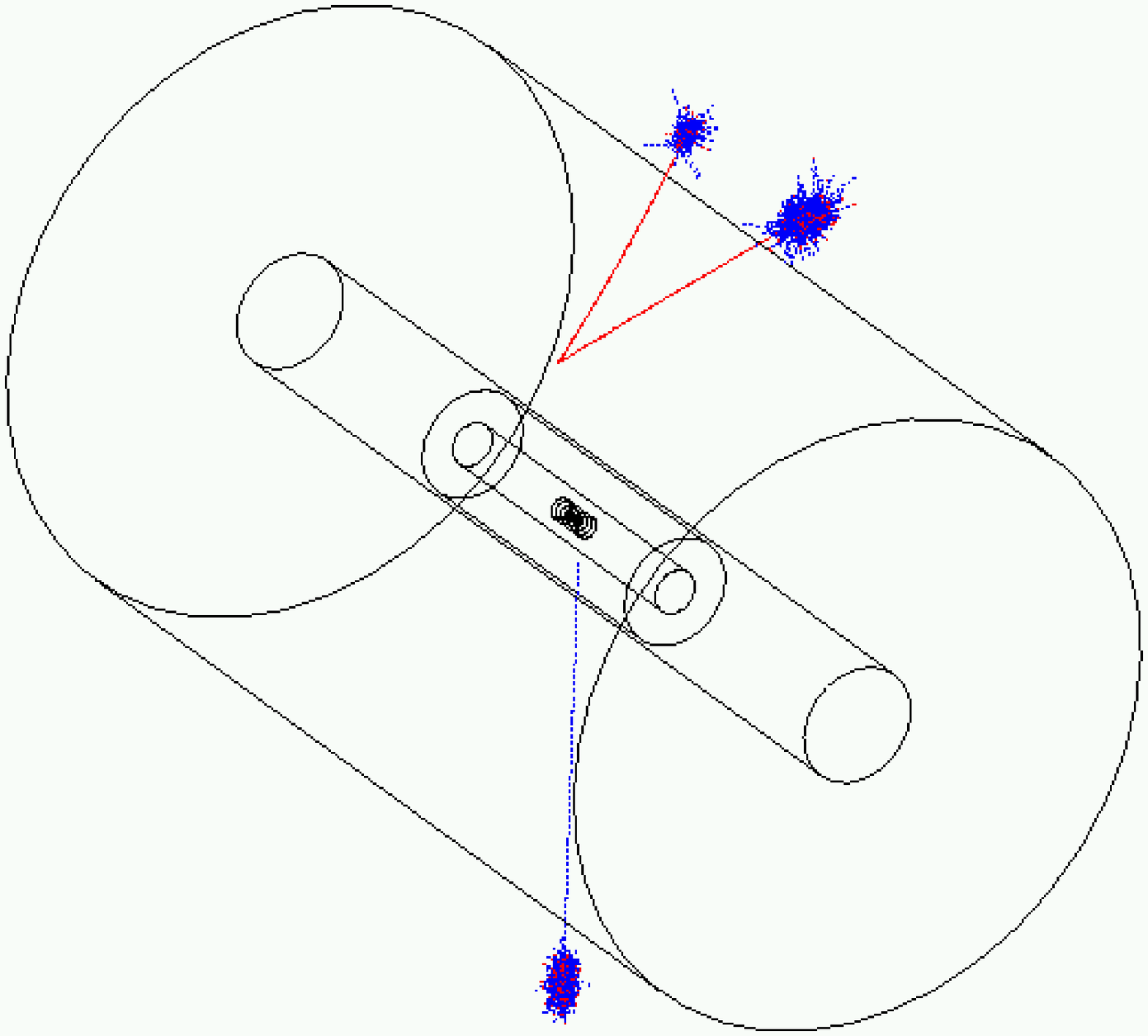} 
\epsfxsize=0.55\textwidth
\epsffile{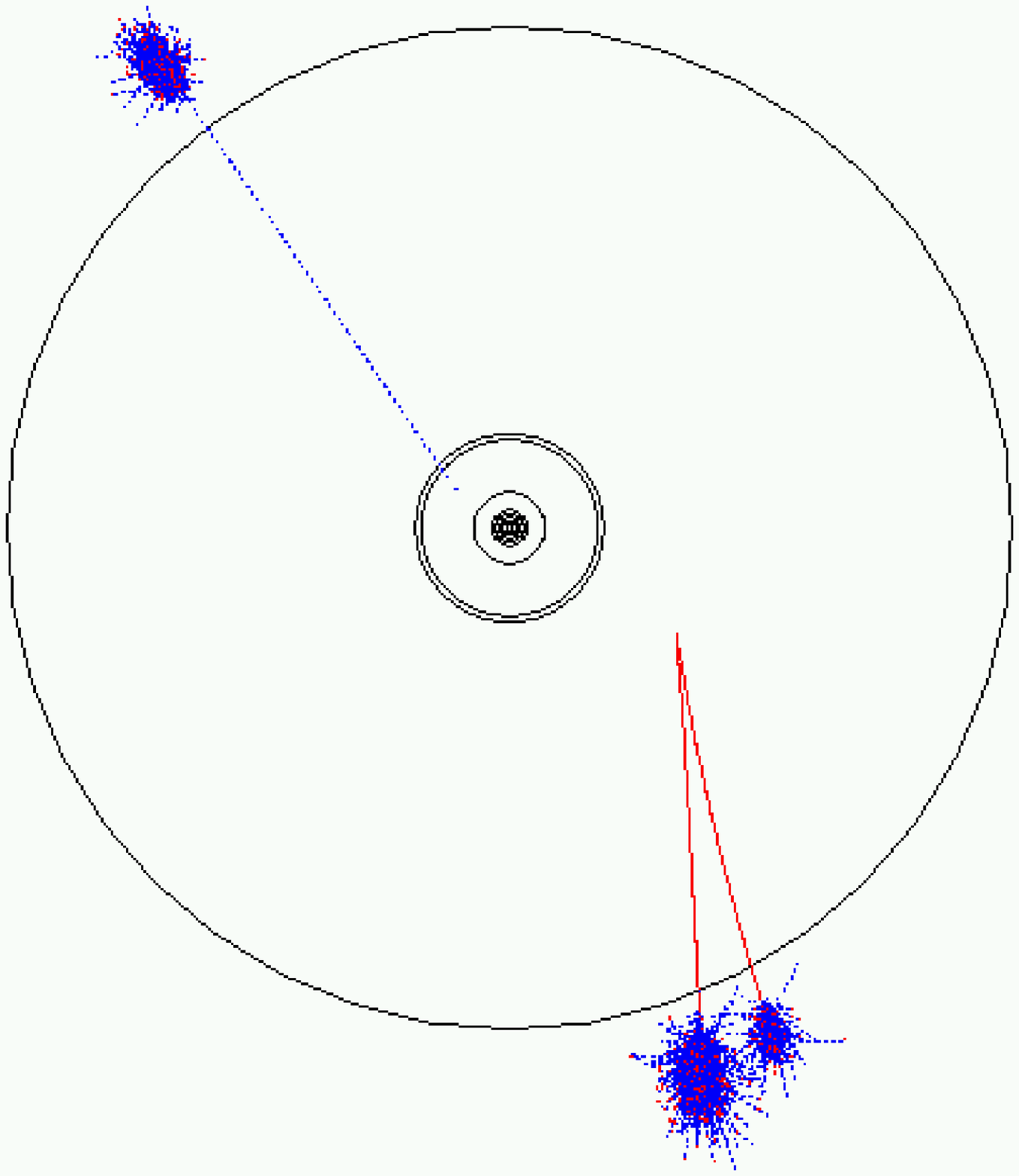} 
}
\caption{\sl 
3D- and end-view of a representative two-track 
($\epem + \gamma + \slashchar{E}$) event, fully simulated in the proposed 
LC detector. 
}
\label{fig:event}
\end{figure}

In Fig.~\ref{fig:event}, we show a representative two-track event
among those we generated and fully simulated in the proposed LC detector
(3D-view on the left, end-view on the right). 
This particular $\NI\NI$ event features one displaced photon and a 
$\epem$ pair coming from a $\NI$ ``charged decay''. The non-zero impact
parameters of all particles are clearly visible; ECAL showers and tracks
are shown (the invisible $\gamma$ path is also indicated). 
Only the vertex detectors and central trackers are displayed. 

The detector track hits are provided by the {\tt BRAHMS} output and
these were then formed into tracks and vertices using a home-made 
reconstruction algorithm. The photon conversion algorithms and multiple
scattering effects are internal to {\tt BRAHMS} and so we have thus
implicitly taken full account of the detector material present.
However, no special provision was made for multiple scattering or for pattern 
recognition effects at the reconstruction stage.

We concentrate first on the case of very short $\NI$ lifetimes, less 
than a few mm, where all the $\NI$ decays take place within the beampipe.  
In this case, any reconstructed $\epem$ pairs will be due to 
``charged decays'' only and so there will be no confusion arising from 
conversions. 
For this region, we must be aware of the beamspot size, which has an rms 
spread in $z$ typically of 400 $\mu$m, meaning that a three-dimensional
decay vertex is no longer useful for lifetime measurements.  Instead
we must project the decay vertex onto the $xy$ plane and determine the
decay length from the resulting distributions of 
$r = \sqrt{\lambda_x^2 + \lambda_y^2}$.

In the following we adopt a conservative approach of using the full
{\tt GEANT} Monte Carlo to generate event samples for a range of true 
(input to {\tt SUSYGEN}) $\NI$ decay lengths and then fit the resulting 
projected lifetime distributions
to a simple exponential, plus a constant to allow for long tails.
The use of this simple fitting function is conservative in that, once
a specific GMSB model is chosen, the projected decay distribution could
be chosen exactly and hence fit the data more accurately.
Only decay length measurements greater than 10 $\mu$m are used in the fits, so
as to eliminate any residual SM backgrounds with tracks originating from
the i.p.

\begin{figure}
\centerline{
\epsfxsize=0.55\textwidth 
\epsffile{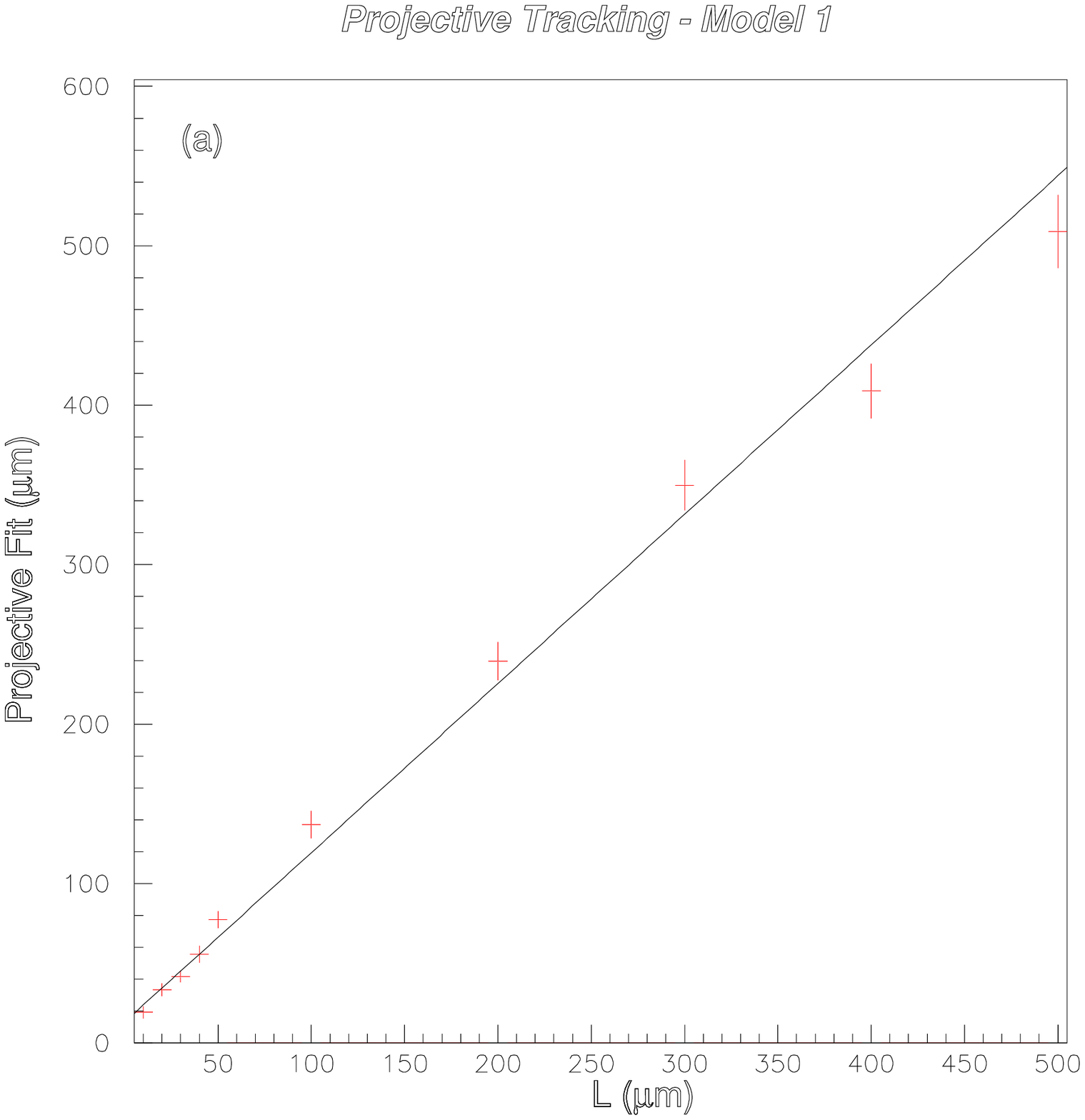} 
\epsfxsize=0.55\textwidth
\epsffile{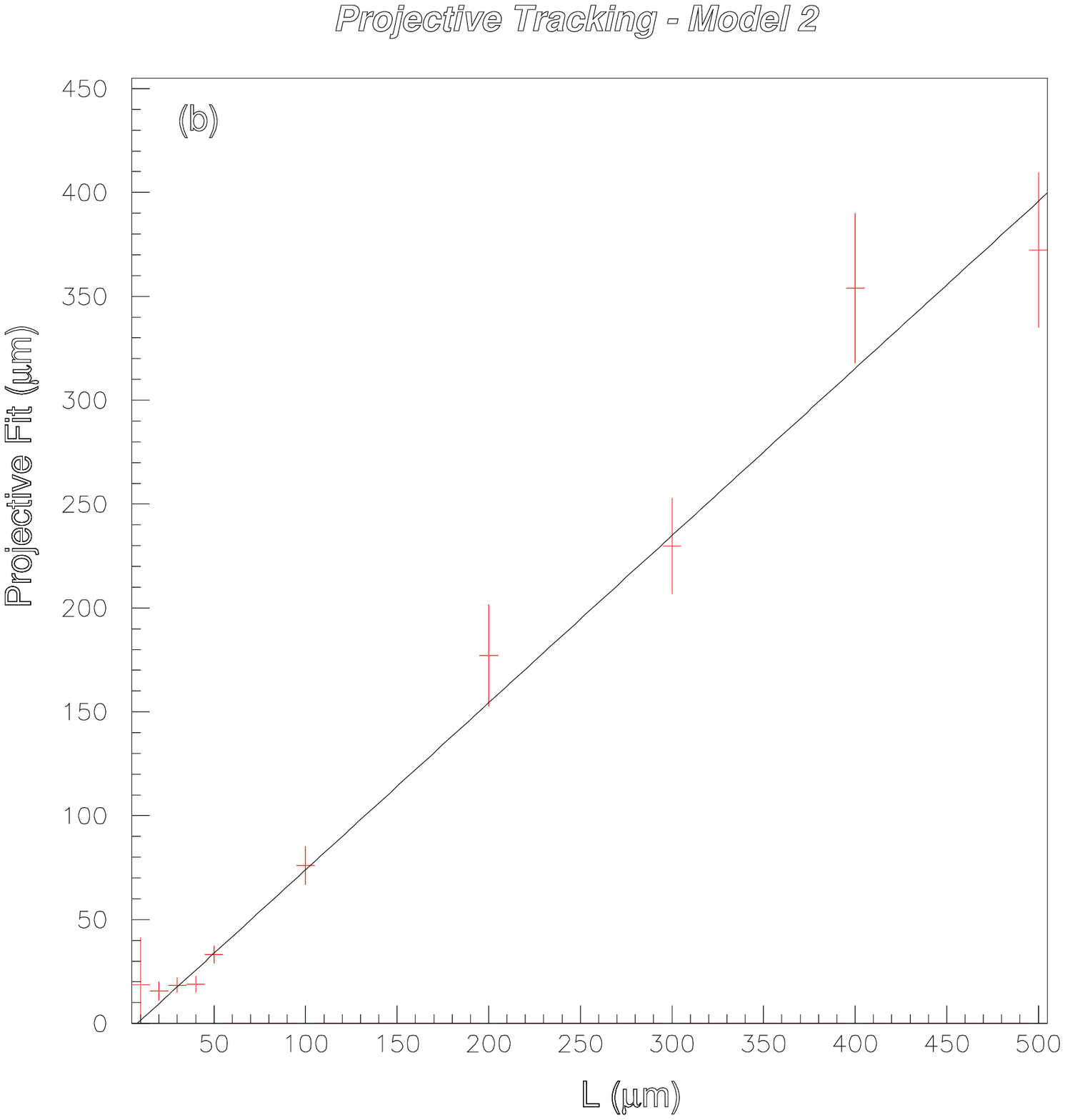} 
}

\caption{\sl 
Results of a fit to the projected lifetime as described in the text
(a) for Model \#~1 and (b) for Model \#~2. In both cases
the error bars correspond to 200 fb$^{-1}$
}
\label{fig:shortlife}
\end{figure}

For Model \#~1, a set of points were generated using 37,600 (corresponding 
to 200 fb$^{-1}$, with $\sigma(\NI\NI) = 188$ fb at $\sqrt{s} = 270$ GeV) 
fully simulated events for each point. Note that only the ``charged decays'' 
are observed here, because any conversions will take place outside
the beampipe, and so the distributions obtained from this sample can be 
applied directly to any model which has a very short-lived NLSP. 

The results are shown in Fig.~\ref{fig:shortlife}(a) where the outcome of 
the fits to the projected lifetime is plotted against the true value of $L$.
The vertical error bars shown are the parabolic errors obtained from the fit
and so correspond to what could be extracted from a run of 200 fb$^{-1}$.
The lifetime is read-off from the fit straight line in order to give the true 
decay length directly; note that any effects due to ISR or vertexing 
systematics are included automatically by such a procedure because all these 
effects are present in the Monte Carlo. This of course also applies to the 
relativistic factor ($\beta\gamma$). 
Reading from the plot, the error bars at the 50 $\mu$m point
correspond to a 10\% error in the decay length measurement whereas the error
bars at 500 $\mu$m correspond to a 4\% error.
It can be seen from the straight-line fit to the points that decay
lengths down to 30 $\mu$m are also well-measured for Model \#~1.

Performing the same fits to the projected distributions according to 
Model \#~2 with 8,460 events (corresponding to 200 fb$^{-1}$, with 
$\sigma(\NI\NI) = 42.3$ fb at $\sqrt{s} = 500$ GeV), gives an error of 13\% 
for a 50 $\mu$m decay length and 10\% for a 500 $\mu$m decay length. 
The results are shown in Fig.~\ref{fig:shortlife}(b).
It should be noted that, although Model \#~2 provides less statistics 
due to a lower neutralino production cross-section, the possibility of 
$\NI$ decay to an on-shell $Z^0$ boson with subsequent decay to 
lepton pairs with significant opening angle somewhat compensates for 
the loss in statistics.

Fig.~\ref{fig:shortlife} also shows that lifetimes shorter than 
30 $\mu$m could also be measured in principle using this technique.
However it should be remembered that, while the intrinsic beamspot
size may be of order 5 nm in $y$ and 500 nm in $x$, the actual position
of the interation point may have to be determined on a pulse-by-pulse
basis which could lead to additional transverse uncertainties. 
It should also be remembered that we are fitting to track pairs 
with essentially zero opening angle, which means that simple assumptions 
about vertex resolution must be avoided.
So in the following we remain conservative in our claim that lifetimes 
as short as 30 $\mu$m could be measured.

Notice that for Model \#~1,
the minimum value of $\sqrt{F}$ allowed by theory is about 110 TeV
(cfr. Sec.~\ref{sec:NLSPdecay}), corresponding to a $\NI$ lifetime 
of about 200 $\mu$m and to a decay length $L \simeq 180$ $\mu$m when 
running at $\sqrt{s} = 270$ GeV, while for Model \#~2, 
$\sqrt{F}_{\rm min} \simeq 212$ TeV 
$\Longrightarrow c\tau_{\NI}^{\rm min} \simeq 70$ $\mu$m 
$\Longrightarrow L_{\sqrt{s} = 500 \; {\rm GeV}} \simeq 53$ $\mu$m. 
So, our result is that at least for these two particular cases, 
this method more than covers the lower end of the range for the 
$\NI$ lifetime. On the other hand, we know that neutralino NLSP models 
exist where $c\tau_{\NI}$ is as short as 5 to 10 $\mu$m 
(cfr. Fig.~\ref{fig:NLSPtau} in Sec.~\ref{sec:NLSPdecay}).
For these cases, an alternative approach would be to increase the 
c.o.m.~energy of the machine, hence increasing the boost of the neutralinos, 
which so far have been close to threshold.  
In this way, shorter decay lengths factors are readily accessible and would
realistically be the preferred approach to measure lifetimes down to 10 
$\mu$m (notice that $(\beta\gamma)_{\NI}$ can never exceed 2.3 for 
$\sqrt{s} \le 500$ GeV). On the other hand, we checked that GMSB models with 
a very short $\NI$ lifetime tend to feature a greater degree of degeneracy 
among the lightest states, so that running at higher energies would also often 
imply opening other SUSY production channels ($R$-slepton pair production). 
As a consequence, it is important to address the problem of being able to 
extract a precise lifetime measurement in the presence of a complex SUSY 
signal, using appropriate selections. 

As an example, we study Model \#~3 at $\sqrt{s} = 470$ GeV with 50,400 
total generated events before cuts, corresponding to 200 fb$^{-1}$, since 
$\sigma(\NI\NI) = 124.6$ fb, $\sigma(\eR\eR) = 64.3$ fb, 
$\sigma(\tauu\tauu) = 31.7$ fb and $\sigma(\muR\muR) = 31.4$ fb. 
The events were then selected at the generator level by the following cuts.
The events had to contain either exactly two 
photons with $\left|\cos\theta\right|<0.95$ 
or exactly one photon with $\left|\cos\theta\right|<0.95$ and one
charged pair of tracks each with $\left|\cos\theta\right|<0.99$.  In
addition to these cuts, an event which contained any additional particle
with $\left|\cos\theta\right|<0.99$ was rejected.  Afer these cuts
the sample contained 21,108 $\NI\NI$ events together with a SUSY 
``background'' of 1,066 selectron pairs, 463 smuon pairs and
336 stau pairs. These events were then passed to the {\tt BRAHMS}
detector simulation followed by the 2D projective fit procedure.

\begin{figure}
\centerline{
\epsfxsize=0.55\textwidth 
\epsffile{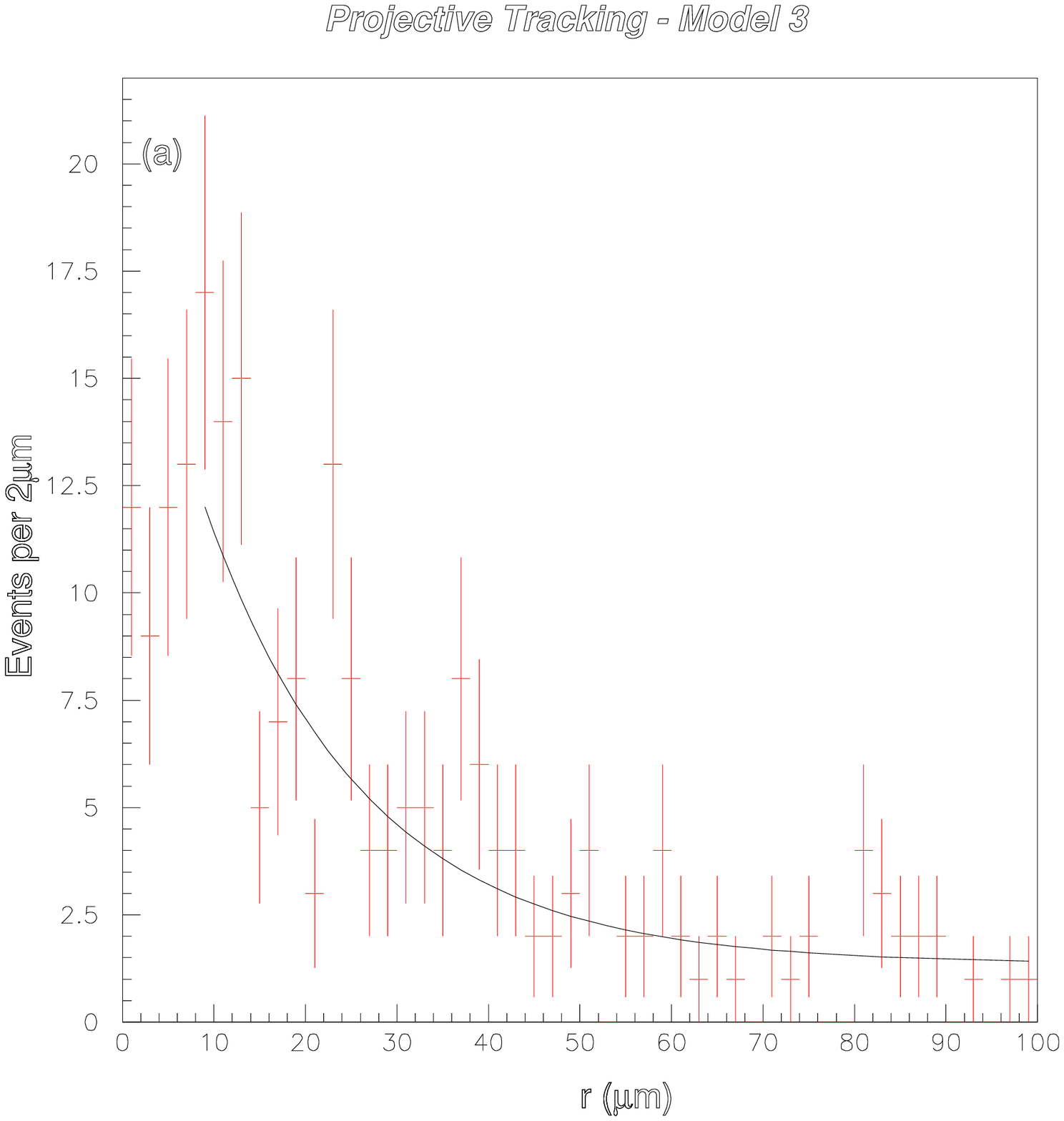} 
\epsfxsize=0.55\textwidth
\epsffile{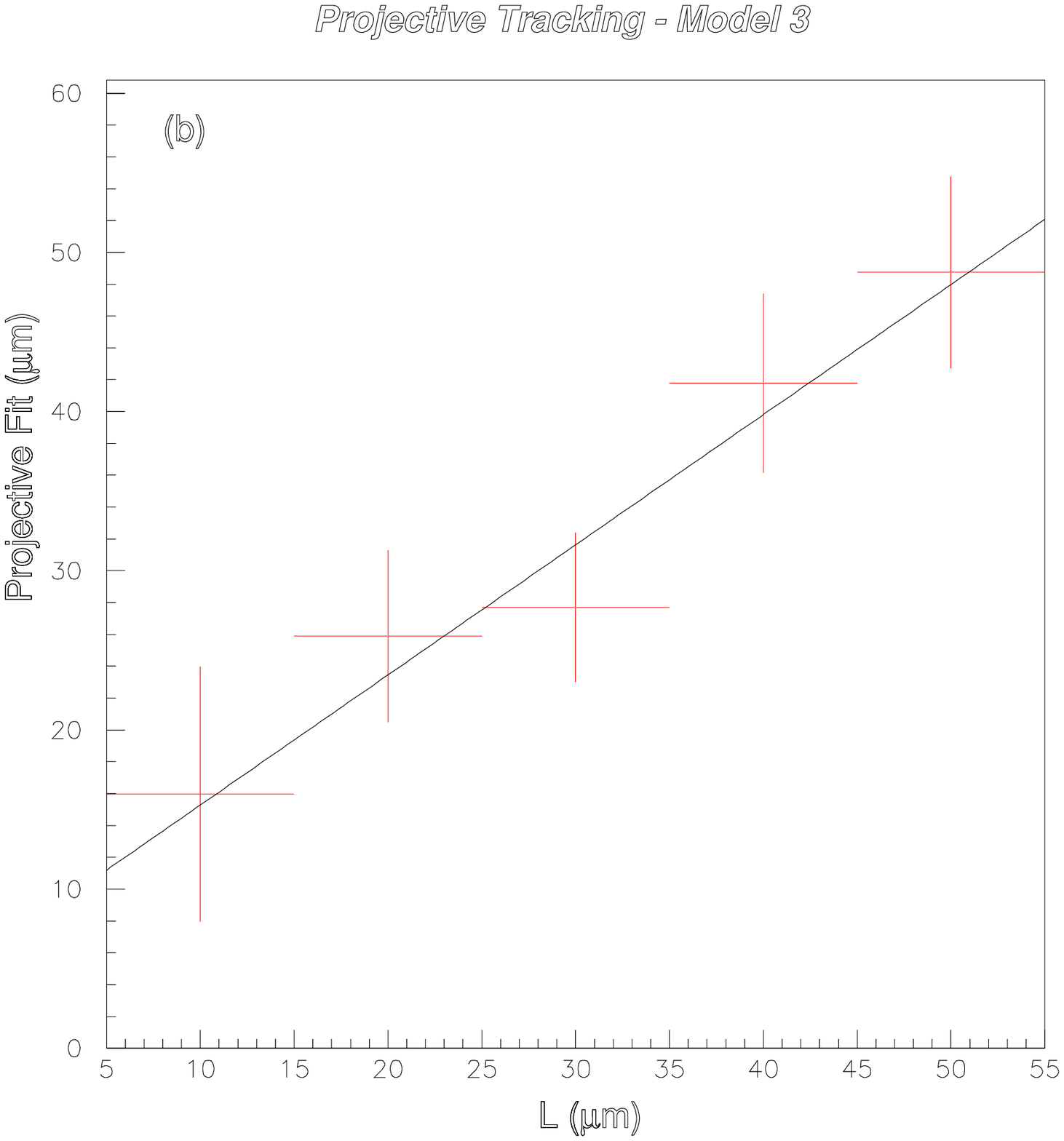} 
}
\caption{\sl 
(a) Reconstructed vertex projective radial distances, $r$, for Model \#~3 
    with $L = 10$ $\mu$m together with a fit to an exponential plus constant.  
(b) Shows the results of the corresponding fits for a range of $L$
    values.
}
\label{fig:model3_plots}
\end{figure}

Fig.~\ref{fig:model3_plots}(a) shows the distribution of
reconstructed vertex projected radii. The remaining SUSY
background consists of only 3 stau events, so its effects 
are negligible in this case. The error bars for $L = 10$ $\mu$m 
correspond to 9.8 $\mu$m. We conclude that for a 200 fb$^{-1}$ run at 
$\sqrt{s} = 470$ GeV, an upper limit on $L$ of approximately 20 $\mu$m 
could be set at the $2\sigma$ level.  The measurement could of course be 
improved with higher luminosity together with running at higher energy to 
utilize a larger ($\beta\gamma$) factor. If we could run with $10^3$ fb$^{-1}$ 
at 800 GeV, then we could expect a gain of $\sqrt{5}$ from the statistics 
together with a gain of 2.2 from an improved ($\beta\gamma$). These would
allow a $c\tau$ measurement to a precision of approximately 20\%.

\subsection{Measuring the Neutralino NLSP Decay Length Using 3D Vertexing}
\label{subsec:3d}

\begin{figure}
\centerline{
\epsfxsize=0.55\textwidth 
\epsffile{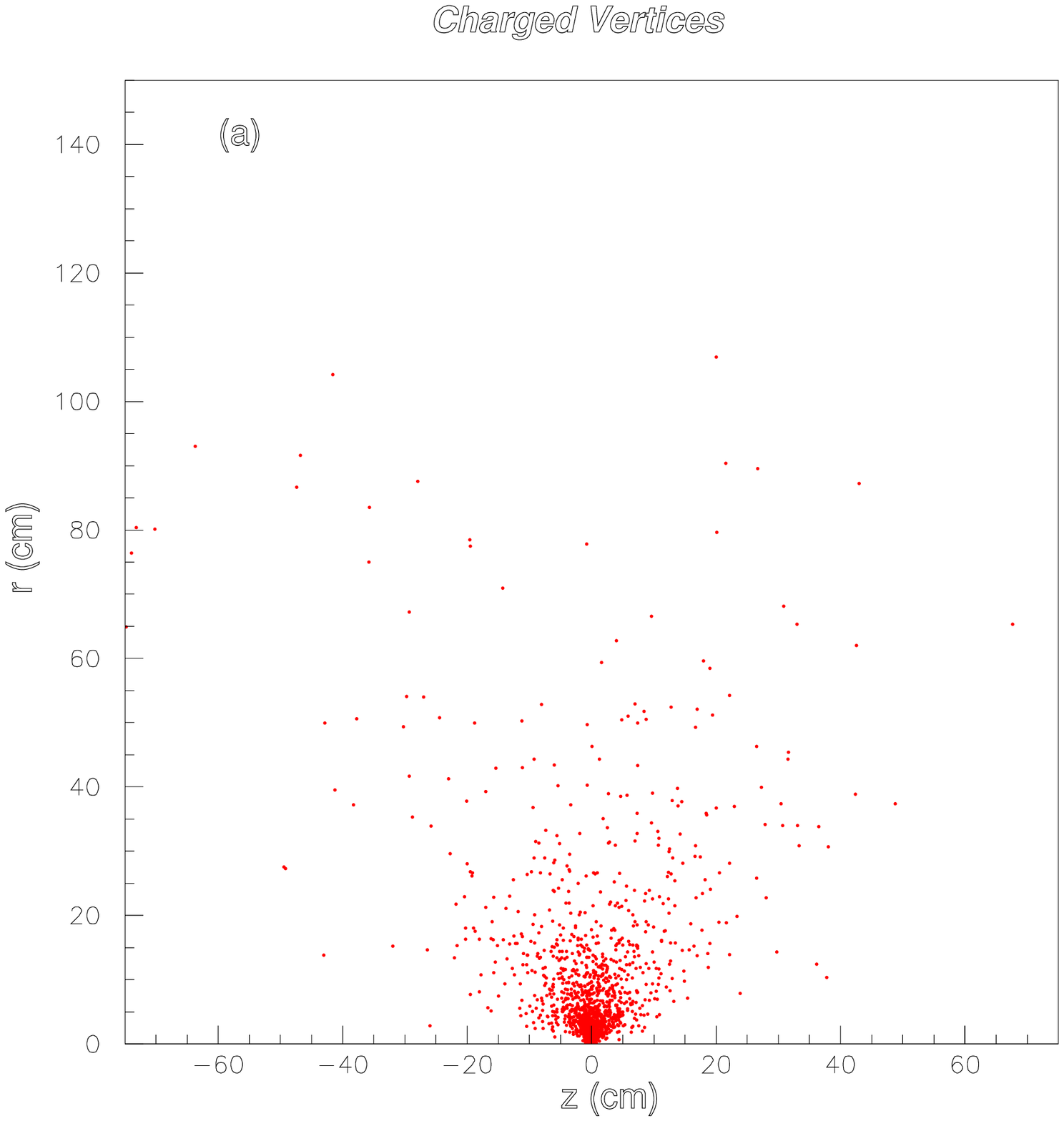} 
\epsfxsize=0.55\textwidth
\epsffile{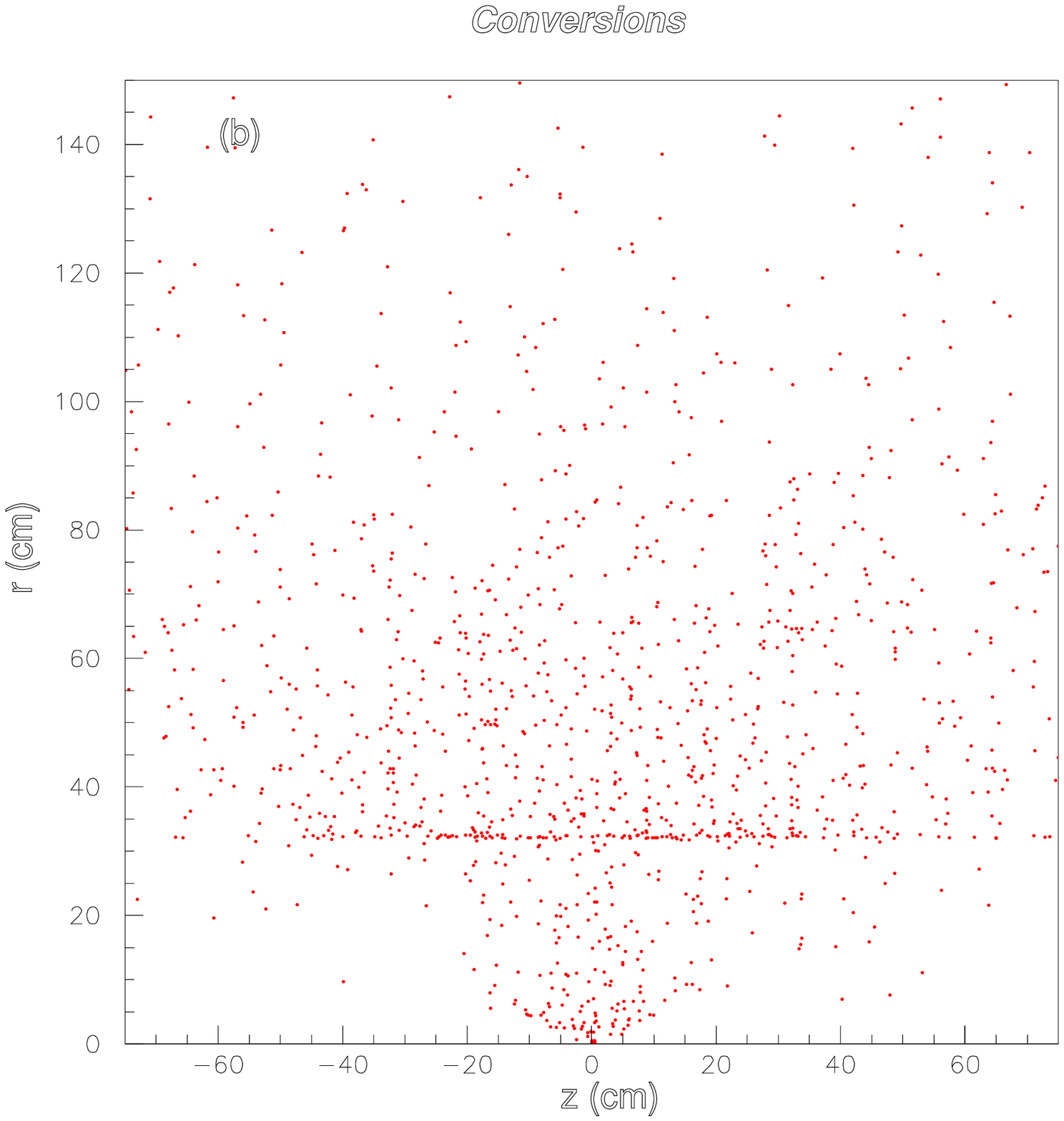} 
}
\caption{\sl 
Reconstructed vertex positions for Model \#~1 with $L = 10$ cm
for (a) ``charged decays'' and (b) conversions of photons
from ``neutral decays''.  The radial coordinate of the vertex $r$ is 
plotted against the longitudinal coordinate $z$.  For case (b) the excess
of conversions in the TPC inner wall and in the inner masks can be seen 
clearly.
}
\label{fig:vertex_distributions}
\end{figure}

\noindent 
For decay lengths greater than about 500 $\mu$m, the three-dimensional 
vertex information is useful because the beamspot size in $z$ is relatively 
less important.  
For this region, photons converting in detector
material must be taken into account because these events will
appear very similar to the ``charged decays''.  The vertex position
of reconstructed tracks from these two processes are shown in 
Fig.~\ref{fig:vertex_distributions} for Model \#~1 with $L = 10$ cm.

\begin{figure}
\centerline{
\epsfxsize=0.55\textwidth 
\epsffile{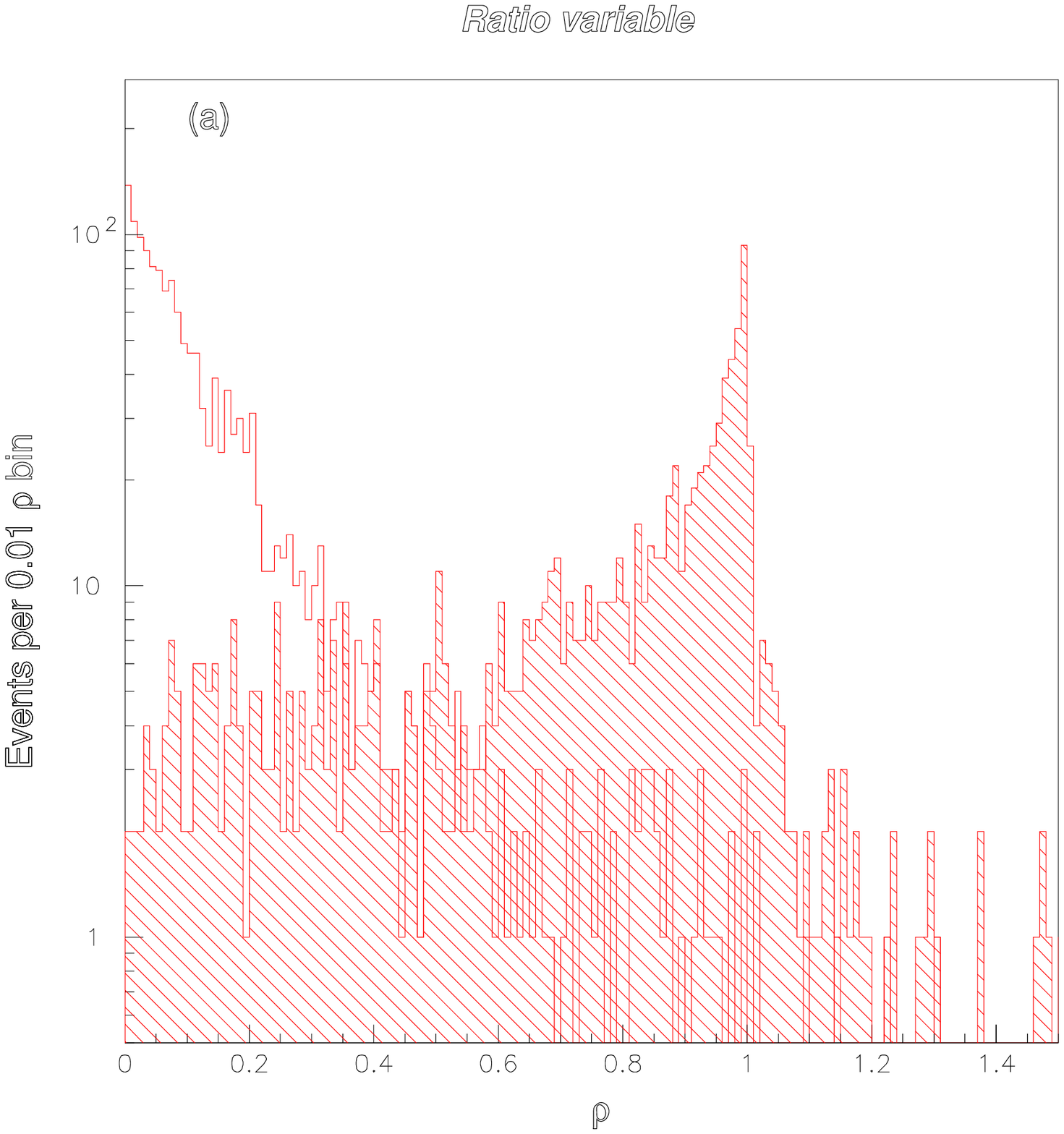} 
\epsfxsize=0.55\textwidth
\epsffile{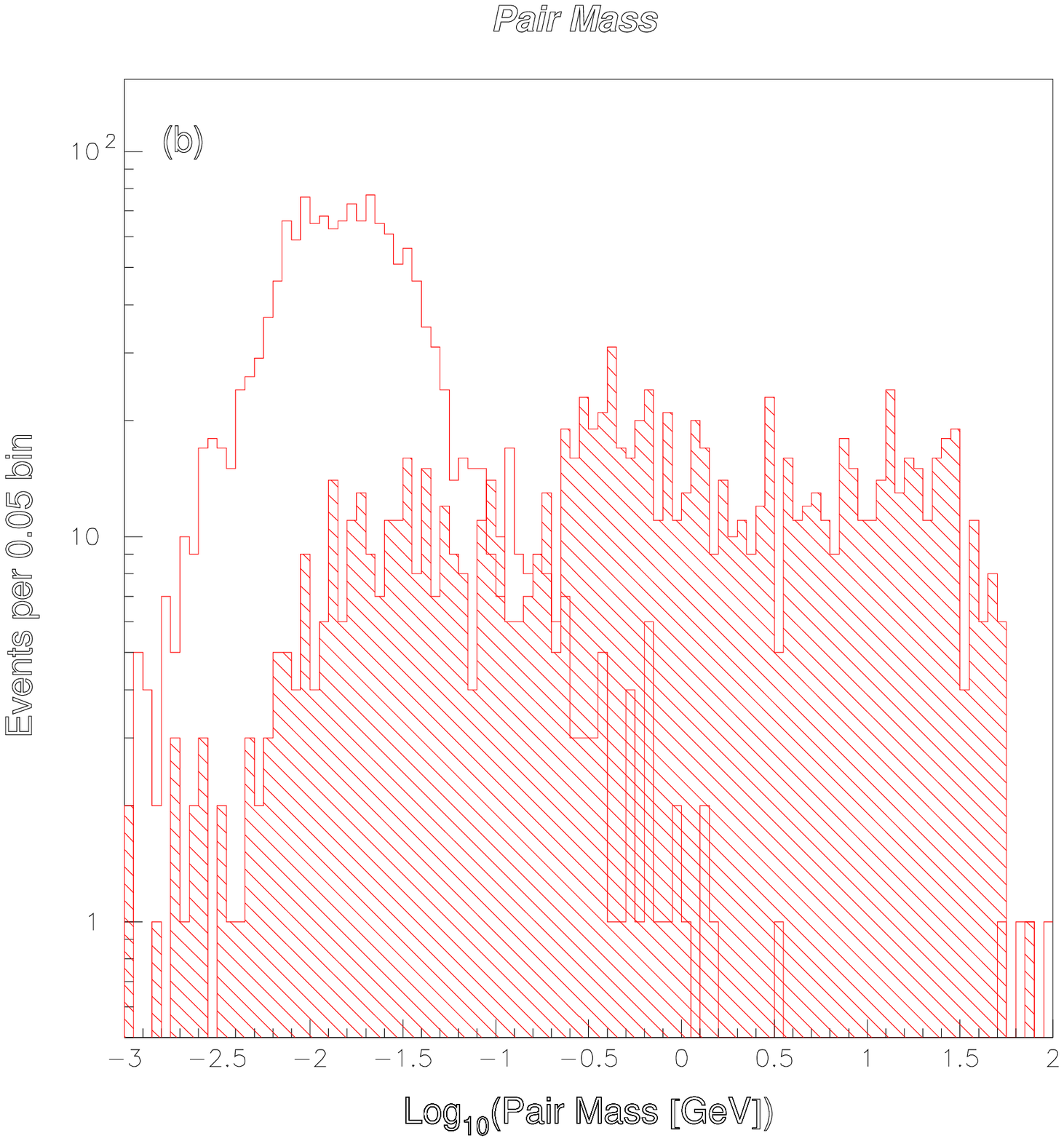} 
}
\caption{\sl 
These figures are for Model \#~1 with $L =10$ cm with statistics corresponding
to 200 fb$^{-1}$. The unshaded histograms are the conversions and the shaded 
histograms are the ``charged'' decays.
(a) Shows the $\rho$ parameter distribution and (b) shows the
invariant mass of the reconstructed track pair.  
Note the logarithmic scales.
}
\label{fig:pair_cuts}
\end{figure}

For both the ``charged decays'' and the conversions,
a line of flight can be defined by the lepton-pair
vertex and the direction of the reconstructed pair momentum. In the case 
of conversions, this line of flight reconstructs that of the photon from the 
neutralino decay.  The ``charged decays'' are also included in this approach 
effectively as  photons which have converted instantly.  
Referring to Fig.~\ref{fig:neutralino_decay}(b), $\cos\phi$ is now determined
by the energy of the reconstructed pair and we have
\be
\lambda = R \frac{\sin\psi}{\sin\phi}  
\ee 
In this way a $\lambda$ measurement can be obtained for
conversions as well as for ``charged decays''.  However, because
we do not wish to make assumptions on the two-track resolution of
the central tracker (and hence the reconstruction efficiency for
conversions in the outer detector),  
we prefer here to use only ``charged decays''
for lifetime determination.  However, as shown in Fig.~\ref{fig:pair_cuts}(a),
we can use the variable $\rho=\lambda/R$ as a discriminant.  In the
following we require $\rho > 0.5$ to improve the ``charged decay'' purity.

The lepton-pair invariant mass distributions for ``charged decays'' and 
conversions are shown in Fig.~\ref{fig:pair_cuts}(b) for Model \#~1
with $L = 10$ cm.
Both peak at very low values, of order 10 MeV, for the detector 
resolutions employed here.  However, the conversions are more sharply
peaked than the ``charged decays'', so cutting on 
20 MeV $ < M_{\rm pair} < 10$ GeV 
improves the purity of the ``charged decay'' sample.  In addition this cut
is desirable because it selects events with a larger opening angle
between the daughter tracks, which results in an intrinsically more
accurate vertex position measurement.  The higher mass cut is to eliminate
any residual SM background from leptonic $W$-pair events.

\begin{figure}
\centerline{
\epsfxsize=10.0cm 
\epsffile{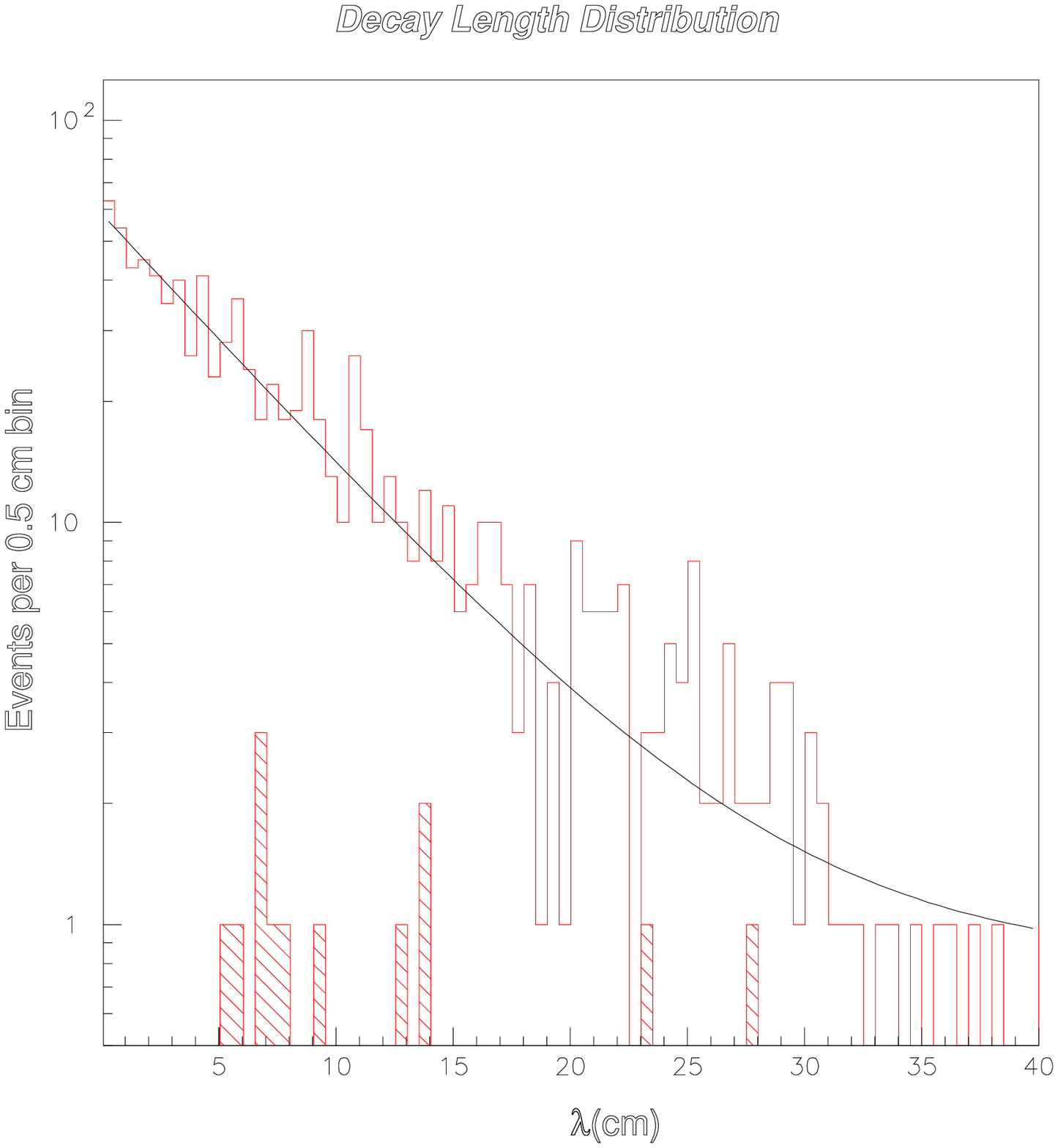} 
}
\caption{\sl 
Distribution of decay length, $\lambda$, for Model \#~1 with $L = 10$ cm
after 200 fb$^{-1}$. 
The 3D vertex reconstruction is used and the distribution is fit to an 
exponential plus constant.  The contributions from conversions 
remaining after cuts are included in the unhatched histogram,
but are also superposed separately as a hatched histogram.
}
\label{fig:exponential_decay_10cm}
\end{figure}

In addition to these cuts we make an additional simple geometrical 
cut of $r < 30$ cm to remove the large number of conversions in the inner 
TPC wall.  After performing these three cuts, the $\lambda$ distribution
is primarily from the ``charged decays'' and this is illustrated in
Fig.~\ref{fig:exponential_decay_10cm}, where the fit to Model \#~1 with
$L = 10$ cm is shown. It is clear from this figure that conversions
have been reduced to a negligible level.

\begin{figure}
\centerline{
\epsfxsize=0.55\textwidth 
\epsffile{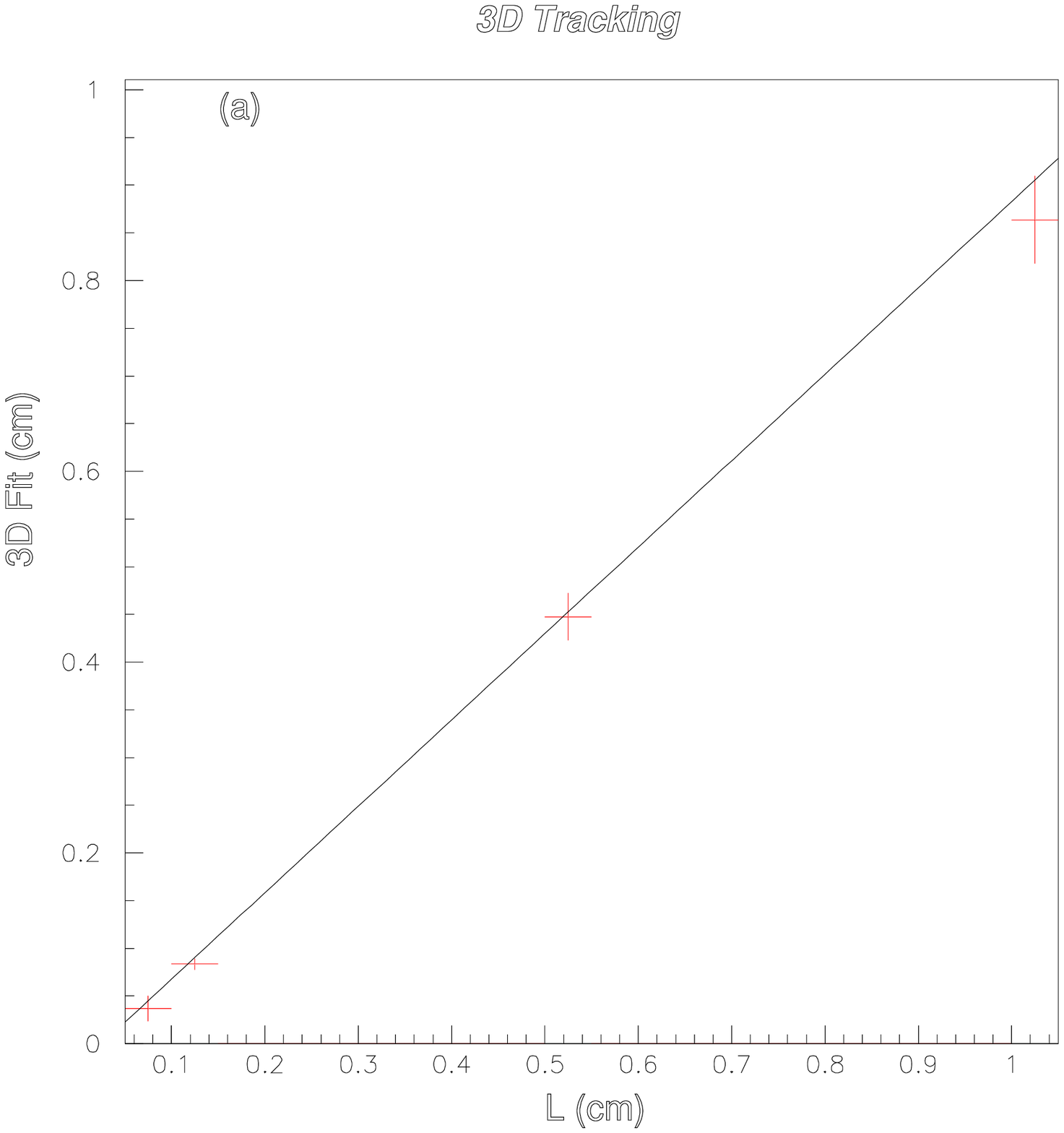} 
\epsfxsize=0.55\textwidth
\epsffile{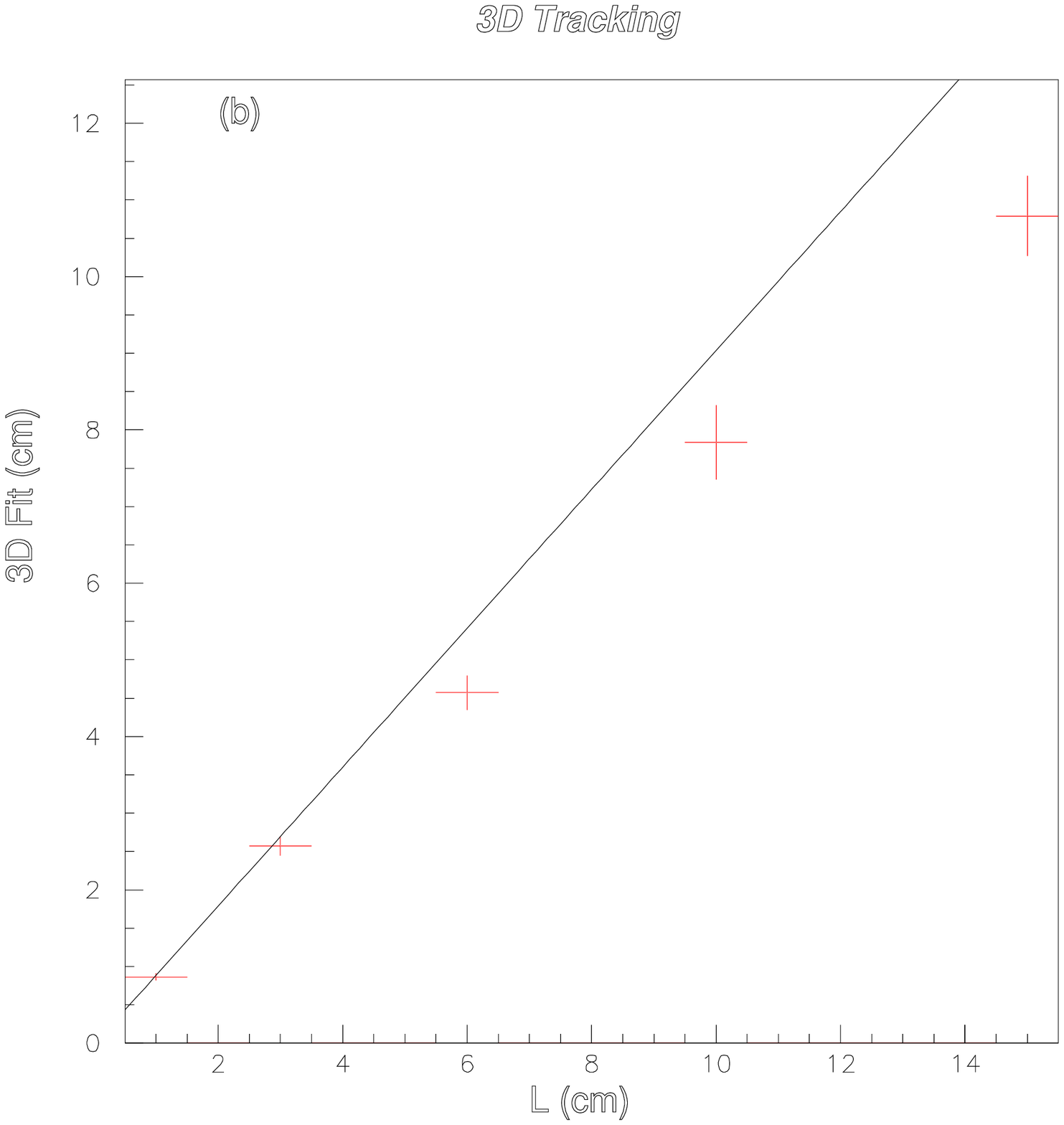} 
}
\caption{\sl 
Results of fits to the 3D decay lengths as a function of the true value
of $L$ using the tracking method for Model \#~1. 
Part (a) for 500 $\mu$m $< L < 1$ cm and (b) is for
the results for 1 cm $< L < 15$ cm.
The error bars correspond to 200 fb$^{-1}$.
The fit lines are described in the text.
}
\label{fig:intermediate_distances}
\end{figure}

This procedure was repeated for a range of $L$ values from 500 $\mu$m
up to 15 cm.
The results are shown in Fig.~\ref{fig:intermediate_distances}
where each point represents the result of the exponential fit,
with corresponding errors for samples consisting of
37,600 generated events  (200 fb$^{-1}$) before cuts.
The Monte Carlo includes the effects of ISR.

Also shown in Fig.~\ref{fig:intermediate_distances} 
is a straight line fit to the data in (a) which is extrapolated
to (b). 
Using this fit we obtain a relative measurement statistical
error of 5\% for $L = 1$ mm and
a relative error of 4\% for $L = 10$ cm.  Our cuts are not
optimised for $L$ greater than about 10 cm.  In this region geometrical
detector effects and variations in track reconstruction efficiency
become important, as evidenced by the deviation from
a straight line in (b).  Rather than correct for these effects in
detail here, we note that this region of $L$ is very well covered by the 
calorimeter pointing method discussed below.

We conclude that it is possible to make a good measurement 
(to 10\% or better) of nominal
decay lengths ranging from 30 $\mu$m up to 10 cm under very conservative
assumptions, using the tracking detectors.  

\subsection{Measuring the NLSP Decay Length Using Calorimeter \\ Pointing} 
\label{subsec:pointing} 
\noindent 
A finely segmented ECAL allows for the possibility of detecting photon
impact parameters with respect to the i.p.  The direction
of the photon is reconstructed by fitting to the distribution of energy 
deposits among the individual ECAL cells which make up the electromagnetic
shower.  The direction finding is improved significantly by the use of 
presamplers to provide a precise point along
the photon direction.  A typical presampler detector consists of 
lead/scintillating fiber structures which both initiate the
shower and measure the point at which the shower begins to a precision
of a few tens of $\mu$m.  The use of more than one presampler
can provide two points along the photon direction and further improve
the direction finding.

In the following, we assume that the angular resolution of the ECAL
is that given in Tab.~\ref{tab:calparms}.   We also need to take
account of the spatial and energy resolutions as listed there.
Referring to Fig.~\ref{fig:neutralino_decay}(a),
the angle $\psi$ and the distance $R$ are determined directly from
the calorimeter shower reconstruction.  The angle $\phi$ is calculated
from the energy of the shower, together with the measured value of the
neutralino mass, as described above.

The effect of the mass measurement error was estimated as follows.
Each Monte Carlo event was reconstructed using the nominal 100 GeV mass
and also using assumed measured masses of 100.1 GeV and 99.9 GeV.
We showed in Sec.~\ref{subsec:N1mass} 
that a mass measurement error of 0.3 GeV could be obtained in a first
run of the detector using the upper end of the photon energy spectrum.
It would be natural to assume, given any such discovery of neutralino
production, that sufficient luminosity would be made available to
perform detailed threshold scans and reduce the mass uncertainty to
the level of 0.1 GeV (cfr. Sec.~\ref{sec:THRESCAN}). 

\begin{figure}
\centerline{
\epsfxsize=0.55\textwidth 
\epsffile{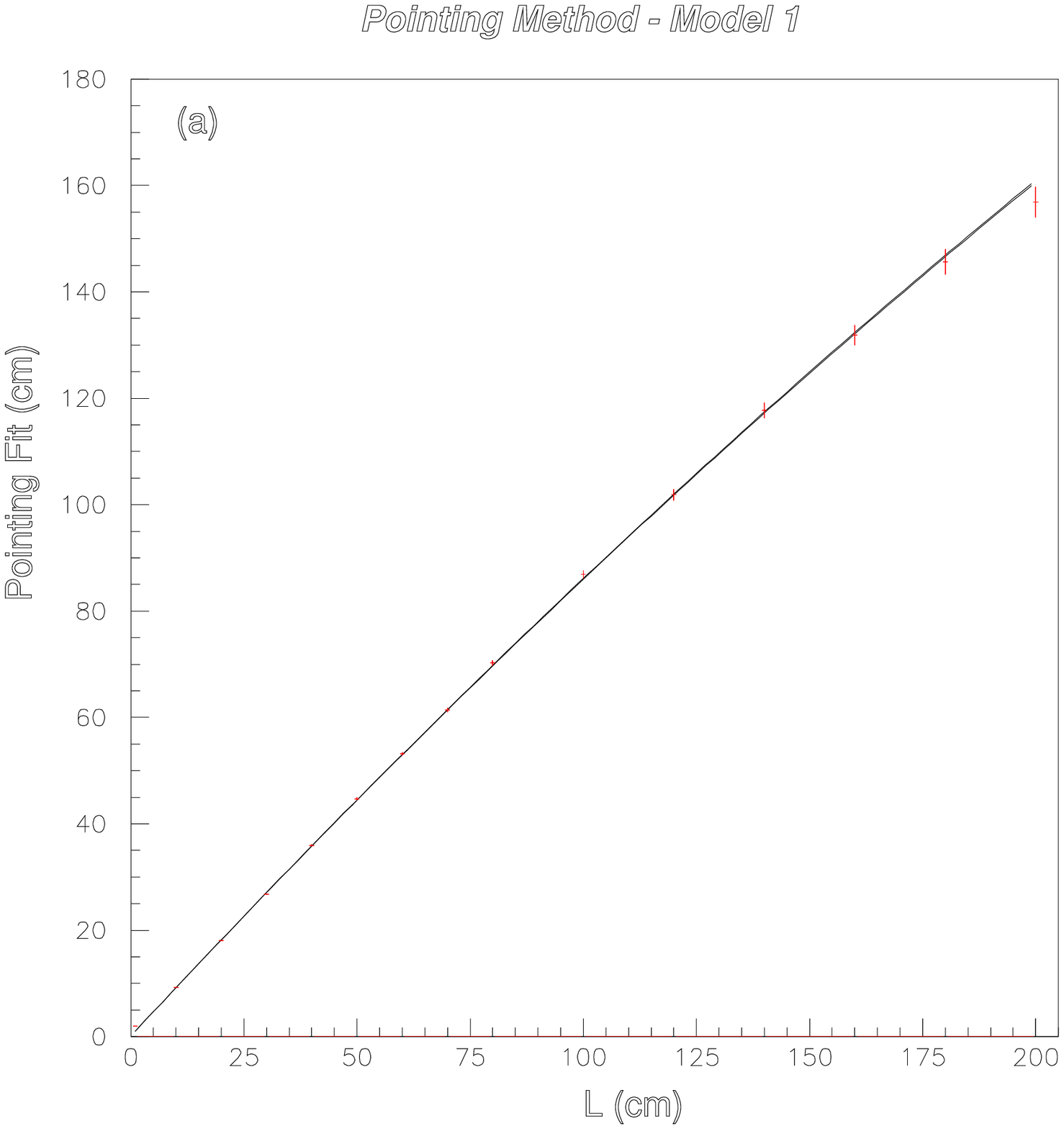} 
\epsfxsize=0.55\textwidth
\epsffile{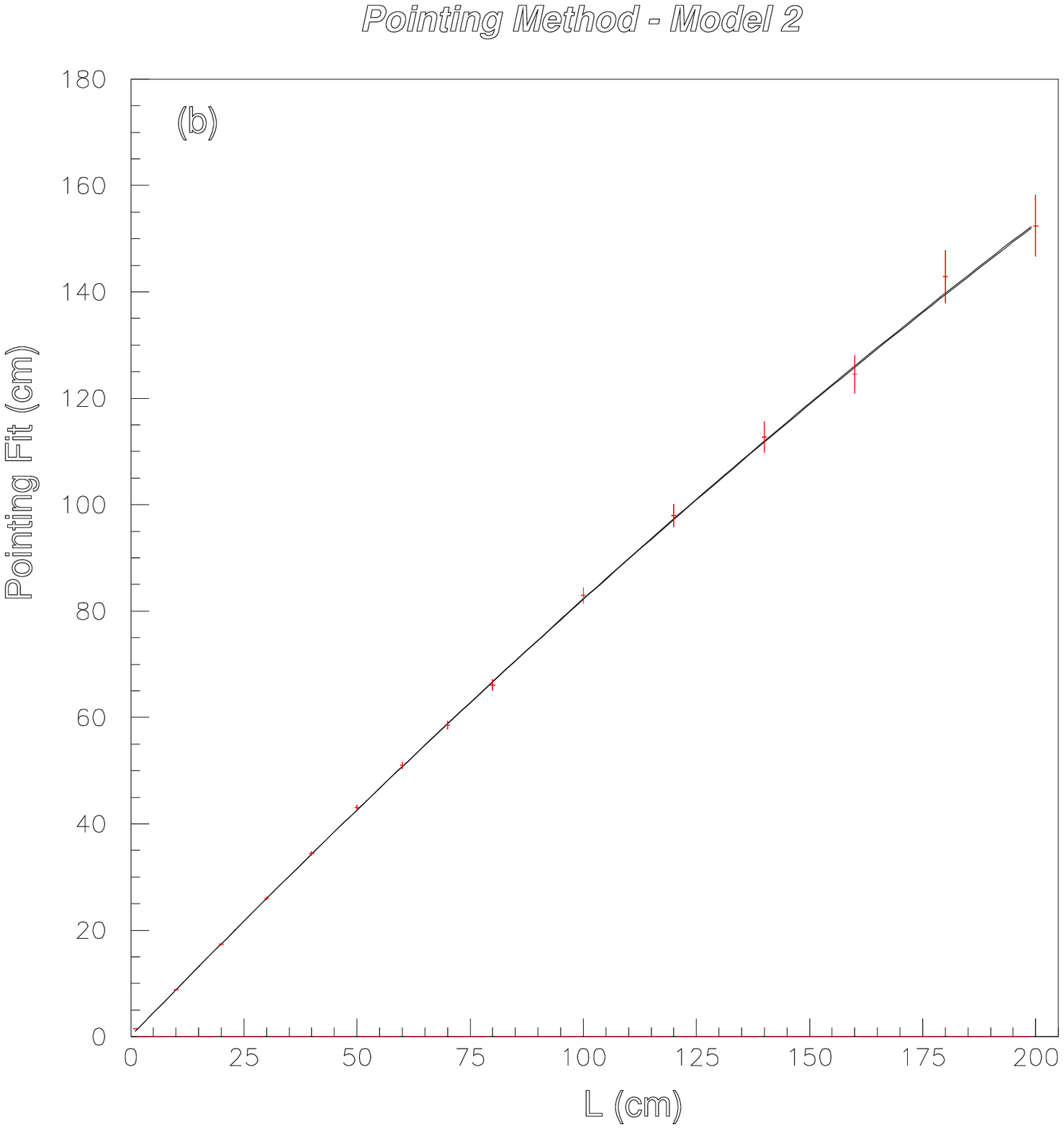} 
}
\caption{\sl 
Results of fits to decay lengths derived from pointing information.
The data points and error bars correspond (a) to a nominal 37,600 
neutralino pairs for  Model \#~1 and (b) 8,460 events for 
Model \#~2 (corresponding to 200 fb$^{-1}$ for each model) . 
The curves correspond to quadratic polynomial 
fits to the data for input neutralino masses of $\pm 100$ MeV from
the true mass. 
}
\label{fig:pointing}
\end{figure}

The results are shown in Fig.~\ref{fig:pointing}. 
The Monte Carlo includes effects of ISR and each point was obtained using
37,600 generated events before cuts for Model \#~1 and
8,460 events for Model \#~2, corresponding to 200 fb$^{-1}$ for each model
running at $\sqrt{s} = 270$ and 500 GeV, respectively. 
At each point, the reconstructed decay lengths $\lambda$ 
were fit to an exponential for $5 < \lambda < 200$ cm and the points show
the fit decay length with corresponding error. The $\lambda > 5$ cm
cut, corresponding to about $2\sigma$ for the resolution of this
method, serves to reduce any SM background to a negligible level.

The resulting calibration curve is well approximated
by a quadratic polynomial and the corresponding fit curves obtained 
assuming an error on the neutralino mass of $\pm 100$ MeV are shown
together with the points obtained using the correct input masses 
of 100 GeV for Model \#~1 and 200 GeV for Model \#~2.  
In this way the systematic error due to the mass uncertainty is shown
to be very small for $L$ less than 200 cm for both models. 
If we examine the lower statistics Model \#~2 
in the region of $L = 200$ cm, we find a statistical error of 6\% 
and in the region of 10 cm we find a statistical error of 1\%.
Our conclusion is that the calorimeter pointing method works very well for 
5 cm $< L < 2$ m.

\subsection{Measuring the NLSP Decay Length Using Calorimeter \\ Timing} 
\label{subsec:timing}

\noindent
Calorimeter timing information is highly desirable in any detector at a
LC in order to reject cosmic rays and many beam-related 
backgrounds as well as for its use in a trigger.  
In this section, we investigate a further use of sub-nanosecond timing for 
neutralino decay length measurement.

For the purposes of this study, we use the energy-dependent
timing resolution given in Tab.~\ref{tab:calparms}.
Referring to Fig.~\ref{fig:neutralino_decay}, a calorimeter timing
measurement gives the quantity $D+\lambda-R$. This timing information
is then combined with the position and energy measurements, all smeared 
according to the resolutions given in Tab.~\ref{tab:calparms}. 

\begin{figure}
\centerline{
\epsfxsize=0.55\textwidth 
\epsffile{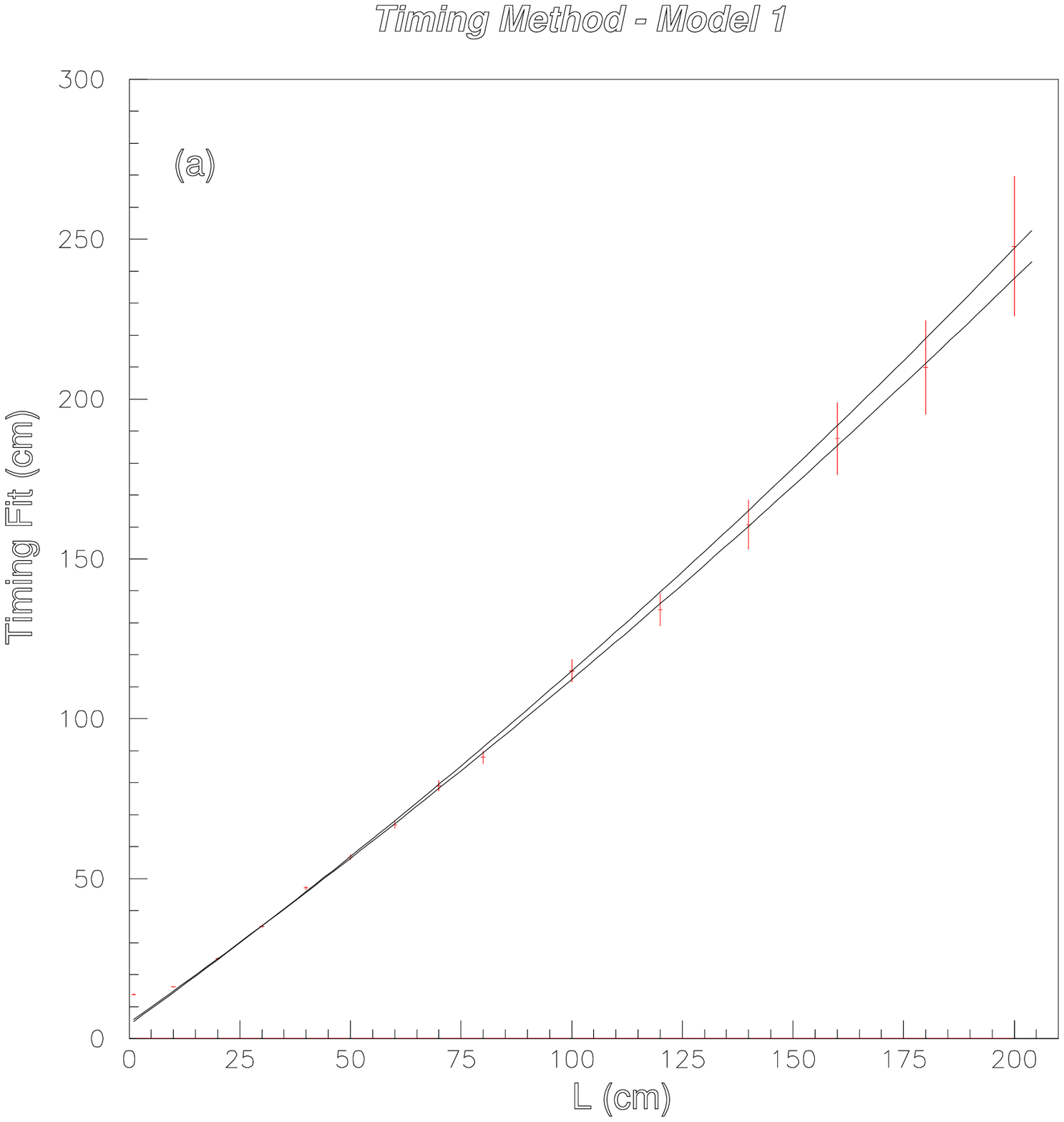} 
\epsfxsize=0.55\textwidth
\epsffile{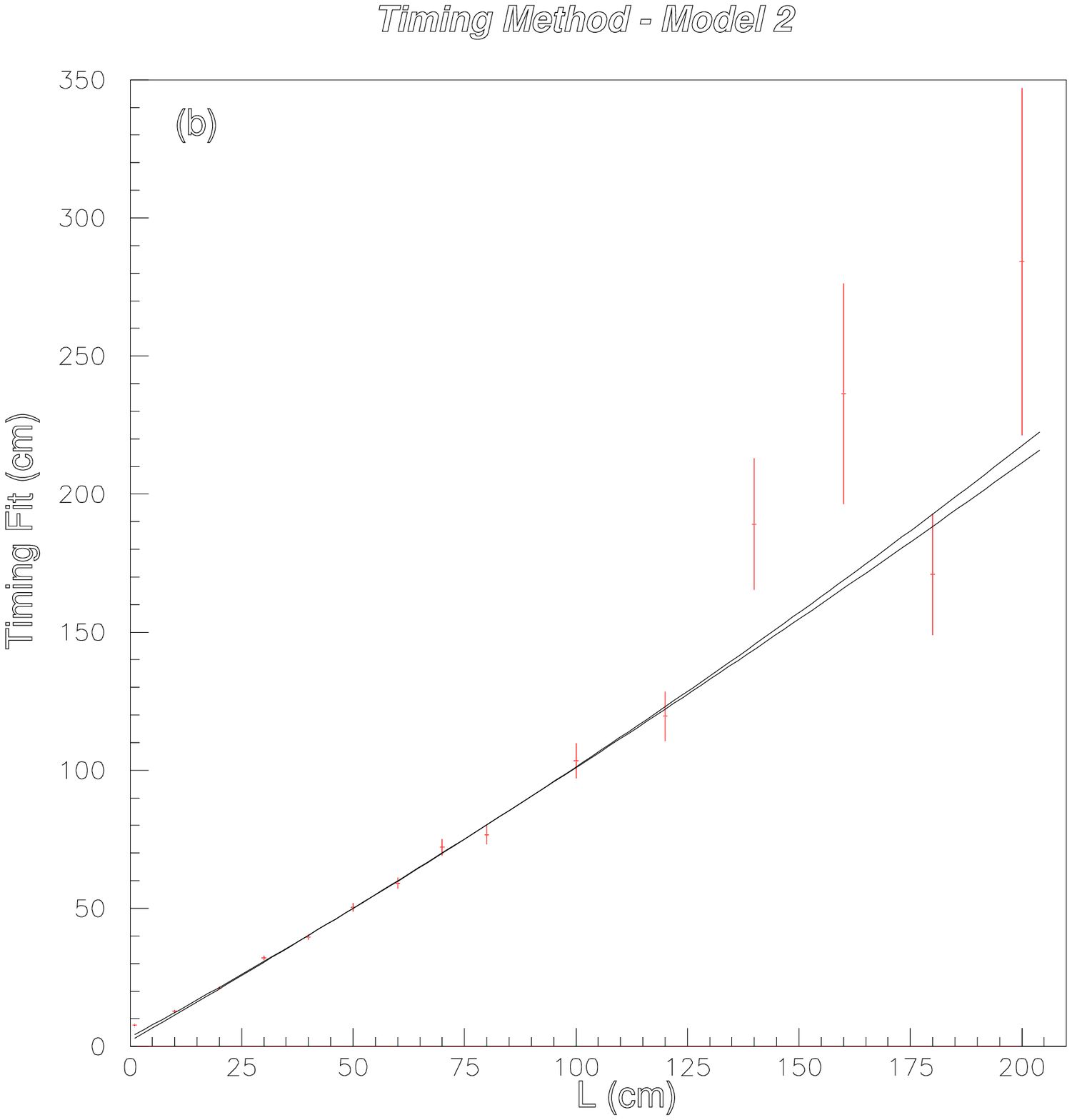} 
}
\caption{\sl 
Results of fits to decay lengths derived from timing information.
The data points and error bars correspond (a) to a nominal 37,600 
neutralino pairs for  Model \#~1 and (b) 8,460 events for 
Model \#~2 (corresponding to 200 fb$^{-1}$ for each model) . 
The curves correspond to quadratic polynomial 
fits to the data for input neutralino masses of $\pm$ 100 MeV from
the true mass. 
}
\label{fig:timing_calib}
\end{figure}

The timing resolution that we use is motivated by what could realistically 
be achieved for a large calorimeter. This is of the order of 0.5 ns, which 
naively implies an intrinsic photon impact parameter resolution of order
15 cm. Data sets for a series of decay lengths were simulated 
using the full Monte Carlo and the reconstructed decay length
distributions were fit to a simple exponential between
the lengths of 30 cm and 120 cm. The lower limit is given by the timing
resolution and the upper limit by the geometrical acceptance of the
detector. The results of the fits are shown in Fig.~\ref{fig:timing_calib}. 
The data points correspond to 200 fb$^{-1}$ 
(37,600 and 8,460 generated neutralino pairs at the usual c.o.m. energies 
for Models \#~1 and \#~2 respectively, before any cuts are applied). 
The decay lengths are determined on an event by event basis assuming the 
exact neutralino mass. The effect of a $\pm 100$ MeV shift in the input 
$m_{\NI}$ is indicated by the curves, which were obtained by fitting the 
resulting data points (not shown) to a quadratic polynomial.  
The results confirm that the effect of neutralino mass uncertainty on a 
lifetime measurement is small.
Values of $L$ below 10 cm are poorly determined due to the intrinsic 
timing resolution and large values of $L$ are limited by statistics;
the worst case is for Model \#~2, where statistics get poor for
$L > 120$ cm, so we take this as the upper limit for this method.

This study shows that timing information alone can not improve
on what could be achieved using calorimeter pointing and tracking 
methods. The timing information could of course be combined with
the tracking and pointing methods to achieve an improved accuracy,
but we do not consider this further here.
We stress, however, that timing should not be neglected in the overall 
detector design because it may be needed to reduce cosmic ray backgrounds 
and beam related backgrounds as well as being of use in the trigger for 
GMSB events. We will further comment on this in the next sections. 
It is possible that the NLSP is so long-lived that the photon decay
would appear in the ECAL several bunch crossings later than expected and
indeed not in time with any bunch crossing. To detect such events
good timing information would clearly be required.

\subsection{Measuring the NLSP Decay Length Using Statistics} 
\label{subsec:stat}

\noindent 
A measurement of the NLSP lifetime can also be made using a simple
counting technique, because the probability that, e.g., a photon from
the $\NI\to\gamma\G$ decay is observed in the ECAL is a function of 
the $\NI$ lifetime. As a consequence, the ratio of two-photon 
to one-photon events observed after SUSY production is a function of the 
$c\tau_{\NI}$.

As anticipated in Sec.~\ref{sec:NLSPdecay} for the spherical detector case, 
this function is in principle a simple combination of exponentials
convoluted with the effects of detector geometry, cuts
designed to eliminate SM backgrounds and calorimeter 
performance. We determine the functional dependence of the 
ratio using Monte Carlo techniques. Neutralinos were generated from 
pair production using our modified version of {\tt SUSYGEN} interfaced to a 
modified version of {\tt CIRCE} \cite{circe} to include full effects of ISR 
and beamstrahlung. 
Any photons coming from $\NI\to\gamma\G$ decays with energy 
loosely included in the range given by Eq.~(\ref{eq:eminmax}) 
and which originated within the detector tracking volume were
extrapolated and required to hit within the acceptance of the ECAL. 
Any event with more than two photons or less than one photon was
rejected at this stage. We also imposed cuts on the missing energy,
where appropriate, based on the expectations for the signal. 

The position of the photons and their energies and angles of pointing
were smeared according to the parameters in Tab.~\ref{tab:calparms}.
The resulting photon impact parameters at the i.p. were then calculated 
and at least one photon in the event was required to have an impact parameter 
greater than 30 cm. If one could believe in gaussian statistics to this level,
this would correspond to a $10 \sigma$ cut and is hence designed to reduce 
any SM background to a negligible level (we will further comment on this 
later in this section).

We define a one-neutralino event as containing
one large impact parameter photon and nothing else, and the
total number of these events is $n_{1\NI}$.  We define a
two-neutralino event as one where there is at least one large
impact parameter photon together with anything else visible in
the detector, and the number of these events is $n_{2\NI}$.
Note that $n_{2\NI}$ does not include two-photon events only, but
also events where both neutralinos decay visibly inside the detector;
one of them through $\NI\to\gamma\G$ and the other through any channel. 
Our choice is designed to maximise statistics, but we note that simple 
experimental cuts based on the event topology could easily remove 
any event which is not purely photonic, should that yield a cleaner analysis.
We checked that this would not result in an important loss of statistics
(up to about 15\% on $n_{2\NI}$ in the worst cases with a heavy $\NI$),
so that the essential conclusions of our analysis would remain unchanged. 
As for $n_{1\NI}$, instead, we only include photonic decays because we
rely on the presence of a non-pointing photon to eliminate the irreducible
radiative SM background, as discussed below.  

\begin{figure}
\centerline{
\epsfxsize=0.55\textwidth 
\epsffile{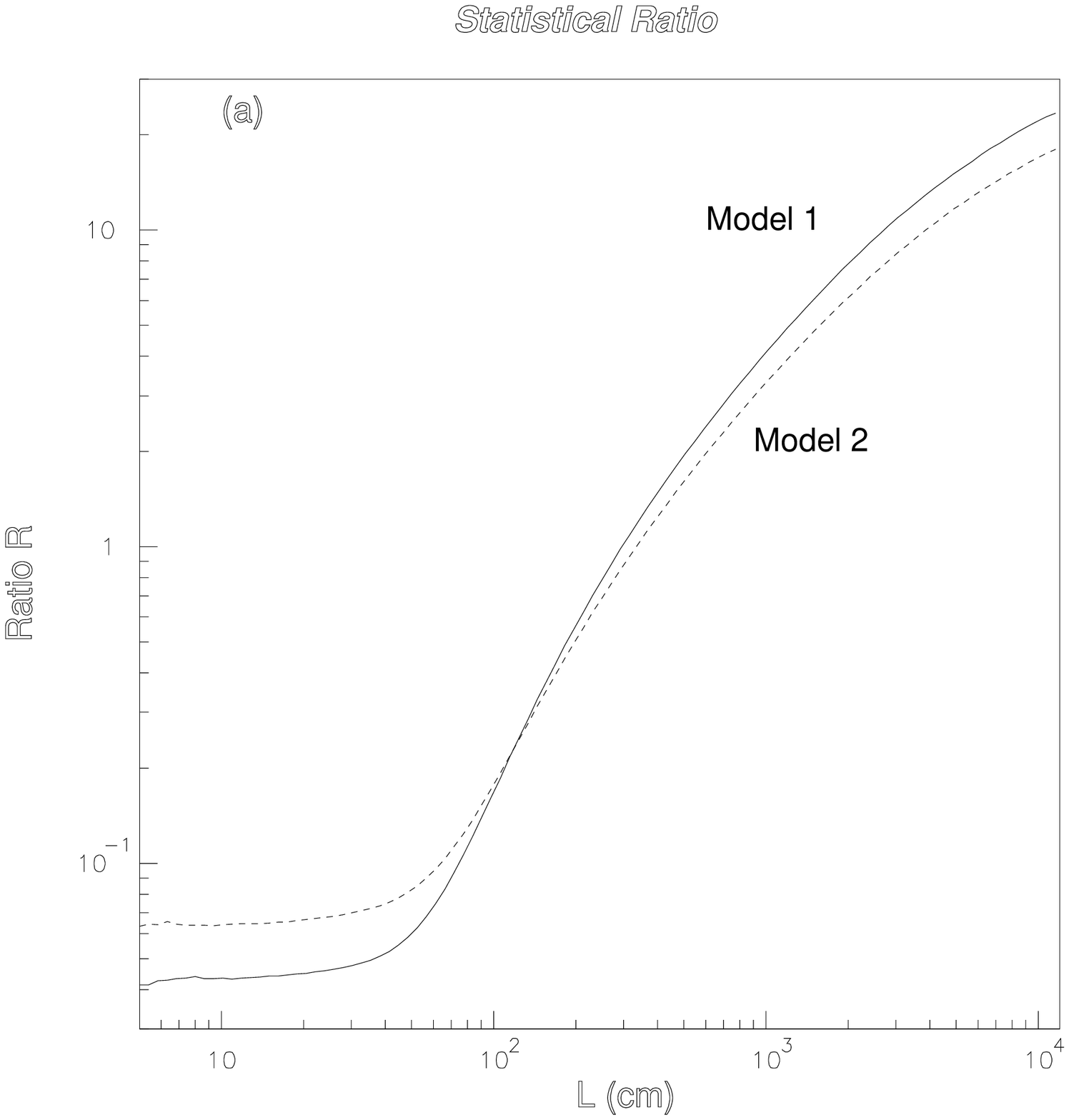} 
\epsfxsize=0.55\textwidth
\epsffile{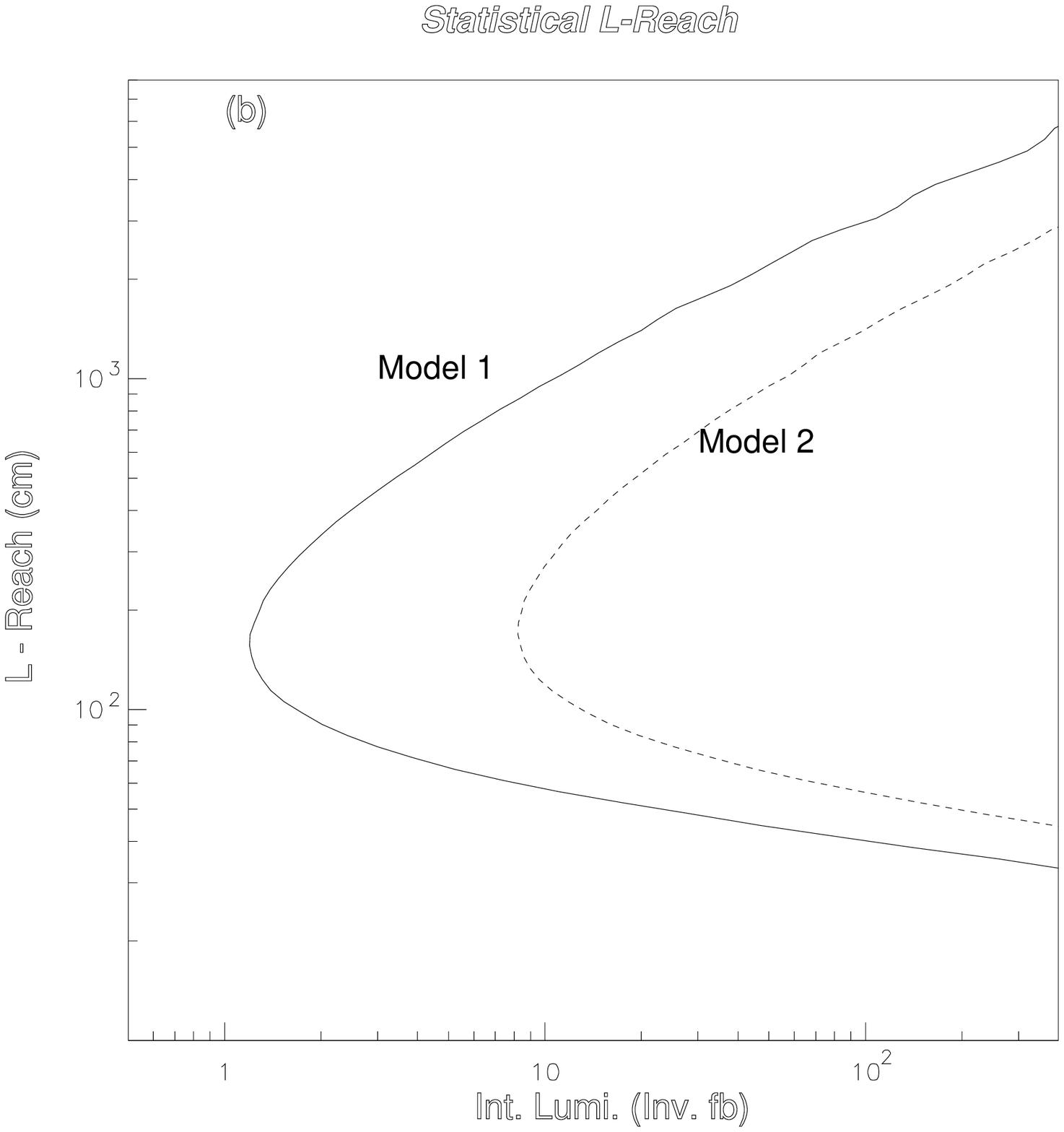} 
}
\caption{\sl
Plot (a) shows the statistical ratio functions for Models \#~1 and \#~2, 
at $\sqrt{s} = 270$ and 500 GeV respectively, as discussed in the text.
Plot (b) shows the decay length reach for the same models
as a function of integrated luminosity; the decay lengths which can be 
measured to an accuracy of 10\% are bounded by the curves.
}
\label{fig:statplots}
\end{figure}

The ratio $R_{\NI} = \frac{n_{1\NI}}{n_{2\NI}}$ 
was then determined as a function of the neutralino decay length for 
Models \#~1 and \#~2 by generating a nominal $10^7$ events running at
 $\sqrt{s} = 270$ and 500 GeV, respectively, to start with. 
The resulting functions are shown in Fig.~\ref{fig:statplots}(a).

In order to extract the precision $\Delta L$ on the measured 
decay length, we take the error $\Delta R^0_{\NI}$ on the measured 
ratio for the $10^7$ events to be:

\be 
\frac{\Delta R^0_{\NI}}{R^0_{\NI}} = \left(\frac{1}{n_{1\NI}} + 
\frac{1}{n_{2\NI}} \right)^{\frac{1}{2}} 
\ee

and then we obtain

\be
\frac{\Delta L}{L} = \frac{1}{L} 
\left(\frac{10^7}{n_{\rm tot}}\right)^{\frac{1}{2}}
\Delta R^0_{\NI} \left(  \frac{\partial R_{\NI}}{\partial L}\right)^{-1}
\ee

\noindent
where $n_{\rm tot}$ is the total number of expected events. Note that 
$n_{\rm tot} = \int\!{\cal L}\sigma(\NI\NI) \: dt$ is not equal to the 
sum of $n_{1\NI}$ and $n_{2\NI}$ because some events will not be 
counted, due to both neutralinos decaying outside the detector acceptance 
and/or to non-selected final states. 

\begin{figure}
\centerline{
\epsfxsize=0.55\textwidth 
\epsffile{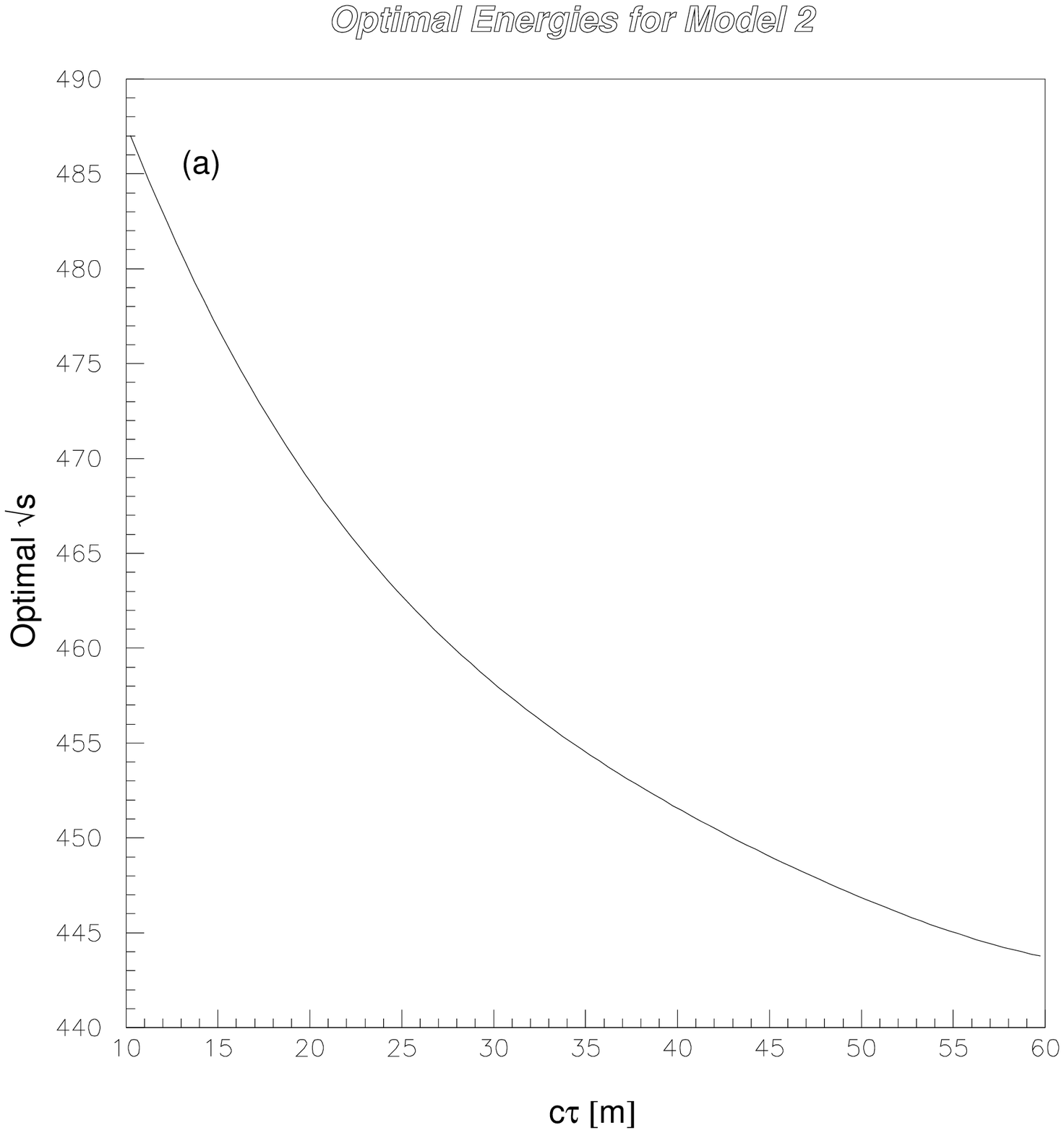} 
\epsfxsize=0.55\textwidth
\epsffile{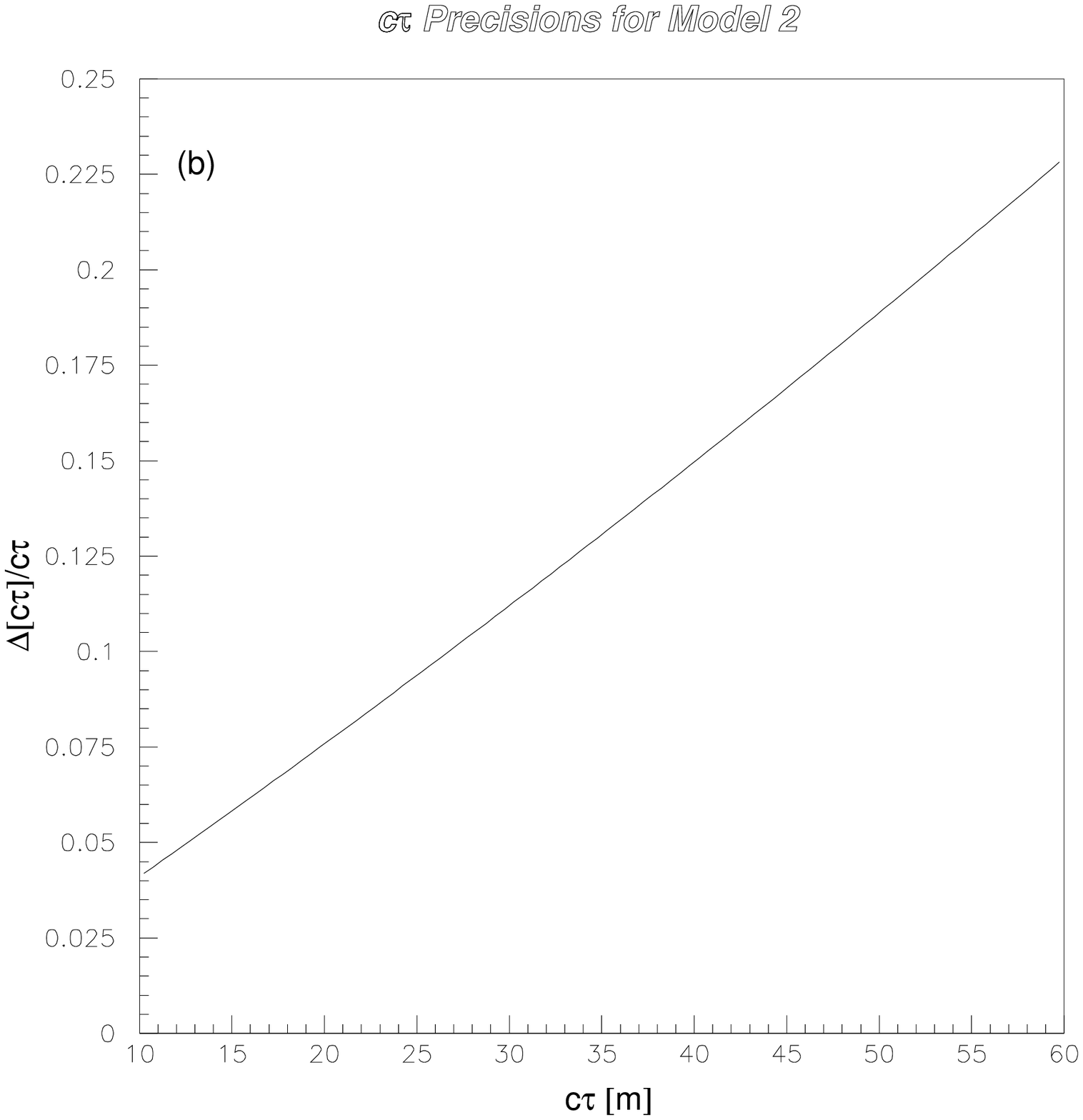} 
}
\caption{\sl 
Figure (a) shows the optimal c.o.m energy at which to run the LC
as a function of $c\tau$ for Model \#~2.  
Figure (b) shows the relative precision
of $c\tau$ obtained after a run of 200 fb$^{-1}$ at the corresponding
optimised c.o.m energy.
}
\label{fig:ctau_optimise}
\end{figure}

The resulting curves giving the decay length reach are shown in 
Fig.~\ref{fig:statplots}(b).
Assuming an integrated luminosity of 200 fb$^{-1}$, the plots show
that laboratory decay lengths are well measured by this technique down to 
approximately 50 cm for both models and that the upper limit ranges
from approximately 20 m for Model \#~2 up to 40 m for Model \#~1, 
corresponding to $c\tau$ values of 27 m and 44 m, respectively.   
It should be noted that the sensitivity of the statistical technique is
dependent on the c.o.m energy.  The statistics are determined by
the quantity $L=\beta\gamma c\tau$, whereas the parameter of physical 
interest is $c\tau$.  We have seen that if we take Model \#~2 with c.o.m. 
energy 500 GeV and allow a maximum measurement error 
$\left(\frac{\Delta L}{L}\right)_{\rm max} = 0.1$
then, as indicated above, we find that we can measure $L$ up to
20 m, corresponding to $c\tau_{\rm max} = 27$ m. 
If we now decrease the c.o.m energy, then we lose in cross section, but 
gain in intrinsic sensitivity.  
In this way, an optimal c.o.m energy can be found for each
value of $c\tau$ and this optimal energy is plotted against $c\tau$ 
for Model \#~2 in Fig.~\ref{fig:ctau_optimise}(a).
The corresponding optimised precision $\frac{\Delta(c\tau)}{c\tau}$ 
is shown in Fig.~\ref{fig:ctau_optimise}(b). 
Unfortunately, it turns out that, at least for Model \#~2, any overall
gain after optimisation is small. The larger value of $c\tau$ that can 
be measured with a precision of 10\% or better is again about 27 m,    
and the corresponding ``optimised'' c.o.m. energy is about 463 GeV. 
While this shows that optimisation gains tend to be small, it also 
tells us that the method is not critically sensitive to the c.o.m. 
energy, even though we are running in a threshold region. 

 To evaluate the level of background reduction we need for a meaningful
analysis, let us consider Model \#~2 with $c\tau = 27$ m and optimise 
the c.o.m. energy to 463 GeV. At this energy, the corresponding laboratory 
decay length is $L = 14.6$ m and the $\NI\NI$ production cross section 
is 25.5 fb. Using 200 fb$^{-1}$, we obtain the expected values of 
$n_{1\NI} = 825$ and $n_{2\NI} = 178$. 
The irreducible backgrounds would need to be subracted from the raw counts 
involving numbers of this order of magnitude, so we need to check 
that after a suitable set of cuts the number of background events left is 
small compared to the statistical errors of about $\sqrt{825} \sim 29$ for one 
neutralino events and $\sqrt{178} \sim 13$ for two neutralino events. 
Experience at LEP (see for example \cite{single_photon}), has shown that 
backgrounds from cosmic rays and detector noise (sparks and radioactive decays)
can be reduced to negligible levels compared to the physics backgrounds by 
requiring the event timing to be consistent with a beam-crossing.
Leaving aside the minor case of events counted in $n_{2\NI}$
not including 2 photons (they are easily treatable anyway), the SM physics 
backgrounds are expected to be dominated mainly by 
$\nu\bar{\nu}\gamma(\gamma)$ events and, to a lesser extent, by 
radiative bhabha's. It thus remains to check that the cuts applied in the 
counting procedure reduce the expected number of $\nu\bar{\nu}\gamma(\gamma)$ 
events to a ``few'' events.
For this purpose, we performed a quick evaluation of the cross sections for 
these processes using {\tt CompHEP}\footnote{At this stage, we did not take 
into account ISR, beamstrahlung effects or any other effects coming from 
undetected radiation} and requiring the following for each photon:
a) $\left|\cos\theta_\gamma\right|<0.95$; b) $E_\gamma$ included 
in the range (\ref{eq:eminmax}). We then found that: 

\beas 
\sum_i \sigma_{\rm CUT}
(\epem\to\nu_i\bar{\nu}_i\gamma) & \sim & 810 \; {\rm fb} \; 
{\rm at} \; {\rm \sqrt{s} = 270} \; {\rm GeV}; \\
                                              &      & 570 \; {\rm fb} \; 
{\rm at} \; {\rm \sqrt{s} = 500} \; {\rm GeV}; \\
\sum_i \sigma_{\rm CUT}
(\epem\to\nu_i\bar{\nu}_i\gamma\gamma) 
                                              & \sim & \;\;31 \; {\rm fb} \; 
{\rm at} \; {\rm \sqrt{s} = 270} \; {\rm GeV}; \\
                                              &      & \;\;\;\;7 \; {\rm fb} \;
{\rm at} \; {\rm \sqrt{s} = 500} \; {\rm GeV}.
\eeas

On top of this, one has to take the 30 cm cut on the non-pointing photon
impact parameter into account. Of course, it is not reasonable to trust the
tails of gaussian statistics so much as to consider this as a real $10\sigma$
cut. Nevertheless, we note that if we were to ``conservatively'' evaluate its  
reduction effect and assume it to be equivalent to an effective $4\sigma$ 
cut, the one photon background in 200 fb$^{-1}$ would still amount to 
10 (7) events only for Model \#~1 (2), while the two photon background
would basically disappear. Hence, we are confident that our set of cuts
is strict enough to reduce any background to the level needed for our
study to be valid. 

 While we are aware that detailed studies on non-gaussian tails for 
calorimeter angular pointing would be highly desirable in order to 
solve more precisely the background subtraction problem in such a 
statistical study, we also note that further background reduction
is possible before appealing to photon non-pointingness. 
First, with reference to the most dangerous $\nu\bar{\nu}\gamma$ 
background, we checked that, after our upper cut on the photon energy
in Eq.~(\ref{eq:eminmax}), the remaining contribution from 
$\epem\to\gamma(Z^*\to\nu\bar{\nu})$ is small and more than 70\% of the 
cross section at $\sqrt{s} = 270$ GeV comes from pure $W$-exchange (even more 
than that at $\sqrt{s} = 500$ GeV). On the other hand, $W$-exchange graphs 
tend to produce relatively soft photons predominantly along the beam 
direction, while the $\NI\NI$ signal gives a flat $E_\gamma$ spectrum between 
the end points (\ref{eq:eminmax}) and does not show an important angular 
structure, due to the isotropy of the $\NI$ decay. Based on this, we 
found that, e.g. for the case of interest for Model \#~1 with $\sqrt{s} = 270$
GeV, imposing stricter cuts $E_\gamma > 40$ GeV and 
$\left|\cos\theta_\gamma\right|<0.9$ would result in a factor 2 to 3 gain
on the signal to background ratio, while the corresponding loss of statistics
would be limited to less than 30\%. 

 Second, and most important, there is here a good chance of exploiting the 
beam polarisation option at the LC. If, on top of the kinematical
cuts mentioned above, one could benefit from, say, an 80\% polarisation 
for the electron and a 60\% one for the positron, one would get an additional 
0.3 or so reduction factor on the dominant $W$-exchange contributions 
to the one photon background, while the $\NI\NI$ signal would get an 
enhancement factor, since an important part of it comes from $R$-selectron 
exchange in the $t$-channel (we remind that our neutralinos are mostly
bino's). If even stronger beam polarisation was available, then one would
probably be able to reduce the one photon SM background to an acceptable 
level by using optimised kinematical cuts and a looser cut on the photon 
impact parameter such that gaussian statistics could still be trusted. 

 On the other hand, we checked that the results we obtained for the upper 
end of our statistical method reach on $c\tau_{\NI}$ is not very sensitive 
to the cut we impose on the photon impact parameter in the 10's of cm range.
The lower end of the $c\tau$ reach does, but we know from the previous section
that the $ L = 10$ cm--1 m range is well covered by the direct calorimeter 
pointing technique. All these considerations make us confident that our
statistical study is meaningful and our claim that a measure of $c\tau$ 
with a precision at the level of 10\% is possible for typical GMSB models 
up to several tens of metres by using such a method is safe and possibly even 
slightly conservative. 

 Finally, we would like to comment on the fact that larger values of $c\tau$ 
might be accessible by considering a statistical analysis based on the ratio 
$R^{\prime}_{\NI} = n_{1\NI}/n_{0\NI}$ 
of one (non-pointing) photon events to events where both neutralinos 
decay outside the detector. Of course, in order to count the latter events, 
it is necessary that other SUSY processes are within kinematical reach in 
addition to $\NI\NI$ production, so that events including the visible products
of the decays to the NLSP are present.
Taking Model \#~1 as an example, one could think of running the LC at 
$\sqrt{s} = 350$ GeV or so, allowing $R$-slepton pair production followed
by $\lR\to\ell\G$ \ ($\ell=e$, $\mu$, $\tau$) decays with 100\% BR. 
The resulting SUSY signal would then be made up mainly of 
$(\gamma\gamma)\slashchar{E}$, 
$\epem(\gamma\gamma)\slashchar{E}$, 
$\mu^+\mu^-(\gamma\gamma)\slashchar{E}$,
and $\tau^+\tau^-(\gamma\gamma)\slashchar{E}$
events, where one or both of the non-pointing photons in parentheses 
coming from $\NI\to\gamma\G$ may or may not originate within the detector, 
depending on $c\tau_{\NI}$ and the $\NI$ boost for the specific process. 
For very large neutralino lifetimes, events including
two photons are very rare and can be neglected. Note that running at 
higher c.o.m. energies would increase the $(\beta\gamma)$ 
factor for $\NI$'s coming directly from pair production, resulting in
a reduction of $n_{1\NI}$ from this source, whereas the other processes 
would produce softer neutralinos with a relatively larger probability of 
decaying within the detector and contributing to $n_{1\NI}$. In addition, 
the latter processes would provide all the visible SUSY events to be 
counted as $n_{0\NI}$. 
Hence, the dependence of $R^{\prime}_{\NI}$ on $c\tau$ would now 
be affected by the sparticle spectrum and model details, which renders
this option less general and more involved than the statistical method
based on $R_{\NI} = n_{2\NI}/n_{1\NI}$.
In addition, while the SM background to events counted in $n_{1\NI}$ 
can still be reduced by requiring large impact parameters for the 
single photon, such a drastic procedure is not available to reduce the SM 
contribution to $n_{0\NI}$.
In the absence of a detailed mSUGRA-like analysis, this would lead to
a reduction in the $R^{\prime}_{\NI}$ sensitivity to $c\tau$.
By quick inspection of Fig.~\ref{fig:decays_in_tpc} for Model \#~1 
at $\sqrt{s} = 270$ GeV, one can get a feeling of the $n_{1\NI}$ dependence 
on $L$. Although one should take into account the precise definition of 
$n_{1\NI}$ and the differences arising from going to e.g. $\sqrt{s} = 350$ GeV 
from 270 GeV, one can still estimate that neutralino lifetimes of the order of
1 km or possibly more might be measurable with some precision by using the 
$R^{\prime}_{\NI}$ statistical method. 
To further increase statistics and the reach in $c\tau$, one could even 
consider going to the highest available c.o.m. energies and include 
all possible processes and cascade decays in the counts. 
However, this would involve very complex analyses which might  
be desirable only at a later stage, once it was clear that either the 
value of $\sqrt{F}$ to be measured is much larger than the bound 
suggested by the simple cosmology condition $m_{\G} \ltap 1$ keV or 
that only an upper limit on $c\tau_{\NI}$ can be set. 
A statistical analysis along these lines, as well as other studies 
concerning measurements of GMSB parameters at the LHC will also appear 
soon \cite{Paige}.

\subsection{Backgrounds} 
\label{subsec:bkgd}

\noindent 
In this section, we address again the issue of background subtraction
and summarize our strategy in this respect for the various techniques 
described above. 

Starting with the very short $\NI$ lifetime and 2D-tracking method case, 
we note that there are no backgrounds from conversions, because we include
only those vertices that occur well within the beampipe with 
$r < 1$ cm ($r$ being the radial coordinate of the vertex)  
in the exponential fit. Also, any tracks originating from the i.p. should 
not contaminate the signal, since we impose a cut $r > 10$ $\mu$m 
when fitting. We also require the invariant mass of the pair of tracks
be either less than 10 GeV or in the ($91\pm 5$) GeV range (to retain the
$\Z$ peak in the case of Model \#~2) to remove any chance of trace 
backgrounds from leptonic $W$-pair decays with a vertex reconstructed 
at a distance greater than 10 $\mu$m from the i.p., due to resolution 
effects. We also require both leptons to be of the same flavour, which
further reduces this background. As for the 
$\Z(\to\nu\bar{\nu})\Z(\to\ell\ell)$ background, we note that in this case the
large $\ell\ell$ invariant mass requires a significant opening angle between 
the tracks, which makes our 10 $\mu$m cut on $r$ very severe.   
Any remaining background must then originate from particles 
with lifetime, such as $\tau$'s. However, $\tau$'s can only mimic our
signal if they decay into 3- (or 5-) prong channels (with 15\% BR), 
which are also mistakenly reconstructed as 2-track events. 
The effect should be small, but any final detector design should be
checked to have sufficient 2-track resolution to ensure that it is 
negligible. We did not address this and the above problems in detail, 
because in our study we can safely require the additional presence of 
a hard [$E_\gamma > 20$ (45) GeV for Model \#~1 (2)] photon coming from 
the decay of the other neutralino and more than 40 (90) GeV of 
missing energy, together with selecting only those events where there is
nothing else apart from the photon and the reconstructed track pair.  

As for the short decay length and 3D-tracking method, the only
additional feature compared with the previous case, is the presence
of $\gamma$ conversions that may or may not be used for the lifetime
measurement. We conservatively decided to remove them and we describe
our procedure for this in detail in Sec.~\ref{subsec:3d}. Note also
that our cuts here are on the full 3D decay length $\lambda$ rather 
than on the projection $r$. 

For intermediate decay lengths (calorimeter pointing/timing methods), 
the main background arises from the $\gamma\gamma\nu\bar{\nu}$ process.
All the photons arising from this channel will point to the i.p.
and so can be reduced by cutting on the reconstructed decay length $\lambda$.
In Sec.~\ref{subsec:pointing}, we achieved this by starting our 
exponential fits at $\lambda=5$ cm, which corresponds to roughly a 
2$\sigma$ cut for the pointing accuracy assumed in our study.  
In Sec.~\ref{subsec:stat}, we found that 
$\sum_i\sigma(\gamma\gamma\nu_i\bar{\nu}_i)$ at $\sqrt{s} = 270$ GeV 
(for Model \#~1) is roughly 31 fb after appropriate cuts, 
which means that about 1.5 fb are left after the $2\sigma$ cut.
This implies that the remaining background is less than 1\% 
of the signal and so we are safe in our claim that a measurement
to 10\% or better in the lifetime measurement is possible using
these methods. It should be noted that we have assumed gaussian angular 
resolution for photon pointing and any effects due non-gaussian tails have 
been neglected. Such residual effects in angular pointing should be included 
in any future detailed calorimeter studies for the LC detector.

For large decay lengths using the statistical method, we have to address
the problem of both  $\gamma\gamma\nu\bar{\nu}$ and the more severe
$\gamma\nu\bar{\nu}$ backgrounds. 
Any backgrounds due to SM processes will be independent
of the neutralino decay length and so the SM contribution can be estimated
from Monte Carlo and be subtracted from the measured values of $n_{1\NI}$ 
(and $n_{2\NI}$) in our curves in Fig.~\ref{fig:statplots}, provided the 
subtractions are relatively small. To this purpose, 
in addition to cuts on the photon and missing energies designed to fit  
the spectrum of the signal, we require a very strict 30 cm cut on the 
non-pointing $\gamma$ impact parameter to reduce the $\gamma\nu\bar{\nu}$ 
background to a negligible level. This is again somewhat dependent on 
neglecting non-gaussian tails in the calorimeter resolution. However, 
in Sec.~\ref{subsec:stat}, we also discussed in detail possible alternative 
strategies involving kinematical cut optimisation and beam polarisation.  

There remains the question of backgrounds from cosmic rays and beam-related
background. As we have mentioned in Sec.~\ref{subsec:stat}, experience 
from LEP has shown that cosmics can be rejected provided one can apply
good timing information and require no hadronic energy in the event.  
In addition, the background from cosmics (and beam-related backgrounds 
or calorimeter noise) can be estimated using random dedicated triggers 
and subtracted if necessary.  Additional veto walls could be constructed
around the detector to kill any remaining muonic background, if the
above measures proved insufficient.

\subsection{Summary} 
\label{subsec:sum}

\noindent 
We have shown in this section that a general purpose detector 
operating at a LC with c.o.m. energies in the 200--500 GeV range
could provide a good $\NI$ lifetime measurement, to 10\% or better,
over 6 orders of magnitude, from tens of $\mu$m to tens of m for a 
few representative scenarios. We have based this on a 200 fb$^{-1}$ 
run, corresponding, e.g., to approximately 1 year at the high luminosity 
proposed for TESLA and a few years at JLC/NLC. 

\begin{figure}
\centerline{
\epsfxsize=10.0cm
\epsffile{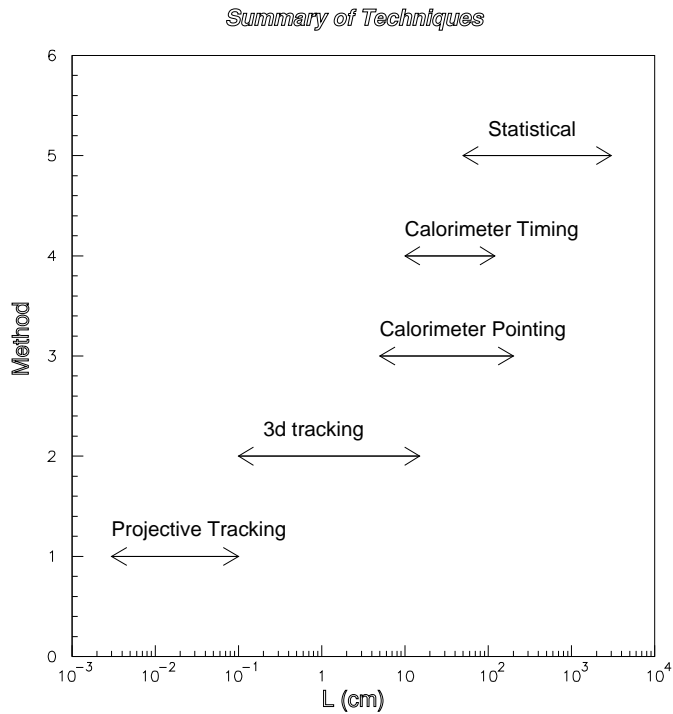}
}
\caption{\sl 
Summary of the various techniques we have proposed to use at a LC
for a $\NI$ lifetime measurement with a precision at the level of 
10\% or better.  
}
\label{fig:detector_summary}
\end{figure}

 The techniques we have used in the study are summarised
in Fig.~\ref{fig:detector_summary}, as a function of the $\NI$ 
average decay length. 
 It can be seen that while a wide $L$ range (e.g. 30 $\mu$m--40 m 
for Model \#~1) can well be covered by using individual techniques, 
there is also significant overlap for many of the intermediate regions. 
This implies important redundancy in the lifetime measurement and the 
opportunity to combine results to achieve an even greater precision.

 For many cases (e.g. Model \#~1), this is more than enough to cover 
the lower end of the theoretically allowed range for $c\tau_{\NI}$ and
the upper end as well, if naive cosmological constraints are imposed.
On the other hand, we recognise that scenarios with very short lifetimes 
at the level of a few $\mu$m are possible in the context of GMSB and 
in this case one can either set an upper limit or push for higher
LC energies to maximise the relativistic factor $(\beta\gamma)_{\NI}$ 
(facing however more complex analyses in the presence of competing SUSY 
signals). 

 We have also considered scenarios where the gravitino mass is heavier 
than $\sim 1$ keV and consequently the neutralino has a very long lifetime
$\gtap 10$'s of metres. For these cases, we have proposed a way to
slightly extend the $c\tau$ reach by running the LC at a c.o.m. energy 
optimised with respect to the relativistic boost factor and a simple 
statistical analysis, but we have seen that a solid improvement  
can only come from increasing the available integrated luminosity. 
Interesting prospects for a measurement of very large values of 
$c\tau_{\NI}$ up to $\sim 1$ km also exist, based on a more complex 
statistical analysis in a SUGRA-like scenario where most produced NLSP's 
appear stable and are invisible, but there is still a non-negligible number 
of neutralinos undergoing a decay within the detector. 

 Referring to Eq.~(\ref{eq:NLSPtau}), we note that a 10\% error
in $c\tau$ corresponds to a 3\% error in $\sqrt{F}$.
This is of the same order of magnitude as the theoretical 
uncertainty on the factor ${\cal B}$ introduced in 
Sec.~\ref{sec:NLSPdecay}. In comparison, the contributing error from 
the neutralino mass measurement using the threshold-scanning
technique or the method of Sec.~\ref{subsec:N1mass} is negligible. 

 Hence we conclude that, for the models considered and under 
conservative assumptions, it is possible to determine $\sqrt{F}$ 
with a precision of approximately 5\% by only performing $\NI$ 
lifetime and mass measurements in the context of GMSB with neutralino
NLSP. Less model dependent results can be obtained by adding information
on the $\NI$ physical composition from other observables, such as 
$\NI$ decay BR's, cross sections etc.

\section{Conclusions} 
\label{sec:conc} 

\noindent 
After introducing the GMSB framework and discussing the region of the 
parameter space of interest for LC searches, we focused on the case
of a neutralino NLSP and demonstrated how measurements made at the LC
can provide information on the detailed structure of the theory at 
both the electroweak and the very high energy scales. 

 We have shown how a study of the SUSY particle mass spectrum,
measured e.g. via threshold-scanning techniques, can allow the 
determination of the fundamental GMSB parameters with high precision.
In an explicit example where many SUSY thresholds can be explored 
running at $\sqrt{s} \le 500$ GeV, we found that accuracies at 
the level of 0.1\% for $\Lambda$ and $N_{\rm mess}$, 1\% for $\tan\beta$ 
and 1--2\% for $M_{\rm mess}$ are achievable. 

 In particular, we stressed the possibility of performing a measurement 
of the $\NI$ lifetime at the LC with the aim of extracting information also 
on the SUSY breaking sector of the theory and its fundamental scale $\sqrt{F}$.
To this purpose, we studied main and rare $\NI$ decay channels in detail and 
set up specific simulation tools based on the TESLA/CDR proposals. 
Using representative GMSB models, we then found that $c\tau_{\NI}$ can be 
measured to 10\% or better over a large range, from tens of $\mu$m to tens 
of m, which in many cases covers the scenarios allowed from the theory and 
suggested by simple cosmological arguments. (We also sketched a possible
way to treat the case where 10's m $\ltap c\tau_{\NI} \ltap 1$ km.)   
We showed that this, together with a $\NI$ mass measurement to $\sim 0.1\%$ 
(whose feasibility at the LC we also demonstrated), yields a 5\% determination of 
$\sqrt{F}$ under conservative assumptions, if minimal GMSB-model constraints 
are used. Better and less model dependent results can also come from analyses 
of other observables such as $\NI$ decay BR's, cross sections etc. 

 Our results in this respect and the reach in $c\tau_{\NI}$ depend on 
details of the detector design, which we assumed to be general-purpose 
oriented. For instance, for intermediate $\NI$ lifetimes, we obtained a 
very high accuracy using calorimeter pointing/timing techniques. 
It should be noted, however, that our calorimeter pointing 
precision has assumed the presence of pre-shower detectors, while simple 
calorimetry alone might not be sufficient for our purpose. 

 On the other hand, extreme cases allowed by the GMSB framework with very 
short (few $\mu$m) or very long ($\gtap 1$ km) $\NI$ lifetime require special 
detector design. For the very short case, one should consider the feasibility 
of upgrading the interaction region to include smaller beampipe radii combined
with ultimate vertexing technology. Better performances for the very long case
can only be achieved by having electromagnetic calorimetry at larger 
distances from the i.p.. This points to either a larger 
detector or dedicated additional devices, e.g. lead scintillator arrays 
or similar, possibly well separated from the main detector 
(a proposal in this direction for the D0 experiment at the Tevatron has 
already been discussed in Ref.~\cite{D0roof}). 
If clear SUSY signals are detected at the LHC and/or during the first
phase of LC operations, then it should be possible to distinguish scenarios 
with a stable $\NI$ (e.g. mSUGRA) from GMSB scenarios with a very long-lived
$\NI$ NLSP and large $\sqrt{F}$ by measuring and studying the sparticle 
spectrum or other observables. If the indications favour a GMSB pattern, then 
the addition of dedicated devices at large distances would be highly
desirable for later LC runs. In this case, the position of the detector in 
the experimental hall as well as the dimensions of the hall itself should be       
cleverly designed to allow for such improvements. 

 Our study clearly shows that high luminosity at the LC, such as 
that proposed for TESLA is very desirable in order to allow both
a better accuracy in extracting the GMSB parameters from detailed 
knowledge of the sparticle spectrum and a good precision measurement 
of the NLSP lifetime in (most of) the range suggested by theory. 
Indeed, we checked that any reduction in luminosity would 
significantly eat away at both ends of the $c\tau_{\NI}$ reach, while 
the situation would improve perceivably if even higher luminosities 
could be available. 

 We have pointed out that the clean environment and the flexibility
of an $\epem$ collider are ideal to achieve precise measurements in
this respect in a variety of GMSB scenarios. In particular, we made 
strong use of the LC ability of running at a variable c.o.m. energy 
to benefit from SUSY thresholds for both measuring the GMSB spectrum 
and for facilitating the $\NI$ lifetime measurement. Also, for the latter 
measurement, the precise knowledge of the $\NI$ production energy was 
essential and we note that this technique is not available at a
hadron collider. 

  Finally, we would like to stress that the present study turned out to
be an ideal benchmark for developing software intended
for the TESLA/CDR detector simulation that is expected to be broadly used 
for future analyses of general interest. We have indeed seen that all
parts of the detector are potentially involved in the measurement of 
the NLSP lifetime and unusual and extreme performance and precision 
have been often required, as e.g. in the analysis based on tracking for
the very short $c\tau$ case. Again, we note that the usefulness of the
software and algorithms we developed for our purposes in GMSB is in many cases 
extendable to other new physics scenarios, such as general LESB models 
or $R$-parity violation. Also, our analysis could hopefully trigger new,
non-standard ideas during the detector design process, in order to improve
the performance for the measurements we described without reducing 
the effectiveness of the apparatus for other tasks. 

\vspace{1.0cm} 

\leftline{\bf Acknowledgements}
\vspace{0.3cm} 
\noindent
We acknowledge useful conversations with 
Alan Caldwell, Thomas Gehrmann, Sven Heinemeyer, Graham Kribs, 
Steve Martin, Frank Paige, Tilman Plehn, Michael Pl\"umacher, 
Sasha Pukhov, Nicholas Walker, Katherine Wipf. 
We would like to thank especially Peter Zerwas for many suggestions 
as well as for constant encouragement and support.


\vspace{0.7cm} 

\begin{thebibliography}{10}

\bibitem{StevePrimer}
  For a recent pedagogical review of supersymmetry and supersymmetry 
  breaking, see S.~P.~Martin, ``A Supersymmetry Primer'', in ``Perspectives
  on Supersymmetry'', G.~L.~Kane ed., World Scientific 1998, hep-ph/9709356 
  and references therein. 

\bibitem{KKRW} See, e.g., G.~L.~Kane, C.~Kolda, L.~Roszkowski, J.~D.~Wells, 
  \PRD{49}{1994}{6173}. 

\bibitem{GR-GMSB}
  For a recent review, see G.~F.~Giudice, R.~Rattazzi, ``Theories with 
  Gauge-Mediated Supersymmetry Breaking'', hep-ph/9801271, submitted to 
  Phys. Rep.

\bibitem{Fayet}
    P.~Fayet, \PLBold{70}{1977}{461}; \PLBold{86}{1979}{272};
    \PLB{175}{1986}{471} and in ``Unification of the fundamental 
    particle interactions", eds.~S.~Ferrara, J.~Ellis,   
    P.~van Nieuwenhuizen (Plenum, New York, 1980) p.~587.

\bibitem{noscalemodels} 
    See, e.g., 
    J.~Ellis, K~Enqvist, D.~Nanopoulos, \PLB{147}{1984}{99};  
    A.~B.~Lahanas, D.~V.~Nanopoulos, \PREP{145}{1987}{1};
    J.~L.~Lopez, D.~V.~Nanopoulos, A.~Zichichi, 
    \PRL{77}{1996}{5168}; hep-ph/9610235; \PRD{55}{1997}{5813};
    J.~L.~Lopez, D.~V.~Nanopoulos, hep-ph/9701264; 
    J.~Kim et al., \PRD{57}{1998}{373}. 

\bibitem{oldGMSB}
    M.~Dine, W.~Fischler, M.~Srednicki, \NPB{189}{1981}{575};
    S.~Dimopoulos, S.~Raby, \NPB{192}{1981}{353};
    M.~Dine, W.~Fischler, \PLBold{110}{1982}{227};
    M.~Dine,  M.~Srednicki, \NPB{202}{1982}{238};
    M.~Dine, W.~Fischler, \NPB{204}{1982}{346};
    L.~Alvarez-Gaum\'e, M.~Claudson, M.~B.~Wise, \NPB{207}{1982}{96};
    C.~R.~Nappi, B.~A.~Ovrut, \PLBold{113}{1982}{175};
    S.~Dimopoulos, S.~Raby, \NPB{219}{1983}{479}.

\bibitem{newGMSB}
    M.~Dine, A.~E.~Nelson, \PRD{48}{1993}{1277};
    M.~Dine, A.~E.~Nelson, Y.~Shirman, \PRD{51}{1995}{1362};
    M.~Dine, A.~E.~Nelson, Y.~Nir, Y.~Shirman, \PRD{53}{1996}{2658}.

\bibitem{GMSBmodels1}
    S.~Dimopoulos, S.~Thomas, J.~D.~Wells, \PRD{54}{1996}{3283};
    \NPB{488}{1997}{39}.

\bibitem{GMSBmodels2}
    J.~A.~Bagger, K.~Matchev, D.~M.~Pierce, R.~Zhang, 
    \PRD{55}{1997}{3188}. 

\bibitem{AKKMM2}
  S.~Ambrosanio, G.~L.~Kane, G.~D.~Kribs, S.~P.~Martin and S.~Mrenna, 
 \PRD{54}{1996}{5395}. 

\bibitem{AKM-LEP2} 
    S.~Ambrosanio, G.~D.~Kribs, S.~P.~Martin, \PRD{56}{1997}{1761}. 

\bibitem{eegg-exp}
  S.~Park, ``Search for New Phenomena in CDF'', 10$^{th}$ Topical Workshop 
  on Proton-Antiproton Collider Physics, R.~Raja and J.~Yoh (eds.), 
  AIP Press, New York (1995); 
  CDF Collaboration, \PRL{81}{1998}{1791}; hep-ex/9806034. 

\bibitem{eegg-th}
  S.~Dimopoulos, M.~Dine, S.~Raby and S.~Thomas, 
  \PRL{76}{1996}{3494}; 
  S.~Ambrosanio, G.~L.~Kane, G.~D.~Kribs, S.~P.~Martin and S.~Mrenna, 
  \PRL{76}{1996}{3498}. 

\bibitem{LEPsearches} 
  For an overview of recent searches at LEP, see e.g.:  
  Joint LEP SUSY Working Group webpage {\tt http://www.cern.ch/lepsusy/}.  

\bibitem{D0searches}
  For an overview of recent D0 searhces at the Tevatron, see e.g.: 
  {\tt http://www-d0.fnal.gov/public/new/analyses/gauge}.

\bibitem{LESB-RunII}
  See, e.g., the LESB-GMSB Working Group webpage for the recent 
  ``Physics at Run II: SUSY/Higgs'' Workshop: 
  {\tt http://fnth37.fnal.gov/gm/gmrun2.html}. 

\bibitem{GMSB-LHC} 
  See e.g.: H.Baer, P.~G.~Mercadante, F.~Paige, X.~Tata, Y.~Wang,
  \PLB{435}{1998}{109}; I.~Hinchliffe, F.~E.~Paige, hep-ph/9812233. 

\bibitem{Cosmo} 
  See e.g.: H.~Pagels, J.~R.~Primack, \PRL{48}{1982}{223}. 

\bibitem{SAsoft}
  For inquiries about this software package, please send e-mail to  
  {\tt ambros@mail.cern.ch}. 

\bibitem{AKM2} 
  A relevant example is discussed in:
  S.~Ambrosanio, G.~D.~Kribs and S.~P.~Martin, \NPB{516}{1998}{55}.

\bibitem{Haber-Kane}
  H.~E.~Haber and G.~L.~Kane, \PREP{117}{1985}{75}, 
  and Erratum, SCIPP 85/47.

\bibitem{higgsinoNLSP} 
  See, e.g.: G.~Dvali, G.~F.~Giudice, A.~Pomarol, \NPB{478}{1996}{31}.

\bibitem{Steve-gen} 
  S.~P.~Martin, \PRD{55}{1997}{3177}.

\bibitem{LCRep} 
  E.~Accomando et al. [The ECFA/DESY LC Physics Working Group], 
  \PREP{299}{1998}{1}. 

\bibitem{LanHEP}
    {\tt LanHEP} was downloaded from the webpage \\ 
    {\tt http://theory.npi.msu.su/\ $\tilde{}$semenov/lanhep.html}. 
    See also A.~V.~Semenov, ``{\tt LanHEP}: A Package for Automatic 
    Generation of Feynman Rules'', hep-ph/9608488, and updates MSU-98-2/503.

\bibitem{CompHEP}
    E.~E.~Boos, M.~N.~Dubinin, V.~A.~Ilin, A.~E.~Pukhov, V.~I.~Savrin,
   ``{\tt CompHEP}: Specialized package for automatic calculation of
    elementary particle decays and collisions'', hep-ph/9503280, and
    references therein; P.~A.~Baikov et al., ``Physical results 
    by means of {\tt CompHEP}'', Proc. of X Workshop on High Energy 
    Physics and Quantum Field Theory (QFTHEP-95), B.~Levtchenko, 
    V.~Savrin eds. (Moscow, 1996), p.~101 (hep-ph/9701412).
    Version 3.3.18 is preliminary and was downloaded from the webpage 
    {\tt http://theory.npi.msu.su/\ $\tilde{}$comphep} 
    (courtesy of A.~E.~Pukhov).

\bibitem{MSSMlag} 
    The MSSM lagrangian created by means of {\tt LanHEP} (version last
    corrected on Nov. 14, 1998) was downloaded from  the webpage 
    {\tt http://theory.npi.msu.su/\ $\tilde{}$semenov/mssm.html}. 
    An older version of the lagrangian was presented in 
    A.~S.~Belyaev et al., hep-ph/9712303.

\bibitem{Steve-priv}
    S.~P.~Martin, private communication, unpublished. 

\bibitem{LCWorld} 
  See, e.g., the webpage: 
  {\tt http://lcwws.physics.yale.edu/lc}. 

\bibitem{CDR} 
  ``Conceptual Design of a 500 GeV $\epem$ Linear Collider
  with Integrated X-ray Laser Facility'', R.~Brinkmann, G.~Martelik, 
  J.~Rossbach, A.~Wagner eds., DESY 1997-048, ECFA 1997-182. 

\bibitem{tesla_parameters} 
  See the webpage \\ 
  {\tt 
  http://www.desy.de/\ $\tilde{}$njwalker/ecfa-desy-wg4/parameter\_list.html} 

\bibitem{threshold_slope}
  See e.g. Ref.~\cite{CDR}, vol.~1, pag.~219--222. 

\bibitem{Sven} 
  See e.g. 
  H.~E.~Haber, R.~Hempfling, A.~H.~Hoang, \ZPC{75}{539}{1997}; 
  S.~Heinemeyer, W.~Hollik, G.~Weiglein, hep-ph/9812472. 

\bibitem{Minuit}
  Consult the webpage \\ 
  {\tt http://wwwinfo.cern.ch/asd/cernlib/minuit}.

\bibitem{SUSYGEN} 
  {\tt SUSYGEN 2.20/03} was downloaded from the webpage \\
  {\tt http://lyohp5.in2p3.fr/delphi/katsan/susygen.html}.

\bibitem{BRAHMS} 
  For inquiries about this software package, please send e-mail to 
  {\tt g.blair@rhbnc.ac.uk} or consult the webpage \\ 
  {\tt http://www.hep.ph.rhbnc.ac.uk/\ $\tilde{}$ blair/detsim/brahms.html}. 

\bibitem{Geant321} 
  Consult the webpage 
  {\tt http://wwwinfo.cern.ch/asdoc/geantold/GEANTMAIN.html}.

\bibitem{circe}
  ``{\tt CIRCE} Version 1/02$\beta$: Beam Spectra for Simulating Linear 
  Collider Physics'', hep-ph/9607454. 

\bibitem{single_photon} 
  The ALEPH collaboration, \PLB{313}{1993}{520}.

\bibitem{Paige} 
  F.~Paige, private communication. These studies will be included 
  in the supersymmetry section of the chapter about the ATLAS 
  detector and physics performance in vol.~II of the 
  ``ATLAS Technical Design Report'', 1999. 

\bibitem{D0roof}
  C.~H.~Chen, J.~F.~Gunion, \PLB{420}{1998}{77}; \PRD{58}{1998}{075005}.

\end{thebibliography}
\end{document}